\documentclass[12pt, oneside, openany, a4paper, afrikaans, english, PhD]{usthesis}
\usepackage[hmargin={30mm,30mm},vmargin={30mm,30mm}]{geometry}
\usepackage{graphicx}
\usepackage{babel}
\usepackage{varioref}
\usepackage{eso-pic}
\usepackage{amsfonts}
\usepackage{amsmath}
\usepackage{pst-all}
\usepackage[toc,page]{appendix}
\usepackage{tikz}
\usepackage{enumitem}
\usetikzlibrary{matrix}

\newcommand{\mytilde}{\raise.17ex\hbox{$\scriptstyle\mathtt{\sim}$}}

\newcommand{\ket}[1]{|#1\rangle}
\newcommand{\rket}[1]{|#1)}

\newcommand{\inp}[2]{\left<#1|#2\right>}

\newcommand{\prm}{\hspace{-0.09cm}'}

\newcommand{\hs}[1]{\hspace{#1 cm}}

\newcommand{\Sc}[1]{$Schr(#1)$} 
\newcommand{\ssc}[1]{$schr(#1)$} 
\newcommand{\cg}[1]{$cgal_{\frac{2}{z}}(#1)$}

\begin{document} 

\title{Constructing Dualities from Quantum State Manifolds}
\faculty{Faculty of Science}
\author{HJR van Zyl}{Hendrik Jacobus Rust van Zyl}
\degree{PhD}{Doctor of Philosophy in Science}
\supervisor{Prof. \ F.G. \ Scholtz}
\cosupervisor{Dr. \ J.N. \ Kriel}
\setdate{12}{2015}
\TitlePage

\DeclarationDate{}
\DeclarationPage

\pagebreak

\begin{abstract}
A constructive procedure to build gravitational duals from quantum mechanical models is developed with the aim of studying aspects of the gauge/gravity duality.  The construction is simplified as far as possible - the most notable simplification being that quantum mechanical models are considered as opposed to quantum field theories.  The simplifications allow a systematic development of the construction which provides direct access to the quantum mechanics / gravity dictionary.  \\
The procedure is divided into two parts.  First a geometry is constructed from a family of quantum states such that the symmetries of the quantum mechanical states are encoded as isometries of the metric. Secondly, this metric is interpreted as the metric that yields a stationary value for the dual gravitational action.  If the quantum states are non-normalisable then these states need to be regularised in order to define a sensible metric.  These regularisation parameters are treated as coordinates on the manifold of quantum states.  This gives rise to the idea of a manifold ``bulk" where the states are normalisable and of a ``boundary" where they are not.  Asymptotically anti-de Sitter geometries arise naturally from non-normalisable states but the geometries can also be much more general.  \\
Time-evolved states are the initial interest.  A sensible regularisation scheme for these states is a simple complexification of time so that the bulk coordinate has the interpretation of an energy scale.  These two-dimensional manifolds of states are dual to models of dilaton gravity where the dilaton has the interpretation of the expectation value of a quantum mechanical operator.  As an example, states time-evolving under an $su(1,1)$ Hamiltonian is dual to dilaton gravity on $AdS_2$, in agreement with existing work on the $AdS_2/CFT_1$ correspondence.  These existing results are revisited with the aid of the systematic quantum mechanics / dilaton gravity dictionary and extended.  As another example, states time-evolving under an $su(2)$ Hamiltonian are shown to be dual to dilaton gravity on $dS_2$.  \\
The higher dimensional analysis is restricted, for computational reasons, to the example of states that possess full Schr\"odinger symmetry with and without dynamical mass.  The time and spatial coordinates are complexified in order to both regularise the states and maintain the state symmetries as bulk isometries.  Dictionaries are developed for both examples.  It is shown that submanifolds of these state manifolds are studied in the existing $AdS/CFT$ and $AdS/NRCFT$ literature.  
\end{abstract}

\begin{abstract}[afrikaans]
'n Konstruktiewe metode word ontwikkel wat swaartekragduale van kwantummeganiese modelle bou met die oog op die ondersoek van die yk / swaartekrag dualiteit.  Die konstruksie word sover moontlik vereenvoudig en spesifiek word kwantummeganiese modelle beskou in plaas van kwantumveldeteorie\"e.  Die vereenvoudigings laat 'n sistematiese ontwikkeling van die metode toe wat dus direkte toegang tot die kwantummeganika / swaartekrag woordeboek verleen. \\ 
Die metode bestaan uit twee dele.  Eers word 'n geometrie saamgestel vanaf 'n familie van kwantumtoestande wat die simmetrie\"e van die toestande as isometrie\"e behou.  Daarna word 'n aksie gedefinieer wat deur hierdie metriek stasion\^er gelaat word.  Indien die kwantumtoestande nie normaliseerbaar is nie moet hul op 'n gepaste wyse geregulariseer word.  Die regularisasieparameters word dan as koordinate beskou.  Dit gee dan aanleiding tot die idee van 'n ``bulk"  waar die toestande normaliseerbaar is en 'n ``rand" waar hulle nie is nie.  Asimptotiese anti-de Sitter geometrie\"e volg op natuurlike wyse vanaf nie-normaliseerbare toestande, maar die geometrie kan egter baie meer algemeen wees as dit.    \\
Tyd-ontwikkelde toestande is die eerste onderwerp.  'n Sinvolle regulariseringsmetode is bloot om tyd kompleks te maak wat dan die radiale koordinaat as 'n energieskaal giet.  Die duale beeld van hierdie twee-dimensionele toestande is 'n model van dilaton-swaartekrag waar die dilaton die interpretasie van 'n kwantumoperatorverwagtingswaarde dra.  As 'n voorbeeld hiervan - die duale beeld van toestande wat ontwikkel onder 'n $su(1,1)$ Hamiltoniaan is dilaton-swaartekrag op $AdS_2$.  Hierdie beeld strook met bestaande restultate uit the $AdS_2 / CFT_1$ literatuur.  Hierdie bestaande resultate word herondersoek en toevoegings word gemaak daaartoe met behulp van die sistematiese kwantummeganika / dilatonswaartekrag woordeboek.  As nog 'n voorbeeld word dit aangetoon dat die duale beeld van toestande wat tyd-ontwikkel onder 'n $su(2)$ Hamiltoniaan 'n model van diltatonswaartekrag op $dS_2$ is.  \\
Die ho\"er-dimensionele ondersoek word, ter wille van eenvoudigheid, beperk tot toestande wat oor volle Schr\"odinger simmetrie beskik met en sonder dinamiese massa.  Die tyd- en ruimtelike koordinate word kompleks gemaak om die toestande te regulariseer en simmetrie\"e te behou.  Woordeboeke word saamgestel vir beide gevalle.  Dit word aangetoon dat submetrieke van hierdie metrieke in die $AdS/CFT$ en $AdS/NRCFT$ literatuur bestudeer word.  
\end{abstract}

\chapter*{Acknowledgements}

This thesis would have not been possible without the academic, financial and emotional support of many people and institutions.  I mention them here though I am sure that I will forget a few through the course of my typing.  To those I forget I wish to ensure you that I reflect on this four year journey constantly and that my gratitude swells when you enter my thoughts.  \\ \\
First and foremost I would like to thank my supervisors Prof Scholtz and Dr Kriel: your guidance of not only this project but of my thinking as a physicist has been of immeasurable value.  I often remark in jest that postgraduate studies is, among others, a continuous exercise in humility.  However, the gentle way in which you have helped me onto the right path, despite my many mistakes, is something that will always stay with me.  \\ \\
I would also like to my lecturers and teachers over the years for the remarkable way in which you gave color and intrigue to the field of physics.  It is largely because of this that I pursued further studies in the field.  This is a decision that I am very grateful to have made.  Special mention in this regard should be made of Mr Hoffman, my high school physics teacher, who was instrumental in me pursuing a career in science and Prof Geyer who encouraged my pursuit of theoretical physics especially.  I must also mention the extraordinary lengths that Prof M\"uller-Nedebock went through at the end of 2010 and 2011.  His assistance and support over that time fills me with immense gratitude and it stands as one of the most formative times in my life.  \\ \\
The undertaking of my studies would not have been possible without the financial contributions of the Wilhelm Frank trust, the National Institute for Theoretical Physics and the Institute of Theoretical Physics at Stellenbosch University.  This is true not only of my PhD but also the many years preceding it.  My sincere gratitude for all the support you have provided.  \\ \\
The interactions with the students and staff at the Department of Physics contributed greatly to the undertaking and conclusion of this thesis.  This is true in the academic sense where I could discuss problems I encountered in my own project, learn great things from the projects others are undertaking and, most importantly, know that there are others who understand the successes and challenges of postgraduate research so very well.  It is also true in simply a social sense.  I consider my years at the Physics Department to be a great privilege because of the wonderful people I have met there.  Thank you to all - lecturers, support staff and students - that help shape the fantastic work environment there.  \\ \\
Lastly, the journey from the start of this project to its finish and the draft of this thesis is an immense one.  As time progressed this fact revealed itself with ever increasing authority.  The people who helped me deal with the many struggles contained therein are countless but I would like to mention a few.  To my family and especially my parents and brothers: your support over this time cannot be summed up in words.  The way you helped me adjust to life back home during the final stretch of the thesis and showered me with love and support is possibly the kindest act that I have ever encountered.  That last bit of time we spent together before my transition into the real world is something I will always cherish.  To Chantel, the love of my life, you have known me since the very first steps of this journey and the happiest thought I have is that we will continue to walk this unpredictable path of life together. \\ \\
To Chris, Hendre, Jandre and Sheree - know that your support and friendship helped carry me through all of this.  I am very fortunate to know people on whose door I can knock at any time of day or night.  For this I am truly grateful.  May we continue to write paragraphs in the chapters of each others lives for many years to come.  

\chapter*{List of Symbols and Abbreviations}

As a guide to the reader to both avoid confusion and to interpret the equations in the text accordingly we provide here a list of abbreviations and commonly used symbols that appear throughout the text.  The reader may note that some symbols are very closely related - in these circumstances the context determines the appropriate interpretation.  Care has been taken to avoid that similar symbols with different meanings appear in the same context.  

\section*{Abbreviations}

\begin{itemize}[label={},topsep=0pt,parsep=0pt,partopsep=0pt,labelwidth=5cm,align=left,itemindent=-0.5cm]
\setlength\itemsep{0.3em}
\item{\makebox[3cm][l]{$AdS$}  Anti-de Sitter as in anti-de Sitter space}
\item{\makebox[3cm][l]{$AdS_d$}  $AdS$ space in $d$ dimensions}
\item{\makebox[3cm][l]{$CFT$} Conformal Field Theory  }
\item{\makebox[3cm][l]{$CFT_d$}  $CFT$ in $d$ dimensions}
\item{\makebox[3cm][l]{$NRCFT$} Non-relativistic Conformal Field Theory}
\item{\makebox[3cm][l]{$SYM$} Super Yang-Mills as in supersymmetric Yang-Mills theory}
\item{\makebox[3cm][l]{BCH} Baker-Campbell-Hausdorff as in the BCH formula}
\item{\makebox[3cm][l]{CQM} Conformal Quantum Mechanics}
\item{\makebox[3cm][l]{BTZ} Ba\~nados-Teitelboim-Zanelli as in the BTZ black hole}
\end{itemize}

\section*{Commonly Used Symbols}

\subsection*{Geometry and Gravity}
\begin{itemize}[label={},topsep=0pt,parsep=0pt,partopsep=0pt,labelwidth=5cm,align=left,itemindent=-0.5cm]
\setlength\itemsep{0.3em}
\item{\makebox[3cm][l]{ $g_{\mu\nu}$} metric tensor}
\item{\makebox[3cm][l]{ $g^{0}_{\mu\nu}$} fixed metric, typically flat space}
\item{\makebox[3cm][l]{ $\sigma_{\mu\nu}$} anti-symmetric two-form, symplectic form in special cases}
\item{\makebox[3cm][l]{ $R$} scalar curvature}
\item{\makebox[3cm][l]{ $R_{\mu\nu}$} Ricci tensor}
\item{\makebox[3cm][l]{ $R_{\mu\nu\alpha\beta}$} Riemann (curvature) tensor}
\item{\makebox[3cm][l]{ $W_{\mu\nu\alpha\beta}$} Weyl tensor}
\item{\makebox[3cm][l]{ $\delta_{\mu\nu}$} Kronecker delta}
\item{\makebox[3cm][l]{ $T_{\mu\nu}$} energy-momentum tensor}
\item{\makebox[3cm][l]{ $L_M$} matter content}
\item{\makebox[3cm][l]{ $\eta$} dilaton field}
\end{itemize}

\subsection*{Operators and Representations}
\begin{itemize}[label={},topsep=0pt,parsep=0pt,partopsep=0pt,labelwidth=5cm,align=left,itemindent=-0.5cm]
\item{\makebox[3cm][l]{ $P$} momentum operator; subscript indicates the component}
\item{\makebox[3cm][l]{ $X$} position operator; the subscript indicates the component}
\item{\makebox[3cm][l]{ $D$} dilitation or scaling operator}
\item{\makebox[3cm][l]{ $K$} special conformal operator; subscript indicates the component}
\item{\makebox[3cm][l]{ $M_{ij}$} rotation operator in the $i-j$ plane}
\item{\makebox[3cm][l]{ $O_j$} an arbitrary ($CFT$) operator}
\item{\makebox[3cm][l]{ $O_\Delta$} a ($CFT$) operator of scaling dimension $\Delta$}
\item{\makebox[3cm][l]{ $\tilde{O}_\Delta$} a primary ($CFT$) operator of scaling dimension $\Delta$}
\item{\makebox[3cm][l]{ $A$} an arbitrary (quantum mechanical) operator}
\item{\makebox[3cm][l]{ $\phi_A$} normalised expectation value of $A$}
\item{\makebox[3cm][l]{ $\Phi$} an arbitrary normalised expectation value}
\item{\makebox[3cm][l]{ $U$} a unitary operator, typically in the context of transformations}
\item{\makebox[3cm][l]{ $U(g)$} unitary representation of the group element $g \in G$}
\item{\makebox[3cm][l]{ $S(g)$} an arbitary representation of the group element $g \in G$}
\item{\makebox[3cm][l]{ $j, N, k, r_0$} commonly used representation labels}
\item{\makebox[3cm][l]{ $[. \ , .]$} commutator of two operators}
\end{itemize}

\subsection*{States and Operators}
\begin{itemize}[label={},topsep=0pt,parsep=0pt,partopsep=0pt,labelwidth=5cm,align=left,itemindent=-0.5cm]
\item{\makebox[3cm][l]{ $|.)$} a state vector, not necessarily normalised or non-normalisable}
\item{\makebox[3cm][l]{ $|.\rangle$} a normalised state vector}
\item{\makebox[3cm][l]{ $\langle . \rangle$} normalised expectation value}
\end{itemize}

\subsection*{Field Theory}
\begin{itemize}[label={},topsep=0pt,parsep=0pt,partopsep=0pt,labelwidth=5cm,align=left,itemindent=-0.5cm]
\item{\makebox[3cm][l]{ $S$} action}
\item{\makebox[3cm][l]{ $\mathcal{L}$} Lagrangian}
\item{\makebox[3cm][l]{ $\phi$} field}
\item{\makebox[3cm][l]{ $\phi_\Delta$, $\Phi_\Delta$} field of scaling dimension $\Delta$}
\item{\makebox[3cm][l]{ $A_\mu$} gauge field}
\item{\makebox[3cm][l]{ $Z$} partition function}
\item{\makebox[3cm][l]{ $g, g_s, g_{YM}$} coupling constants}
\end{itemize}

\subsection*{Derivatives and Vector Fields}
\begin{itemize}[label={},topsep=0pt,parsep=0pt,partopsep=0pt,labelwidth=5cm,align=left,itemindent=-0.5cm]
\item{\makebox[3cm][l]{ $\frac{\partial}{\partial_x}, \partial_x$} partial derivative with respect to $x$}
\item{\makebox[3cm][l]{ $\frac{\delta}{\delta f(x)}$} functional derivative with respect to $f(x)$}
\item{\makebox[3cm][l]{ $\nabla_\mu, \nabla_{(x^\mu)}$} covariant derivative with respect to the $\mu$'th coordinate, $x^\mu$ }
\item{\makebox[3cm][l]{ $\nabla^2$} Laplace operator, Laplacian}
\item{\makebox[3cm][l]{ $\chi, \chi^\mu\partial_\mu$} vector field}
\end{itemize}

\subsection*{Variables}
\begin{itemize}[label={},topsep=0pt,parsep=0pt,partopsep=0pt,labelwidth=5cm,align=left,itemindent=-0.5cm]
\item{\makebox[3cm][l]{ $z, \tau, \theta$} complex variables}
\item{\makebox[3cm][l]{ $\overline{z}, \overline{\tau}, \overline{\theta}$} conjugate complex variables}
\item{\makebox[3cm][l]{ $\beta, t, x, y, \zeta, s$} some examples of real variables}
\end{itemize}
\vspace{0.5cm}
As a final convention: when we end a series expansion with the symbol $O(x^m)$ we mean that the next-leading term may be of order $m$. 

\tableofcontents

\chapter{Introduction}

Since the publication of the famous Maldacena conjecture \cite{Maldacena} the study of the $AdS/CFT$ correspondence (or more generally the gauge/gravity duality) has grown into a substantial field of research.  Indeed, this has lead to the paper \cite{Maldacena} becoming one of the most cited works in history.  Though not the first work that probed the equivalence of gravitational theories and gauge theories (see e.g. \cite{Donoghue}), the conjecture provided the first explicit example of the so-called gauge/gravity duality.   The duality is a conjectured correspondence between certain gauge theories and theories of gravity i.e. the physical information contained in each is equivalent, only packaged differently.  If a physical quantity in the one theory can be calculated in some domain of the theory's parameters then the value for a physical quantity of the dual theory can be extracted from it.  In order to apply such a procedure one requires the dictionary i.e. how the physical quantities of the one theory are related to the physical quantities of the other.  The seminal works in the development of the dictionary \cite{Witten}, \cite{Polyakov} still form the cornerstone of it \cite{AdSCFTbook}, \cite{AdSUserguide}.  \\ \\
The gauge/gravity duality is significant on a conceptual level \cite{AdSCFTbook}.  If understood properly it holds the promise of reformulating theories of quantum gravity in terms of their dual gauge theories.  This would be of great benefit since even the most well-studied model of quantum gravity, string theory, can only be formulated consistently as a perturbative theory \cite{AdSCFTbook}.  Reformulating it in terms of its gauge theory dual would thus allow one to go beyond the perturbative expansion and formulate it consistently for all parameters. \\ \\
This only explains part of the great interest that was sparked by the Maldacena paper \cite{Maldacena}.  The conjecture goes further to claim that, at least in some cases, the gauge/gravity duality is a strong/weak duality.  This means the dual theory is solvable in a region of parameter space where the original theory is not i.e. the dual theory is weakly coupled when the original theory is strongly coupled.  It thus provides one of the few (if, in some cases, not the only) tool to study gauge theories at strong coupling.  This remarkable feature of the conjecture lies at the core of its power and has been applied in problems varying from the quark-gluon plasma \cite{PedrazaRev} to holographic superconductors \cite{Sunandan} and condensed matter physics \cite{AdSCondensedMatter}, \cite{Hartnoll}.  \\ \\
As we will illustrate with reference to the Maldacena conjecture \cite{Maldacena}, symmetries play a key role in the formulation and application of the gauge/gravity duality.  The intuitive reasoning for this is reasonably clear.  Critical points of a quantum model cannot be treated perturbatively due to a vanishing energy gap which implies a strongly coupled problem \cite{AdSCondensedMatter}.  However, the vanishing gap also implies a high degree of symmetry at the critical point.  If the treatment of the quantum model is thus rearranged around the symmetries then the problem may possibly become simpler.  \\ \\
This idea that the treatment of a problem can be made simpler by rearranging it has at least one rigorous example.  It is known that gauge theories permit a $\frac{1}{N}$ expansion in the large $N$ limit where $N$ is the dimension of the gauge group \cite{LargeN}, \cite{LargeNW}.  This simplification is, at first glance, counter-intuitive.  One would expect that a gauge theory becomes more complicated as the dimension of the fundamental representation increases.  The results \cite{LargeN}, \cite{LargeNW} shows the opposite is true as long as the theory is repackaged or rearranged appropriately.  In particular the large $N$ expansion is a topological rearrangement of Feynman diagrams.  It is also interesting to note that this sidesteps the issues of a perturbative expansion in the coupling constant precisely because the expansion parameter in this topological series, $\frac{1}{N}$, is independent of the coupling constant.  \\ \\
Though a very powerful and widely used calculational tool the gauge/gravity duality remains largely unproven.  This is due, in no small part, to the theories involved in the duality being difficult to work with in their own right.  The applications of the duality are numerous and consequently proving the duality is of great importance.  As one may expect this is not a simple task and it is a sensible strategy to target smaller goals aimed at an eventual proof.  In this regard we note, as least as far as this writer's knowledge of the literature is concerned, that there is a lack of fully systematic procedures that can construct the appropriate gravitational dual from a given quantum model.  \\ \\
In this thesis we undertake what can be viewed as a first step to realising this goal of a systematic procedure.  Specifically we will investigate how gravitational duals can be constructed from quantum mechanical models in a systematic way.  Such a procedure holds the great benefit of granting us direct access to a quantum mechanics / gravity dictionary.  This would allow us to address pertinent questions.  Under what circumstances does a quantum mechanical model permit a dual description?  Is the dual description a unique theory?  Can we find evidence that repackaging a quantum mechanical theory in a dual description is useful?  Of course, it is likely that numerous systematic constructions can be made.  A question that we will be dealing with regularly in this thesis is whether the construction we choose reproduces existing results in the literature.  If so then the systematic procedure may allow us to progress beyond these existing results in a natural way.  \\ \\
It should be emphasised that our focus in this thesis will be on quantum mechanical models and not field theories.  The most notable difference is that we do not consider models with gauge symmetry so that only the global symmetries will feature in our construction.  At first glance this may appear to be an oversimplification as the large $N$ expansion \cite{LargeN}, \cite{LargeNW} ($N$ is the dimension of the gauge group) is no longer applicable.  Nonetheless, this simplification will allow the construction of simple gravitational duals and, when we focus on the simplest quantum models, we will find very good agreement with existing works in the $AdS/CFT$ literature.  The construction we will employ must be seen as a toy model of holography, but one that may hopefully be extended to the more intricate setting of quantum field theory in future. \\ \\
We will start with a discussion of the gauge/gravity duality with specific reference to the famous Maldacena conjecture \cite{Maldacena} in chapter \ref{BackChap}.  Our discussion will be basic and only highlight the aspects of the conjecture that will be relevant to this thesis and some proposed future generalisations.  \\ \\
In chapter \ref{ConstrChap} we will introduce the construction that takes as input a family of quantum states and produces as output a metric and anti-symmetric two-form that encodes the symmetries of the states as isometries.  We will motivate why this metric, and not some other geometric construction, is a sensible first choice for a systematic procedure.  In a natushell it is a relatively simple construction that respects the symmetries of the quantum model.  One of the first features that will be appealing with this construction is that, if the family of quantum states is non-normalisable, we have to include additional parameters that regularise the states.  These additional parameters will have the natural interpretation of ``bulk" coordinates.  The quantum states then live on the ``boundary" of this manifold.  Both of these features fit well with the conventional gauge/gravity duality.  \\ \\
We will proceed to apply the construction to the simplest family of quantum states in chapter \ref{GeomChap}, time-evolved states.  These are the states generated by some time-evolution of a reference state.  If the reference state is non-normalisable, the Hamiltonian is time-independent and we regularise the states by complexifying time then the geometry is asymptotically $AdS_2$.  The geometries can be much more general than this, however.  We will show how de Sitter and flat geometries result from the appropriate coherent states.  A general feature of the geometries is that states with the same set of dynamical symmetries produce metrics that are the same up to coordinate transformation.  This will imply, for example, that the duals of the free particle and harmonic oscillator states are geometrically equivalent, a rather counterintuitive result that we will discuss in further detail.  At this point, without any gravitational content, we will have acquired enough results to extend one of the existing results in the $AdS_2/CFT_1$ literature \cite{Jackiw}. \\ \\
Chapter \ref{GravChap} contains the most well-developed of our results.  We proceed from the two-dimensional families of states to a gravitational dual description.  By using properties of the geometric reformulation of quantum mechanics \cite{Ashtekar} we are able to write down equations of motion for the expectation values of quantum mechanical operators.  We show that, in general, these equations of motion can be matched with the on-shell field equations of a model of dilaton gravity.  Depending on the manifold symmetries and the expectation value being solved for, an appropriate energy momentum tensor may have to be included.  \\ \\
As a specific example we examine the dilaton gravity duals of the $SU(1,1)$ class of Hamiltonians.  We find very good agreement with the work of \cite{Cadoni1}-\cite{CadoniCQM} and find interpretations for the dilaton black hole mass in terms of the $su(1,1)$ operators.  With our machinery we are able to reproduce these results very naturally.  We draw particular attention to the calculation of the $CFT_1$ central charge from the dilaton gravity description which, in our construction, can be related directly to conformal transformation which are in turn related to the unconstrained field equation solutions.  This picture of the calculation makes matters very clear.  We will, furthermore, be able to extend these existing results beyond the expectation values of symmetry generators.  We will also briefly explore the dual descriptions of states that lead to a de Sitter geometry.  The results are not as well-developed as the $SU(1,1)$ class of Hamiltonians, but interesting nonetheless. \\ \\
Our attention will then move to the higher dimensional families of states.  In chapter \ref{HighDim} we add spatial translations to the time-evolved states and examine their dual descriptions.  The generators of dynamical symmetry of the simplest model, the free particle, are generated by the so-called Schr\"odinger algebra.  Even for this simple case we encounter several difficulties in putting together the dictionary.  Firstly, the metrics we find are no longer conformally flat.  The non-zero Weyl tensor complicates the equations of motion.  We show how this can be remedied by only considering the trace of the equations of motion thereby exchanging the equations we do not consider for boundary conditions.  A second difficulty is more problematic. In the regularisation scheme we employ in the chapter, the manifolds are also not Einstein, even for the free particle, so that the expectation values require quite a bit of calculational maneuvering to recover.  The scheme is presented at the end of chapter \ref{HighDim} but further work is needed to understand it fully.   \\ \\
We proceed to centrally extend the Schr\"odinger algebra and consider this central extension (the mass) as a dynamical variable in chapter \ref{McGChap}.  This will allow us to write down a simple dictionary for the $d$-dimensional Schr\"odinger algebra Hamiltonians dual to a massive scalar field action on an appropriately chosen background.  \\ \\
These results do not have analogues in the $AdS/CFT$ literature, however.  The most obvious departure from the conventional approach is that we have too many dimensions added in the bulk.  We show that when we restrict ourselves to only a submanifold then we again recover a number of geometries studied in the literature \cite{McGreevyGravDual}, \cite{Son}.  Unfortunately what we lose by focussing on the submanifold is the developed dictionary itself since we rely throughout on the fact that the family of quantum states is parametrised by complex coordinates in order to put it together.  The states that live on the submanifold do not, in general, possess this property.  We propose that one may possibly use the existing dictionary to extract the submanifold dictionary.  The chapter concludes with speculations as to how this may be done.  \\ \\
What we hope to achieve in the chapters that follow is twofold.  First, we showcase how a systematic procedure to build gravitational duals from quantum mechanical models is possible.  Even if our construction is only applicable to the simple models we study in this thesis we hope that it shows that the development of a systematic procedure for building duals is an attainable goal.  Secondly, we intend to show that the construction we have chosen is, at least for the problems we study, an applicable and beneficial one.  The evidence for this will be the many works in literature we may add clarity to and extend.  This should serve as good motivation to investigate the generalisations of this construction in future.

\chapter{Overview of the $AdS/CFT$ Correspondence}

\label{BackChap}

The duality first formulated by Maldacena \cite{Maldacena}, that of type $IIB$ string theory on $AdS_5 \times S_5$ dual to $\mathcal{N}=4$ Super Yang-Mills ($SYM$) with an $SU(N)$ gauge group on the boundary, remains the most famous example of the gauge/gravity duality, \cite{Witten}, \cite{Polyakov}, \cite{AdSCFTbook}, \cite{AdSUserguide}, \cite{PedrazaRev}, \cite{Zaffaroni}.  Indeed, this duality is responsible for the conjecture's historical name, the $AdS/CFT$ correspondence.  The historical name originates from the type $IIB$ string theory living on anti-de Sitter space (the $AdS$ part) and from the $SYM$ theory being a conformal field theory (the $CFT$ part).  In this chapter we will define all the concepts mentioned in this paragraph concretely, all in due course.   Good reviews on the gauge/gravity duality and its applications can be found in \cite{AdSCFTbook}, \cite{AdSUserguide}, \cite{PedrazaRev}, \cite{Zaffaroni} and \cite{LargeN2}.  A good discussion can also be found in \cite{McGreevy}.  These examples cover a very small fraction of the available literature on the $AdS/CFT$ correspondence but will be sufficient for our purposes in this thesis.  \\ \\
The conjectured correspondence is remarkable first and foremost since both of these theories (string theory and Super Yang-Mills) are difficult to work with in their own right.  Consequently it is also a very hard (and still an unaccomplished) task to prove the conjecture in full, even for this well-studied example \cite{AdSCFTbook}.  This famous example is exceptionally well understood and it thus still serves as a means to lay out the holographic dictionary in a clear way.  We will proceed to do exactly that in this chapter.  The purpose of this exercise is to illuminate the status of the construction that will be made in this thesis as a toy model of holography.  This will allow us, firstly, to show which aspects of dualities may be understood and learned from by means of this toy model and secondly to identify its limitations.  These limitations are important to take note of especially for future generalisations.    \\ \\
It is important to emphasise that the power of the construction we will employ lies not in its ability to capture all aspects of dualities (consequently its status as a toy model).  Rather the power of the construction lies in its systematic nature.  Many familiar features of the gauge/gravity duality arise naturally in this toy model and, we believe, to a sufficient extent to warrant future attempts to generalise the construction.  \\ \\
The aim of this chapter is to partition the correspondence, with specific reference to the Maldacena conjecture \cite{Maldacena}, into what will become aspects included in the toy model and aspects not included in the toy model.  We will then develop our construction in the course of the ensuing chapters with this background knowledge and context in mind.   

\section{Global Symmetries}

One of the key motivations for the $AdS/CFT$ correspondence is the coincidence of the isometries of $AdS_{d+1}$ and the symmetries of $CFT_d$ where $d$ refers to the spacetime dimension \cite{AdSCFTbook}.  Indeed, it is hard to imagine that two physical models can be equivalent if they do not share the same symmetries.  This matching of symmetries can thus easily be seen to be a necessary condition for duals.  The coincidence of symmetries is even more significant.  Indeed, a ``trick'' may be employed to generate conformally invariant partition functions starting from gravity actions defined on $AdS$ \cite{AdSCFTbook}.  A (consistent) theory of gravity defined on $AdS$ thus carries a consistently defined conformal field theory on its boundary.  It is not clear though whether all $CFT$s can be generated in this way \cite{AdSCFTbook}.  \\ \\
It is important for the purpose of the discussion we now undertake that we distinguish between global symmetries and local symmetries.  We first discuss the global symmetries as these will be of particular relevance later.  By global symmetry we mean that the action remains invariant if we perform the same transformation at every point.  This is typically associated with a unitary operator $U = e^{i \alpha T}$ where $T$ is the generator and the parameter $\alpha$ does not have coordinate dependence.  Local symmetries, where the coefficient can have coordinate dependence, play a different role in the conjecture.\\ \\
In the Maldacena conjecture \cite{Maldacena} the global symmetry corresponds to the $\mathcal{N}=4$ $SYM$ part.   This means that the field theory is superconformal with four supercharges (for supersymmetry).  For the gravitational theory (the type $IIB$ string theory) the symmetries are manifest as the isometries of $AdS_5 \times S_5$.  We will now examine these global symmetries on both sides of the Maldacena conjecture more closely.  

\subsection{The Conformal Algebra (in $d>2$ Dimensions)}

As the name suggests conformal field theory ($CFT$) is a quantum field theory that is invariant under conformal group transformations.  The $d$-dimensional conformal group, $conf_d$ i.e. $SO(d-q+1, q+1)$, can be defined as the transformations that leave the $d$-dimensional flat metric in arbitrary signature, 
\begin{equation}
g^0_{\mu\nu} = \left\{ \begin{array}{cc} -\delta_{\mu\nu} & \mu = 1,2,...,q \\ \delta_{\mu\nu} & \mu = 1,2,...,d-q   \end{array} \right.   \label{ConfTransDef}
\end{equation}
invariant up to a local scale factor i.e. a conformal factor \cite{AdSCFTbook}.  Note that the integer $q$ is related to the signature of the metric.  By $\delta_{\mu\nu}$ we mean the Kronecker delta function.  Throughout this discussion we refer to the coordinates of the conformal field theory as $\left\{ x_0 \equiv t, x_1, x_2, ..., x_{d-1} \right\}$.  In $d > 2$ dimensions (we will discuss the $d\leq2$ case separately) these are $\frac{1}{2}d(d-1)$ Lorentz transformations, $d$ translations, $d$ special conformal transformations and one dilatation or scaling.  The corresponding generators are $\tilde{M}_{\mu\nu}$,  $\tilde{P}_{\mu}$, $\tilde{K}_{\mu}$ and $\tilde{D}$ respectively, given by \cite{CFTbook}
\begin{eqnarray}
\widetilde{P}_\mu & = & -i \partial_\mu \nonumber \\
\widetilde{K}_\mu & = & -i(2 x_\mu x^\nu \partial_\nu - x^\nu x_\nu \partial_\mu) \nonumber \\
\widetilde{D} & = & -i x^\mu \partial_\mu \nonumber \\
\widetilde{M}_{\mu\nu} & = & i(x_\mu \partial_\nu - x_\nu \partial_\mu).  \label{ConformalTrans}
\end{eqnarray}
where $\partial_\mu \equiv \frac{\partial }{\partial x^\mu}$.  The lowering and raising of indices is done by contracting with the flat metric $g^0_{\mu\nu}$ (\ref{ConfTransDef}) and its inverse $g_0^{\mu\nu}$ respectively. It can be verified that the transformations (\ref{ConformalTrans}) leave the metric $g^0_{\mu\nu}$ invariant up to a local scale factor i.e. if the coordinates transform as $x^\mu \rightarrow y^\mu(x^\nu)$ then
\begin{equation}
g^0_{\mu\nu}(x^\mu) \rightarrow \lambda(y^\mu) g^0_{\mu\nu}(y^\mu). \label{MetricTrans}
\end{equation}
One can define the algebra purely in terms of their commutation relations and the coordinate forms (\ref{ConformalTrans}) may be recovered as a specific representation.  The $d$-dimensional conformal algebra (for $d>2$) is the set of $\frac{1}{2}(d+1)(d+2)$ operators that satisfy the following commutation relations (see Appendix \ref{HolsteinApp} for a summary of all the algebras that appear in this thesis)
\begin{eqnarray}
\left[\widetilde{D}, \widetilde{K}_\mu \right] & = & i \widetilde{K}_\mu \nonumber \\ 
\left[\widetilde{D}, \widetilde{P}_\mu \right] & = & -i \widetilde{P}_\mu \nonumber \\ 
\left[ \widetilde{P}_\mu, \widetilde{K}_\nu \right] & = & 2 i \widetilde{M}_{\mu\nu} - 2 i g^0_{\mu\nu} \widetilde{D} \nonumber \\
\left[\widetilde{K}_{\alpha}, \widetilde{M}_{\mu\nu} \right] & = & i(g^0_{\alpha\mu}\widetilde{K}_\nu - g^0_{\alpha\nu}\widetilde{K}_\mu) \nonumber \\
\left[\widetilde{P}_{\alpha}, \widetilde{M}_{\mu\nu} \right] & = & i(g^0_{\alpha\mu}\widetilde{P}_\nu - g^0_{\alpha\nu}\widetilde{P}_\mu) \nonumber \\ 
\left[ \widetilde{M}_{\alpha\beta}, \widetilde{M}_{\mu\nu} \right] & = & i( g^0_{\alpha\mu} \widetilde{M}_{\beta\nu} + g^0_{\beta\nu} \widetilde{M}_{\alpha\mu} - g^0_{\alpha\nu} \widetilde{M}_{\beta\mu} - g^0_{\beta\mu}\widetilde{M}_{\alpha\nu}   ).   \label{ConformalComm}
\end{eqnarray}
For $d>2 $, the conformal transformations leave the free, massless Klein Gordon equation in flat space form invariant \cite{Henkel} i.e.
\begin{equation}
\partial_{(x_\mu)} \partial^{(x^\mu)} \psi(x) = 0 \ \ \ \rightarrow \ \ \  \partial_{(y_\mu)} \partial^{(y^\mu)} \Psi(y) = 0 \ \ \ \textnormal{if} \ \ \ x^\mu \rightarrow y^\mu(x^\nu)\label{KGEquation}
\end{equation}
where $\Psi(y) = e^{i \alpha(x)}\psi(x)$ so that the wave function may pick up a phase $\alpha(x)$ where $\alpha$ is an arbitrary function of $x = x_0, x_1, ..., x_{d-1}$.  This provides another useful way to visualise these symmetries.  

\subsection{$CFT$ Correlation Functions}

\label{CFTCorr}

The requirement of conformal symmetry for a field theory places significant restrictions on the form of the correlation functions \cite{Francesco}, \cite{CFTbook}
\begin{equation}
\langle 0_{CFT} | O_{1}(t_1, \vec{x}_1) O_{2}(t_2, \vec{x}_2) ... O_{n}(t_n, \vec{x}_n) | 0_{CFT}  \rangle   \label{CorrFuncsDef}
\end{equation}
where 
\begin{equation}
O_{m}(t, \vec{x}) \equiv e^{i t H} \  O_m(\vec{x}) \  e^{-i t H}  \label{DynOpDef}
\end{equation}
and $|0_{CFT}\rangle$ refers to the conformal field theory vacuum.  The operator $O_m(\vec{x})$ has spatial dependence.  By $H$ we mean $P_0$ for the case of conformal field theory but we make the distinction to allow generalisations (of the time evolution operator).  A basis for the enveloping conformal algebra are those operators of definite scaling dimension, $O_{\Delta}$, which we define by \cite{AdSCFTbook}
\begin{equation}
[\tilde{D}, O_\Delta(0, \vec{0})] = -i \Delta O_{\Delta}(0, \vec{0})
\end{equation}
where $\Delta$ is the scaling dimension.  We may simplify this even further by only considering the primary operators defined by \cite{AdSCFTbook}
\begin{equation}
\tilde{O}_{\tilde{\Delta}}(0, \vec{0}) \in \left\{ O_\Delta(0, \vec{0}) \right\} \ \ \ \textnormal{such that} \ \ \ [\tilde{O}_{\tilde{\Delta}}(0, \vec{0}), \tilde{K}_\mu ] = 0.  \label{Primaries}
\end{equation}
This is precisely because the commutator of $O_\Delta$ with $\tilde{P}_\mu$ increases scaling dimension while the commutator with $\tilde{K}_\mu$ decrease scaling dimension.  The primary operators (\ref{Primaries}) can thus be viewed as the lowest tiers of the ladder of scaling dimension operators and one can ladder up by means of differentiation with respect to $t$ and $x_i$ from (\ref{DynOpDef}).  The operators obtained by this differentiation process are known as descendants \cite{AdSCFTbook}.  \\ \\
The desired quantities from our calculations are thus the 2- and 3-point correlation functions of primaries which take a very specific form \cite{PolyakovCorr}, \cite{Freedman} for $CFT$s due to the very restrictive symmetry requirements, namely
\begin{eqnarray}
& & \langle 0_{CFT} | \tilde{O}_{\tilde{\Delta}_1}(t_1, \vec{x}_1) \tilde{O}_{\tilde{\Delta}_2}(t_2, \vec{x}_2) | 0_{CFT}  \rangle \nonumber \\
& = & \delta_{\tilde{\Delta}_1,\tilde{\Delta}_2}\prod_{i < j}^2 |t_i - t_j + |\vec{x}_i - \vec{x}_j ||^{-(\tilde{\Delta}_i + \tilde{\Delta}_j)} \nonumber \\ 
& &\langle 0_{CFT}  | \tilde{O}_{\tilde{\Delta}_1}(t_1, \vec{x}_1) \tilde{O}_{\tilde{\Delta}_2}(t_2, \vec{x}_2) \tilde{O}_{\tilde{\Delta}_3}(t_3, \vec{x}_3) | 0_{CFT}  \rangle \nonumber \\
& = & c_{123} \prod_{i<j}^3 |t_i - t_j + |\vec{x}_i - \vec{x}_j||^{\tilde{\Delta}_1 + \tilde{\Delta}_2 + \tilde{\Delta}_3 - 2\tilde{\Delta}_i - 2\tilde{\Delta}_j}  \label{PrimaryCorrelators}\footnote{}
\end{eqnarray}
where the coefficients $c_{ijk}$ are dependent on the model under consideration.  The symbol $\delta_{\Delta_1, \Delta_2}$ again refers to the Kronecker delta function.  The coefficients, $c_{ijk}$, of three-point functions completely determine the theory since higher point functions are determined by these \cite{Rychkov}.  This is by virtue of the operator product expansion \cite{Rychkov} where the product of two primary operators at different points may be expressed as the sum of primary operators (and descendants).  This allows one to reduce higher point functions to a function of the two- and three-point functions.  \\ \\
The $AdS/CFT$ dictionary provides a prescription for calculating these correlation functions on the gravity side of the duality.  We will present this shortly, but for now the important point is that the global symmetries of the conformal field theory determine the form of the two- and three-point functions (\ref{PrimaryCorrelators}). 

\subsection{$AdS$ Space} 
\label{AdSSpace}

On the other side of the $AdS/CFT$ correspondence we have a theory of gravity defined on $AdS$.  By this we mean that fields and matters fields may introduce fluctations around $AdS$.  We assume, though, that these fluctuations tend to zero towards the boundary of the space.  This allows for a dynamic geometry.  One of the main motivations for considering an $AdS$ background is that, as mentioned, the $d$-dimensional conformal symmetry can be matched exactly to the isometries of $(d+1)$-dimensional $AdS_{d+1}$ geometry (on the boundary) \cite{AdSCFTbook}.  This high degree of symmetry constrains the possible physical models greatly so that this matching is significant.  We will now discuss this matching of symmetries explicitly.  \\ \\
First it is necessarily to point out that the requirement for a transformation to be an isometry of a metric is different to the requirement for it to be a symmetry of some scalar function.  Specifically, a transformation is an isometry if
\begin{equation}
x^\alpha \rightarrow y^\alpha(x^\beta) \ \ \ \Rightarrow \ \ \  g_{\mu\nu}(x^\alpha) dx^{\mu} dx^{\nu} = g_{m n}(y^\alpha) dy^m dy^n \label{Isometry}
\end{equation}
i.e. the metric in the new coordinates has the same functional dependence on these new coordinates as the metric in the old coordinates had on the old coordinates. \\ \\  It is important to highlight the difference between conformal symmetries and isometries, see  (\ref{MetricTrans}) compared to (\ref{Isometry}).  Conformal symmetry allows the transformation up to a conformal factor whereas isometries require this conformal factor to be $1$.  Consequently the isometries is a subset of the conformal symmetries.  Indeed, the largest number of continuous isometries that a $d$-dimensional metric can possess is $\frac{1}{2}d(d+1)$ while the conformal group for $d \geq 2$ consists of $\frac{1}{2}(d+1)(d+2)$ continuous conformal symmetries.  The conformal group is defined in terms of conformal symmetries of a metric (\ref{MetricTrans})  but in the correspondence we require it to be true isometries of a metric.  In the $AdS/CFT$ correspondence the $AdS$ side of the duality must thus be (at least) one dimension higher than the $CFT$ side in order to capture all the $CFT$ symmetries as isometries.  \\ \\
The metrics that contain their full compliment of continuous isometries are called maximally symmetric.  Indeed, this condition is so highly restrictive on the metric that there are only three possible candidates - de Sitter space, flat space and anti-de Sitter space \cite{AdSCFTbook} (for a given signature).  These three metrics can be distinguished by the sign of their scalar curvature which is positive, zero and negative respectively (see Appendix \ref{AppGeo} for definitions of the geometric quantities used in this thesis).  However, though these metrics share the same number of isometries, the explicit form of these isometries are different.  \\ \\
It is only anti-de Sitter space which contains all the appropriate isometries in the sense that they match the symmetries (\ref{ConformalTrans}) of the conformal group \cite{AdSCFTbook}.  This exact matching can be done on the conformal boundary of $AdS$.  A convenient form for the $AdS_{d+1}$ metric is the so-called Poincar\'e patch given by
\begin{equation}
ds^2 = \frac{L^2}{\beta^2}\left( d\beta^2 + d\vec{x}\cdot d\vec{x} \right) \label{AdSmetric}
\end{equation}
where $x_\alpha$ has $d$ components and for which the scalar curvature, $R_S = -\frac{(d+1)(d)}{L^2}$, is constant.  Note that while the $AdS_{d+1}$ metric (\ref{AdSmetric}) always possesses $\frac{1}{2}(d+1)(d+2)$ isometries, it is only on the $\beta \rightarrow 0$ boundary that the explicit coordinate form of these isometries corresponds exactly to the $d$-dimensional conformal group \cite{AdSCFTbook}.  The metric (\ref{AdSmetric}) is in Euclidean signature.  We will be working in Euclidean signature throughout this thesis.  

\subsection{The $d \leq 2$ Conformal Group}

\label{VirasSection}

As promised we need to discuss the conformal group for dimension $d \leq 2$ separately.  We borrow greatly from \cite{Francesco}, \cite{CFTbook} in this section.  We discuss the case where $d = 2$ explicitly, but the case $d=1$ is treated in very similar fashion.  \\ \\
As before the conformal group is defined in terms of the transformations that leave the flat space metric (\ref{ConfTransDef}) invariant up to a conformal factor.  We consider the Euclidean signature flat space metric and transform to complex coordinates $z = x_0 + i x_1$, $\overline{z} = x_0 - i x_1$.  We then have
\begin{equation}
ds^2 = g^0_{\mu\nu} dx^\mu dx^\nu = dz d\overline{z}.  
\end{equation}
Consider an arbitrary coordinate transformation $w = w(z)$ and the corresponding $\overline{w} = \overline{w}(\overline{z})$.  The metric is transformed to
\begin{equation}
ds^2 = \frac{d z}{d w}\frac{d \overline{z}}{d \overline{w}} dw d\overline{w}
\end{equation} 
so that it is clear that this arbitrary coordinate transformation is a conformal transformation of the metric.  The conformal group in two dimensions is thus infinite dimensional.  By an almost identical argument one can show that the conformal group in one dimension is also infinite dimensional.  \\ \\
A subset of the transformations $w = w(z)$ are of special interest namely
\begin{eqnarray}
w = \frac{\alpha z + \beta}{\gamma z + \delta}  \label{globalConf}
\end{eqnarray}
where $\alpha, \beta, \gamma$ and $\delta$ are complex numbers that satisfy $\alpha \delta - \beta \gamma = 1$.  These transformations are called the ``global conformal transformations" and correspond exactly to $SO(3,1)$.  This is what we would have gotten if we simply substituted $d = 2$ in $SO(d+1, 1)$.  These transformations (\ref{globalConf}) are of special interest because they are the only transformations of the form $w = w(z)$ that are globally defined invertible mappings.  What this means is that there are no essential singularities and that the complex plane is mapped onto itself.  The consequence of this is that the transformations $\overline{w} = \overline{w}(\overline{z})$ that are not of the form (\ref{globalConf}) must be thought of as being performed only locally i.e. infinitesimally.  \\ \\
Consider then the infinitesimal version of the transformations i.e. $w = z + \epsilon(z)$ where $\epsilon(z)$ is small.  Note that $\epsilon(z) = a$, $\epsilon(z) = a z$ and $\epsilon(z) = a z^2$ correspond to the global conformal transformations (\ref{globalConf}).  We may expand the arbitrary function $\epsilon(z)$ in a power series and we find that
\begin{equation}
w(z) = z + \sum_n a_n z^{1-n}.  
\end{equation}
The differential operators $l_n \equiv -z^{1-n}\partial_z$ (and the corresponding $\overline{l}_n \equiv -\overline{z}^{1-n}\partial_{\overline{z}}$) are the generators and satisfy the commutation relations
\begin{equation}
[l_n, l_m] = (n-m)l_{n+m} \ \ \ ; \ \ \ [\overline{l}_n, \overline{l}_m] = (n-m)\overline{l}_{n+m} \ \ \ ; \ \ \ [l_n, \overline{l}_m] = 0.  \label{WittAlgebra}
\end{equation}
The algebra (\ref{WittAlgebra}) is two copies of the Witt algebra. In one dimension the conformal symmetry generators form only one copy of the Witt algebra.  \\ \\
The Witt algebra permits a central extension to the Virasoro algebra which satisfies
\begin{equation}
\left[V_n, V_m \right] = (n-m)V_{n+m} + \frac{c}{12}(m^3-m)\delta_{n, -m} \ \ \ \ n,m \in Z  \label{Viras}
\end{equation}
where $c$ is referred to as the central charge  and $V_n$ is the $n$'th Virasoro algebra element.  Note that the elements $V_{-1}, V_0, V_1$ close on an $su(1,1) \cong so(2, 1)$ algebra, regardless of center.  The centerless, $c=0$, Virasoro algebra is the Witt algebra.  \\ \\
The central charge features most prominently when the energy momentum tensor is considered.  The energy momentum tensor is equal to the variation of the field theory action by the inverse metric
\begin{equation}
T_{\mu\nu} = \frac{\delta S}{ \delta g^{\mu\nu}}
\end{equation}
The appropriate way to now extract the central charge is to elevate the fields to operators, normal order the energy momentum tensor and apply the operator product expansion to the product of the energy momentum tensor with itself.  The generic form of this expansion for $CFT$'s is \cite{Francesco}
\begin{equation}
T(z)T(w)  \ \ \mytilde \ \  \frac{2 T(w )}{(z - w)^2} + \frac{\partial_{w} T(w)}{z - w} + \frac{1}{2}\frac{c}{(z - w)^4} \label{EMTensor}
\end{equation}
where $c$ is the model-dependent central charge and $T(z) \equiv -2\pi T_{zz}$.  \\ \\
We take note of an important consequences of (\ref{EMTensor}) and the conformal Ward identity \cite{Francesco} applied to the energy momentum tensor
\begin{equation}
\delta_{\epsilon} T(w) = -\frac{1}{2\pi i} \oint_C dw \epsilon(z) T(z) T(w).   \label{ConfWard}
\end{equation}
The expression (\ref{ConfWard}) calculates the change of $T(z)$ under a shift $z \rightarrow z + \epsilon(z)$.  The contour integral picks up the residues of the integrand.  Substituting (\ref{EMTensor}) into (\ref{ConfWard}) yields
\begin{equation}
T(z) \rightarrow T(z) - \epsilon(z) \partial_z T(z) - 2 \partial_z \epsilon(z) T(z) - \frac{c}{12} \partial_z^{3}\epsilon(z).   \label{EMCoordTrans}
\end{equation}
The finite version of the transformation (\ref{EMCoordTrans}), where $w = w(z)$ is given by
\begin{equation}
T(w) = \left( \frac{d w}{d z} \right)^{-2}\left[ T(z) - \frac{c}{12} \left\{ w; z\right\} \right]  \label{EMTensorTrans}
\end{equation}
where $\left\{ w; z\right\}$ is the Schwarzian derivative
\begin{equation}
\left\{ w; z\right\} = \partial_z \left( \frac{\frac{d^2 w}{d z^2}}{\frac{d w}{d z}}  \right) - \frac{1}{2}\left( \frac{\frac{d^2 w}{d z^2}}{\frac{d w}{d z}}  \right)^2.  \label{Schwarzian}
\end{equation}
We will use the transformation property (\ref{EMTensorTrans}) in section \ref{CQMSection} to identify a central charge of a one-dimensional conformal field theory.  

\section{Correlation Functions From Generating Functionals}

\label{CorrGenF}

Now that we have discussed the symmetries in some detail we turn our attention to how correlation functions are calculated in the field theory and, through the use of the dictionary, in the gravitational dual.  \\ \\
We specify the coordinates of the $CFT_d$ as $\left\{ x_0 \equiv t, x_1, x_2, ..., x_{d-1} \right\}$ which is matched with the boundary of $AdS_{d+1}$.  The gauge/gravity dictionary provides a very particular prescription \cite{Witten}, \cite{AdSCFTbook}, \cite{AdSUserguide},  \cite{Skenderis} for calculating correlation functions (\ref{PrimaryCorrelators}) in which the generating functional
\begin{equation}
Z_{CFT}[\phi_{\Delta_i}] = \left\langle 0_{QFT} \right| \exp\left\{ \int d^d x \sum_i \phi_{\Delta_i}(x)\tilde{O}_{\Delta_i}(x) \right\} | 0_{QFT} \rangle  \label{PartFunc}
\end{equation}
features prominently.  The correlation functions (\ref{PrimaryCorrelators}) can be found by taking appropriate functional derivatives of the generating functional (\ref{PartFunc}) with respect to the sources $\phi_{\Delta_i}(x_i)$ and afterwards setting the sources to zero.  The operators $\tilde{O}_{\Delta_i}$ are primary operators, as discussed in section \ref{CFTCorr}.  The generating functional (\ref{PartFunc}) thus represents the single quantity one needs to compute in order to find the quantities of interest, the correlation functions.  \\ \\
Now, the correspondence states that, for an appropriately chosen theory of quantum gravity with fields $\Phi_1, \Phi_2, ..., \Phi_n$, one can relate the partition functions of the $CFT$ and the theory of gravity.  The theory of gravity is in one dimension higher and we indicate this extra dimension by $\beta$ i.e. the fields have argument $\Phi_{\Delta_i}(\beta, x)$.  The correspondence is now   
\begin{equation}
Z_{CFT}[\phi_{\Delta_i}] = Z_{qg}[\Phi_{\Delta_i}(\beta, x)] \ \ \ \textnormal{with} \ \ \ \phi_{\Delta_i}(x) \  \mytilde \ \Phi_i(0, x).   \label{PathInt}
\end{equation}
By $\phi_{\Delta_i}(x) \ \mytilde \ \Phi_{\Delta_i}(0, x)$ we mean the boundary values of the fields $\Phi_{\Delta_i}$ act as the sources of the partition function (\ref{PartFunc}).  This can be best visualised if we write the quantum gravity partition function (if it may be written as such) as
\begin{equation}
Z_{qg}[\Phi_{\Delta_i}(0, x)] = \int_{\vec{\phi}(x) \mytilde \vec{\Phi}(0, x) } D[g_{\mu\nu}] D[\vec{\Phi}(\beta, x) ] \  e^{-S'[\vec{\Phi}(\beta, x) \ , \ g_{\mu\nu}]}.    \label{grav2}   
\end{equation} 
The partition function now only depends on the boundary values of the fields and the asymptotic behaviour of the metric.  The boundary condition for the metric must be such that, on the boundary, the appropriate symmetries are encoded.  For $CFT$'s this is the requirement that the theory of gravity is defined on $AdS$.  \\ \\
The claim is thus that, for the appropriate action, differentiating with respect to the boundary values of the fields will generate correlation functions so that the correlation functions of the quantum theory may be calculated fully on the gravitational side of the duality.  Two of the key aspects that need answering is whether such a gravitational dual exists for every quantum model and how one would go about finding this dual in a systematic way.  \\ \\
One may ask furthermore which field boundary values do you associate with which generating functional sources i.e. which fields are associated with which operators?  For this a set of quantum numbers (like scaling dimension, as discussed, or spin) are required which labels the different operators.  The dictionary states that the appropriate field shares the same set of quantum numbers with its associated operator \cite{Witten}, \cite{AdSUserguide}.  \\ \\
Equation (\ref{PathInt}) is the formal expression of the correspondence.  Two simplifications are customary and are relevant for our analysis ahead.  Firstly, a saddle point approximation for the metric yields an action of the form
\begin{equation}
Z_{f}[\Phi_0(x)] = \int_{\vec{\Phi}(0, x) \mytilde \vec{\Phi}_0(x) } D[\vec{\Phi}(\beta, x)] \ e^{-S'[\vec{\Phi}(\beta, x) \ , \ g^0_{\mu\nu}]}.  \label{gravPart}
\end{equation}
where the metric is fixed on $g^0_{\mu\nu}$.  This is then a model of semi-classical gravity.  If the metric can only fluctuate slightly then this field theory (on a fixed background) is a good approximation to the partition function (\ref{grav2}) \cite{AdSCFTbook}, but is a simplification we will have to motivate later.  Note that the metric is no longer dynamic so that, for instance, we don't take into account the backreaction.  A second simplification is also useful \cite{AdSCFTbook} namely to make a saddle point approximation in the fields also.  Varying the action with respect to the fields yields a differential equation for the fields $\vec{\Phi}$ of which the solutions are $\vec{\Phi}_{cl}$ i.e. 
\begin{equation}
\left. \frac{\delta S}{\delta \vec{\Phi}} \right|_{\vec{\Phi} = \vec{\Phi}_{cl}} = 0   \label{ClassSols}
\end{equation}
The correspondence then becomes
\begin{equation}
Z_{CFT}[\phi(x)] \approx Z_{cl-qg}[\phi(x)] = \sum_{ \vec{\Phi}_{cl} } e^{-S[\Phi_{cl}]} \label{ClassPathInt}
\end{equation}
where we mean $\sum_{ \vec{\Phi}_{cl} }$ to be a sum over all the possible solutions of (\ref{ClassSols}).  In this notation it is slightly more hidden, but the generating functional is still determined by the boundary values for the fields (only now their classical solutions). 

\section{Local / Gauge Symmetries}

The discussion of the previous sections may be viewed as the most basic outline of the correspondence and we have not yet in any way specified how the gravitational theory may be chosen.  In order to discuss further aspects of the correspondence we have to consider more specifics of conformal field theories. \\ \\
In the Maldacena correspondence \cite{Maldacena} there is, in addition to the conformal global symmetry, also the $SU(N)$ gauge symmetry on the $CFT$ side of the duality.  This is a matter we have not addressed yet precisely because gauge symmetry will not be a feature of our ensuing construction.  This aspect is, however, very important both as evidence for the duality as well as for the role played by the gauge group dimension $N$ in defining the strong and weak coupling regimes. Future generalisations of our construction that include these local symmetries are thus very important.  \\ \\
A local transformation, when represented as a unitary operator means $U = e^{i \alpha(t, \vec{x}) T}$ where $T$ is the generator \footnote{We assume that the operator $U$ is well-defined as it illustrates the gauge transformations more clearly.  If not, then our notation means the infinitesimal version of these transformations.}.  Note that, unlike a global symmetry, the coefficient $\alpha$ is now a function of the coordinates.  To best illustrate this difference consider the following action density
\begin{equation}
\mathcal{L}(\phi, \phi^\dag) = \partial_\mu \phi^{\dag} \ \partial^\mu \phi
\end{equation}
with matrix valued fields $\phi$.  Global transformations, where $\phi \rightarrow U\phi$ and $\phi^\dag \rightarrow \phi^\dag U^\dag$ leave $\mathcal{L}$ invariant.  Local transformations, on the other hand, are affected by the derivative and $L$ will thus not retain its form.  In order to allow local transformations one needs to augment $\mathcal{L}$ and consider
\begin{equation}
\mathcal{\mathcal{L}}'(\phi, \phi^\dag, A_\mu) = (\partial_\mu + i A_\mu) \phi^{\dag} (\partial^\mu - i A^\mu)\phi.  
\end{equation}
Local transformations can now be included as a symmetry if the $A_\mu$'s transform as
\begin{equation}
A_\mu \rightarrow U A_\mu U^\dag - i U^\dag \partial_\mu U. 
\end{equation}
A distinguishing property of the global and local transformations thus is that global transformations affect the quantum states or fields only while the local transformations affect the states but also the gauge.  This observation fits well with our construction that will be made in chapter \ref{ConstrChap} - the geometry is constructed from the quantum states and thus can only take note of the global symmetries.  \\ \\
It is useful (and quite typical) to consider these theories in the fundamental representation i.e. the $N \times N$ matrix representation of the gauge group $SU(N)$.  In vector valued theories, such as higher spin \cite{higherSpin}, the fields $\phi$ then represent $N$-index vectors and the inner product $\phi^\dag \phi'$ is simply the dot product while in matrix-valued theories (such as $SYM$) the fields are $N \times N$ matrices with the trace inner product.  The usefulness of this representation is that the $N$-dependence of the inner product of fields becomes explicit.  \\ \\
It was shown by t'Hooft \cite{LargeN} and Witten \cite{LargeNW} that gauge theories permit a $\frac{1}{N}$ expansion for the (many-point) correlators, with each term in this expansion corresponding to a class of diagrams that have a specific topological character.  The Feynman diagrams of the leading order terms, for instance, are planar i.e. they can be drawn without crossings on the surface of a sphere.  The next leading order term can be drawn on a 1-torus (a torus with a single hole), the next on a 2-torus et cetera.  These results are remarkable since one may intuitively expect that increasing $N$ adds complexity to the problem - somehow the converse is true and the theory can be rearranged so that it is in fact simpler in this large $N$ expansion.   \\ \\
This classification scheme and particularly its topological character, is reminiscent of Feynman diagrams for string theory and thus a hint that these gauge theories may be described by string theories \cite{AdSCFTbook}.  The expansion parameter $\frac{1}{N}$ of the topological series is crucial to the convergence properties of this series.  For the Maldacena case of Super Yang-Mills, for instance, the limit needs to be taken in a very specific way \cite{Maldacena}.  The t'Hooft limit is $N \rightarrow \infty$ and $g_{YM} \rightarrow 0$ while keeping $\lambda \equiv g_{YM}^2 N$ constant.  The constant $\lambda$ is called the t'Hooft coupling and $g_{YM}$ is the Yang-Mills coupling.  This limit permits the large $N$ topological expansion of the gauge theory.  

\subsection{The Maldacena Correspondence}

In \cite{Maldacena} the relevant parameters on the side of the of the $\mathcal{N}=4$ $SYM$ are the dimension of the gauge group $N$ and the Yang-Mills coupling $g_{YM}$.  On the gravity side the relevant constants are the string coupling, $g_s$ and the string length $l_s$.  The other parameters, such as the $AdS$ radius, $L$, forms part of the geometry as already discussed.  \\ \\
The Maldacena conjecture relates these quantities explicitly \cite{AdSCFTbook}, \cite{AdSUserguide}
\begin{eqnarray}
g_{YM}^2 & = & g_s \nonumber \\
\lambda & = & \frac{L^4}{l_s^4}   \label{stringy}
\end{eqnarray} 
or alternatively, using the relation between the string scale, Planck scale and the coupling \\ $g_s = \frac{ (l_s)^4}{(l_p)^4}$, we find
\begin{equation}
N = \frac{L^4}{l_p^4}.     \label{planky}
\end{equation}
These conjectured relations between the two theories provide a powerful insight.  Firstly, the conformal field theory can be solved perturbatively if the t'Hooft coupling, $\lambda$ is small.  Conversely the string length is much larger than the length scale of the $AdS$ space from eq. (\ref{stringy}).  The string theory on the $AdS$ background is consequently hard to analyse.  Conversely, if $N$ is large we have that the $AdS$ radius is large compared to the Planck scale from eq. (\ref{planky}).  This implies that quantum effects will play a small role in the string theory so that we may consider a model of semi-classical gravity.  Note that this holds for any value of the string coupling $g_s$ so that we have not specified the t'Hooft coupling.  \\ \\
In other words, the strongly coupled string theory may be described by a weakly coupled field theory and the field theory for large $N$ may be described by semi-classical gravity.  This is a so-called strong/weak duality and it promotes the duality to a powerful tool to calculate physical quantities in the strongly coupled regime. \\ \\
For our purposes we take note of the fact that there exists a limit in which the gauge theory can be accurately described by a semi-classical model of gravity.  In the analysis ahead we will be working with semi-classical gravity since it is the simplest case.  

\section{Summary}

There are many aspects of holography that have been omitted and may be considered in future generalisations.  The aim of this chapter was simply to illuminate some essential aspects of the foundation of the gauge/gravity duality.  The role of symmetries as the core of the correspondence was highlighted along with intuitive arguments for how a quantum theory may be repackaged as a theory of gravitation.  \\ \\
The dictionary of the Maldacena correspondence \cite{Maldacena} was stated and particular note must be taken of the role of the boundary values of fields acting as sources.  It should be noted that the dictionary we will construct in the chapters that follow will not attach this interpretation to the fields of the gravitational model.  For our purposes there is a simpler choice that can be made (and one that relates remarkably well to existing work in the literature).  We will point out exactly where this choice of interpretation is made in the procedure so that one may in future investigate other possibilities.  \\ \\
The simplification from conformal field theory to quantum mechanics will come at a price - we will not be working with gauge theories and will thus apparently lack a large $N$ expansion.  We acknowledge that incorporating gauge symmetry is a layer of complexity that warrants an extensive look in future.  \\ \\
We will show, in the chapters ahead, that this simplification does yield great value in that it is possible to build dual descriptions of quantum mechanical models systematically and explicitly.  This will allow us to investigate the dictionary for the dual theories in a very direct way and we will show how many existing results, especially of the $AdS_2/CFT_1$ correspondence, come about very naturally from this systematic machinery.

\chapter{ Constructing Metrics From Quantum States  }

\label{ConstrChap}

We will be constructing geometries from quantum mechanical models as a first step to finding a systematic, constructive and efficient way to repackage these quantum models as gravitational theories.  We choose this geometric perspective for several reasons.  The matching of (global) symmetries between gauge theories and theories of gravity is one of the main motivations for conjecturing the existence of gauge/gravity dualities.  In numerous works examining candidate duals for quantum mechanical models \cite{Jackiw}, \cite{McGreevyGravDual}, \cite{Son}, \cite{Alisha}, \cite{Wen} a metric possessing the appropriate isometries is taken as a starting point for investigations.  If the symmetries of the two models match, and the sets of symmetries are large (and thus restrictive) enough, then, at least in this sense, a significant part of the dual matching is done. Not only can a procedure be devised that guarantees that the appropriate symmetries of a quantum model are encoded as isometries of a geometric structure but this procedure can be systematic and explicit. \\ \\
As a study of the literature will point out \cite{Ashtekar}, \cite{xpModel}, \cite{Schilling}, \cite{Provost}, \cite{QuantGeom}, \cite{Anandan}, \cite{Bures}, \cite{Bures2},  there are, in fact, many ways to construct geometries from quantum models.  It is thus of critical importance that a sensible choice of geometry is made.  In \cite{Ashtekar} it was shown that appropriate geometric structures allow a given quantum mechanical model to be reformulated entirely in terms of these structures.  This aspect has to be treated with some care - what quantum mechanical quantities can we calculate from our dual description?  Is knowledge of the geometry sufficient to calculate the quantities of interest and, if so, how does one do this?  If not, what is needed in addition to the geometry?\\ \\
In this chapter we will introduce the construction of a metric and anti-symmetric two-form that we will use throughout this thesis.  We will give some motivations for why this construction is chosen.  We elaborate briefly on other intriguing constructions that can be investigated in future.  

\section{Dynamical Symmetries}

Before we present the construction it is important to clarify what is meant by a symmetry of a set of quantum mechanical states as these states are our starting point.  The terminology we will be using is that of dynamical symmetries \cite{Alfaro}, \cite{Henkel} which can also be found in the literature under the name of the kinematical invariance group \cite{NiedererFree}, \cite{NiedererHO}.  As far as this author can tell these names refer to the same symmetries.  \\ \\ 
Throughout this thesis we will use the notation $| \cdot \rangle$ for normalised kets and $|\cdot )$ for kets that aren't necessarily normalised.  Now, consider a family of states labelled by a set of coordinates $\vec{\alpha}$ which may or may not be real.  If there exists a unitary transformation, $U_g$ whose action on the states can be absorbed as a reparametrisation and normalisation of the states $|\vec{\alpha})$ i.e.
\begin{equation}
U_g |\vec{\alpha} )=[f_g(\vec{\alpha})]^{-1}|g(\vec{\alpha}) )  \label{DynamicSymm}
\end{equation}
then the transformation $\vec{\alpha} \rightarrow g(\vec{\alpha})$ is what we will call a dynamical symmetry of the states.  The motivation for this terminology will be more apparent in the next section.  Note that if the states are normalised in (\ref{DynamicSymm}) then the normalisation factor $f_g(\vec{\alpha})$ will simply be a phase.  Since we will be working with both normalised and unnormalised states we keep it as a general normalisation factor.  \\ \\
One may ask why start with the symmetries of quantum states and not, for instance, the symmetries of a Lagrangian or action.  The reason for this will become apparent in section \ref{JackiwSection}.  The transformation properties of the quantum states under unitary transformations will allow us to speak to the properties of state overlaps and expectation values.  These are, for this thesis, the quantum mechanical analogue of correlation functions.   

\subsection{The Schr\"odinger Equation as a Specific Example of Dynamical Symmetries}

\label{SchrDynSection}

The dynamical symmetries are often discussed \cite{NiedererFree}, \cite{NiedererHO} on the level of the time-dependent Schr\"odinger equation.  We would like to stress that, though instructive, this is a specific example of the definition (\ref{DynamicSymm}).   \\ \\  The dynamical symmetries may be visualised, if applicable to the problem under consideration, as the transformations that leave the time-dependent Schr\"odinger equation invariant up to a scale factor i.e. there is a transformation $\left\{x, t \right\} \rightarrow \left\{x'(x, t), t'(x, t) \right\}$ such that 
\begin{eqnarray}
0 & = & i\frac{\partial }{\partial t} \psi( t, x) + \frac{1}{2}\nabla^2 \psi( t, x)  - V(x) \psi( t, x)  \nonumber \\
 \rightarrow \ \  0 & =& i\frac{\partial }{\partial t'} \Psi( t', x') + \frac{1}{2}\nabla'^2 \Psi(t', x')  - V(x') \Psi( t', x') \label{coord}
\end{eqnarray}
where $ \Psi( t'(t, x), x'(t, x)) = f(t, x) \psi( t, x)$ with $f(t, x)$ some scalar function.  We have chosen units such that $\hbar = 1$ and $m = 1$.  \\ \\
We can recast the symmetries of (\ref{coord}) into the form of definition (\ref{DynamicSymm}) by rewriting the wave function in bra-ket notation.  The conjugate of the wave function is given by
\begin{equation}
\psi^*(t, x) = \langle \psi | t, x ) \ \ \ ; \ \ \ \textnormal{where} \ \ \  |t, x) = e^{i t H} e^{i x P}|x=0).  
\end{equation}
The dynamical symmetries, applying definition (\ref{DynamicSymm}), are associated with unitary transformations $U_g |x, t) = [f_g(t, x)]^{-1}|g(t, x) ) $.  The dynamical symmetries thus leave the propagator unchanged up to a normalisation of the states
\begin{equation}
(x', t'| x, t) = (x', t'| U_g^\dag U_g |x, t) = [f_g(t, x)]^{-1}[f_g^*(t', x')]^{-1}(g(t', x')| g(t, x))   \label{FreePartProp}
\end{equation}
but not necessarily the wavefunction.  

\subsection{The Dynamical Symmetries of the Free Particle}

\label{DynSymFreeP}

An important and illustrative example of dynamical symmetries is that of the free particle \cite{NiedererFree}.  The dynamical symmetry generators of the free Schr\"odinger equation in 1+1 dimensions ((\ref{coord}) with $V(x) = 0$) closes on the 1+1 dimensional Schr\"odinger algebra $schr_{1+1}$.  The algebra can be represented in many different ways - for instance as creation and annihilation operators \cite{RepMatrixPoly} or $4\times 4$ matrices  \cite{RepAppell}.  For the purposes of this thesis we will represent them in terms of position and momentum operators
\begin{eqnarray}
I & = & -i(XP - PX) \nonumber \\
H & = & \frac{1}{2}P^2 \nonumber \\ 
D & = & -\frac{1}{4}(XP + PX) \nonumber \\
K & = & \frac{1}{2}X^2  \label{Schr11Comm}
\end{eqnarray}
along with position, $X$, and momentum, $P$.  The operators $H, D, K$ (\ref{Schr11Comm}) are in the $k=\frac{1}{4}$ irrep of $su(1,1)$.  See Appendix \ref{HolsteinApp} for more detail.  The $schr_{1+1}$ algebra closes on the following set of commutation relations
\begin{eqnarray}
 \left[X, P \right] = -i  \ \ \ & ; & \ \ \ \left[K, H\right] = -2 i D   \nonumber \\
\left[ P, D \right] = \frac{i}{2} P \ \ & ; & \ \ \ \left[ X, D \right] = -\frac{i}{2} X \nonumber \\
\left[ P, K  \right] = iX 
 \ \ & ; & \ \ \ \left[ X, H \right] = -iP \nonumber \\
 \left[K, D \right] = -iK & ; & \ \ \ \left[ H, D\right] = i H \nonumber \\
0 & & \textnormal{otherwise} \label{Schr11Alg}
\end{eqnarray}
and is the semi-direct sum of $su(1,1)$ (spanned by $\left\{ H, D, K \right\}$) and the Heisenberg algebra (spanned by $\left\{ P, X, I\right\}$).  These operators derive their names from the coordinate transformation induced on the free particle states, $|t, x) \equiv e^{i t H} e^{i x P}|x=0)$,  namely
\begin{eqnarray}
e^{i a H} |t, x) & = & |t+a, x)  \nonumber \\
e^{i a P} |t, x) & = & |t, x+a) \nonumber \\
e^{i a D} |t, x) & = & e^{\frac{a}{2}}|e^{a}t, e^{\frac{a}{2}}x) \nonumber \\
e^{i a X} |t, x)  & = & e^{i a x + ia^2 t} |t, x+2at) \nonumber \\
e^{i a K} |t, x)  & = & e^{i \frac{\alpha x^2 }{2(1 - \alpha t)}} \left(1 - a t\right)^{-\frac{1}{2}} \left|\frac{t}{1 - a t}, \frac{x}{1 - a t} \right) \label{Schr11Coord}
\end{eqnarray}
which is time translation, space translation, scaling, Galilean boost and special conformal transformations respectively.  These transformations can be calculated using the BCH formulas outlined in appendix \ref{BCHApp}.  The special conformal transformation of (\ref{Schr11Coord}) is shown explicitly in (\ref{SCTrans}). \\ \\
A comment here is in order.  The reader may pick up that the operators (\ref{Schr11Alg}) do not have an explicit time dependence while the generators of dynamical symmetry in \cite{NiedererFree} do.  The time and spatial dependence of the operators come about when they act on the state $|t, x)$.  Their action on the state can be viewed as a differential operator where $H$ represents a time-derivative and $P$ represents a spatial derivative.  Of course, the time-dependent operators $e^{-i t H} U e^{i t H}$ are also symmetry generators of the state $|t, x)$.  \\ \\
We can verify (\ref{FreePartProp}) by explicitly applying the transformations (\ref{Schr11Coord}) to the $1+1$ dimensional free particle propagator
\begin{equation}
(t', x'| t, x) = (2\pi i (t - t'))^{-\frac{1}{2}} e^{-\frac{(x-x')^2 }{2 i (t - t')}}.  \label{FreePartOLap}
\end{equation}
As a final comment, if we restrict ourselves to the free particle symmetries that involve time, $\left\{ H, D, K \right\} $ from eq. (\ref{Schr11Coord}), we explicitly have the $su(1,1) \cong so(2,1) $ algebra.  In section \ref{AdSSpace} we identified the $SO(2,1)$ group as the isometry group of $AdS_2$ (\ref{ConfTransDef}) so that, if the metric we construct encodes dynamical symmetries as isometries, one may anticipate that the geometry will be $AdS_2$.

\section{The Construction of our Metric and Anti-symmetric Two-Form}

The construction we will be employing in this thesis is the metric \cite{Cheng} as studied by Provost and Vallee \cite{Provost}.  This metric is closely related to the work of \cite{Ashtekar}, \cite{Schilling} which will be the topic of section \ref{GeoReform}.  We will show explicitly that this construction can be used to encode the dynamical symmetries of a family of quantum states as isometries of the resulting metric (\ref{DynamicSymm}). We will specifically be considering states that are parametrised by continuous coordinates.  This may seem strange at first sight since in quantum mechanics one typically considers states that are labelled by discrete quantum numbers.  The states of continuous parameters must be viewed as superpositions of these states of discrete quantum numbers where the superposition coefficients are continuous.  \\ \\
We begin by defining the metric \cite{Provost} which will be used in the ensuing construction.  We define it here for an arbitrary family of states and we will apply it to specific physical situations later.  Let $\{\ket{s}\}$ be a manifold of normalized states parametrized by a set of real coordinates $s=(s_1,s_2,s_3,\ldots)$. As mentioned we use $\rket{s}$ to denote a state proportional to $\ket{s}$ which need not be normalized. In the construction we set
\begin{equation}
	\beta_j(s)=-i\partial_j\prm \inp{s}{s'}|_{s=s'}\hs{1.5}{\rm and}\hs{1.5}\gamma_{ij}(s)+i\sigma_{ij}(s)=\partial_i\partial_j\prm\inp{s}{s'}|_{s=s'}, \label{ProvDef}
\end{equation}
where $\gamma_{ij}=\gamma_{ji}$ and $\sigma_{ij}=-\sigma_{ji}$ are related to the real and imaginary parts of the inner product 
\begin{equation}
(\langle s + ds_i | - \langle s |) ( | \vec{s} + ds_j \rangle - | \vec{s} \rangle).  
\end{equation}
By $s + ds_i$ we mean an infinitesimal shift in $s_i$.  The metric \cite{Provost} is then defined as
\begin{equation}
	g_{ij}(s)=\gamma_{ij}(s)-\beta_i(s)\beta_j(s), \label{Prov1}
\end{equation}
which may be rewritten as
\begin{equation}
	g_{ij}(s)=[\,\partial_i\partial_j\prm\log|(s|s')|\,]_{s=s'}. \label{Prov2}
\end{equation}
The definitions (\ref{Prov1}) and (\ref{Prov2}) are completely equivalent.  The subtraction of the $\beta_i \beta_j$ combination in (\ref{Prov1}) ensures that the distance between state vectors that only differ by a phase is zero i.e. the metric (\ref{Prov2}) can be thought of as a ``distance" between physical states.  We can thus refer to the metric and anti-symmetric two-form as being defined on the manifold of rays.  In other words state vectors that differ by a phase factor (or normalisation) are represented by the same point on the manifold.  This can also be seen in (\ref{Prov2}) which has an additional useful property - the metric is no longer sensitive to whether the states are normalised.  Note, importantly, that in the definition (\ref{Prov2}) we have the freedom to use either the normalised or unnormalised states to calculate the metric.  \\ \\
A similar formula to (\ref{Prov2}) exists for the anti-symmetric two-form $\sigma_{ij}$ (\ref{ProvDef}) namely
\begin{equation}
\sigma_{ij} = \frac{1}{2 i} [\,\partial_i\partial_j\prm\log\frac{(s|s')}{(s'|s)}\,]_{s=s'}.  \label{ProvSym}
\end{equation}

\subsubsection{A Quick Example}

As an example of calculating the metric and anti-symmetric two-form consider the application of (\ref{Prov2}) and (\ref{ProvSym}) on the free particle overlap (\ref{FreePartOLap}).  We have that
\begin{equation}
\log (t', x'| t, x) = -\frac{1}{2}\log(2\pi i) - \frac{1}{2}\log(t - t') - \frac{(x - x')^2}{2 i (t - t')}.  \label{logLap} 
\end{equation}
If we apply the formulae (\ref{Prov2}) and (\ref{ProvSym}) directly the metric and two-form will clearly be divergent when we try to set $x' = x$ and $t' = t$.  A way to rectify this is to complexify the coordinates (we will employ regularisation schemes throughout when these situations arise).  We thus alter $t \rightarrow t + i\beta$ and the corresponding $t' \rightarrow t' - i\beta$ on the right hand side of (\ref{logLap}).  The reason for the different sign is the conjugation involved when considering bras vs. kets.  \\ \\
These changes yield
\begin{eqnarray}
& & \log (t'+i\beta', x'| t+i\beta, x) \nonumber \\
& = &-\frac{1}{2}\log(2\pi i) - \frac{1}{2}\log(t - t' + i(\beta + \beta')) - \frac{(x - x')^2}{2 i (t - t' + i(\beta + \beta'))}.  \nonumber 
\end{eqnarray}
One can now readily calculate
\begin{eqnarray}
\partial_t \partial_{t'} \log (t'+i\beta', x'| t+i\beta, x) & = & -\frac{1}{2(t - t' + i(\beta + \beta'))^2} - \frac{i(x-x')^2}{(t-t' + i(\beta + \beta'))^3} \nonumber \\
& = & \partial_\beta \partial_{\beta'} \log (t'+i\beta', x'| t+i\beta, x) \nonumber \\
\partial_t \partial_{\beta'} \log (t'+i\beta', x'| t+i\beta, x) & = & \frac{i}{2(t - t' + i(\beta + \beta'))^2} - \frac{(x-x')^2}{(t-t' + i(\beta + \beta'))^3} \nonumber \\
& = & -\partial_\beta \partial_{t'} \log (t'+i\beta', x'| t+i\beta, x) \nonumber \\
\partial_x \partial_{x'} \log (t'+i\beta', x'| t+i\beta, x) & = &  -\frac{i}{t - t' + i(\beta + \beta')}.  
\end{eqnarray}
The remaining derivatives are omitted since they will yield entries that are zero.  From the above expressions we find
\begin{eqnarray}
g_{tt} = g_{\beta\beta} & = & \frac{1}{8\beta^2} \nonumber \\
g_{xx} & = & -\frac{1}{2\beta}.  
\end{eqnarray}
The non-zero entries for the $\left\{ \beta, t\right\}$ derivatives do not reflect in the metric because they are anti-symmetric.  They do, however, feature in the anti-symmetric two-form
\begin{equation}
\sigma_{\beta t} = -\sigma_{t \beta} = \frac{1}{8\beta^2} 
\end{equation}

%==========================================================================================================
%==========================================================================================================
\subsection{Dynamical Symmetries and Isometries}

The construction is chosen precisely because it encodes the dynamical symmetries of a family of states as the isometries of a metric.  Consider a unitary transformation $U_g$ which produces a mapping $s\rightarrow g(s)$ on the manifold $\{\ket{s}\}$ as in (\ref{DynamicSymm}).  The transformation constitutes a dynamical symmetry as discussed.  In particular, this implies that 
\begin{equation}
 \inp{g(s)}{g(s')}=\inp{s}{s'}f^*_g(s)f_g(s').
\end{equation}
Now consider $s\rightarrow u(s)$ as a coordinate transformation and let $t=u(s)$ denote the new coordinates. Inserting 
$\inp{s}{s'}=\inp{t}{t'}[f^*_g(s)f_g(s')]^{-1}$ into eq. (\ref{Prov2}) reveals that
\begin{equation}
	g_{ij}(s)=[\,\partial_{s_i}\partial_{s'_j}\log|\inp{t}{t'}|\,]_{s=s'}=\frac{\partial u_k}{\partial s_i}\frac{\partial u_l}{\partial s_j}[\,\partial_{u_k}\partial_{u'_l}\log|\inp{t}{t'}|\,]_{t=t'}=\frac{\partial u_k}{\partial s_i}\frac{\partial u_l}{\partial s_j}g_{kl}(t)
\end{equation}
and thus
\begin{equation}
 ds^2=g_{ij}(s)ds_i ds_j=g_{kl}(t)\frac{\partial t_k}{\partial s_i}\frac{\partial t_l}{\partial s_j}ds_i ds_j=g_{kl}(t)dt_kdt_l.
\end{equation}
We conclude that the mapping $s\rightarrow g(s)$ is an isometry of the metric.  

\subsection{Family of States Generated by Group Elements}

\label{GroupFamily}

To get an idea of the physical content of the metric, consider a family of states generated by the action of a unitary representation of some group, $G$, on a reference state $|\phi_0 )$ in the Hilbert space \cite{Perel1}, \cite{Perel2}.  We are thus considering states of the form
\begin{equation}
|\phi_g) = U(g)|\phi_0) \ \ \ ; \ \ \ g \in G
\end{equation}
where $U$ is the (unitary) representation of the group.  The stationary subgroup, $H$ of the state $|\phi_0)$ is defined as the group elements that satisfy
\begin{equation}
U(h)| \phi_0) = e^{i \alpha(h)} |\phi_0) \ \ \ h \in H \subseteq G  \label{HDefine}
\end{equation}
where $\alpha$ is some arbitrary function.  In other words the stationary group does not map a physical state of the Hilbert space onto a different physical state.  We define the factor space $G/H$ as the equivalence classes of group elements related by right multiplication of elements of $H$.  Let $g_h \in G/H$ and let $U(g_h)$ be parametrised by a set of real coordinates $s=(s_1,s_2,s_3,\ldots)$.  We then define the states
\begin{equation}
|s) \equiv U(g_h(s_1, s_2,...)) |\phi_0). \label{alphaStates}
\end{equation} 
The states (\ref{alphaStates}) are thus precisely generalised coherent states \cite{Perel1}, \cite{Perel2}.  Note that in section \ref{SchrDynSection} when we discussed the dynamical symmetries of the Schr\"odinger equation we defined the states $|x, t)$ in terms of unitary transformations of the state $|x=0)$.  The state $|x=0)$ is thus the chosen reference state (in the language of coherent states).  \\ \\
The generators feature in the expression $U^{\dag}(\vec{s}) \partial_{s_i} U(\vec{s}) = \sum_n f_{i, n}(\vec{s}) K_n $ where the $K_n$'s are the generators of the group and the $f_{i, n}$'s are coordinate dependent coefficients.  If we use the states (\ref{alphaStates}) and calculate the metric (\ref{Prov2}) we find
\begin{equation}
g_{ij} = \sum_{m, n}(f_{i, m})^* f_{j,n} \left( \langle s| K^\dag_m K_n |s\rangle - \langle s| K^\dag_m |s\rangle  \langle s| K_n  | s \rangle\right) \equiv \sum_{m,n}(f_{i, m})^* f_{j, n} P_{m n}  \label{GroupMetric}
\end{equation}
so that the metric is formed from the expectation values of products of algebra elements w.r.t. the normalised state $|s \rangle$ \cite{Provost}.  Note that combinations of algebra elements that have $|\phi_0)$ as an eigenvector correspond to zero distance entries.  This is precisely why, to produce sensible metrics, we need to restrict ourselves to the factor space (\ref{alphaStates}).  Note that, due to (\ref{HDefine}), a choice of reference state with a larger stationary group will lead to a metric of lower dimension. 
\\ \\
As a last, but very important remark, note that we can only produce sensible metrics (\ref{GroupMetric}) if the family of states is normalisable.  If it is not normalisable then it will be essential to regularise the reference state i.e. we require a family of states $| \phi_0(\vec{\beta}) ) $, parametrised by real coordinates $\vec{\beta} = (\beta_1, \beta_2, ...)$, so that $ ( \phi_0(\vec{\beta})  | \phi_0(\vec{\beta}) ) < \infty$ for some domain of $\vec{\beta}$ and $| \phi_0(\vec{0}) ) = |\phi_0 )$.  This regularisation will almost certainly break (some of) the symmetries of the family of states but these are recovered on the $\vec{\beta} \rightarrow \vec{0}$ boundary, where the original family of states are defined.  We interpret these regularisation parameters as bulk coordinates so that we have a clear ``bulk" and ``boundary" region.  For the examples that we will be considering it is possible to choose the regularisation in such a way that all the symmetries are retained as isometries in the bulk.

\section{The Geometric Reformulation of Quantum Mechanics}

\label{GeoReform}

Thus far we have identified the construction (\ref{Prov2}) which encodes the dynamical symmetries of a set of quantum states as isometries of a metric, whilst being a relatively straightforward calculation.  These features make the construction an appealing first step towards building dual descriptions.  We will now point out that this metric and anti-symmetric two-form (\ref{Prov2}) is related to the geometric reformulation of quantum mechanics of Ashtekar and Schilling \cite{Ashtekar}, \cite{Schilling}.  \\ \\
First, some context is necessary.  This geometric reformulation of quantum mechanics starts from an interesting premise: can quantum mechanics be formulated entirely in terms of measurable quantities?  In other words, can quantum mechanics be formulated entirely in terms of the projective Hilbert space?  The result to the ensuing analysis is a geometric reformulation of quantum mechanics.  For our purposes the precise form of, for instance, the postulates of quantum mechanics in this reformulation are unimportant (these can all be found in \cite{Ashtekar}, \cite{Schilling}). \\ \\
We will here only give a brief summary of the aspects that will be important for our purposes.  We borrow their notation almost identically.  Their conventions for symplectic geometry is that the Hamiltonian vector field $X_f$, generated by the function $f$ satisfies $i_{X_f} \Omega = d f$, where $\Omega$ is the symplectic form.  \\ \\
The starting point of the reformulation \cite{Ashtekar} is to take a complex Hilbert space and decompose it into its real and imaginary parts.  In terms of the inner product of two arbitrary states in the Hilbert space $|\psi), |\phi)$ this is
\begin{eqnarray}
(\phi| \psi) &=& \operatorname{Re}((\phi| \psi) ) + i \operatorname{Im}((\phi| \psi)) \nonumber \\
&\equiv& \frac{1}{2}G(\phi, \psi) + \frac{i}{2} \Omega(\phi, \psi).  \label{AshDef} 
\end{eqnarray}
The properties of the inner product imply that $G$ is positive definite and that $\Omega$ is a symplectic form.  Now one interprets multiplication by $i$ as a complex structure $J$ and splits the complex Hilbert space into two real parts, connected by $J$.  This implies the relation
\begin{equation}
G(\phi, \psi) = \Omega(\phi, J\psi) = -\Omega(J\phi, \psi)  \label{GOmega}
\end{equation}
so that the Hilbert space has the structure of a K\"ahler space \cite{Ashtekar}.  One may associate with each quantum observable $\hat{F}$ the vector field
\begin{equation}
Y_{\hat{F}}(\psi) \equiv -J \hat{F}|\psi )  \label{SchrVField}
\end{equation}
which is called the Schr\"odinger vector field.  We can also calculate the expectation value function
\begin{equation}
F(\psi) = (\psi| \hat{F}|\psi ).  
\end{equation}
One can now readily prove that, if $\eta$ is any tangent vector at $\Psi$, then
\begin{eqnarray}
dF(\eta) & = & \left.\frac{d}{d\lambda} \langle \psi + \lambda \eta | \hat{F} | \psi + \lambda \eta \rangle \right|_{\lambda = 0} \nonumber \\
				 & = & \langle \psi | \hat{F} | \eta \rangle + \langle \eta | \hat{F} \psi \rangle \nonumber \\
				& = & G(\hat{F}\psi, \eta) \nonumber \\ 
				& = & G(J Y_{\hat{F}}(\psi), \eta) \nonumber \\
				& = & \Omega(Y_{\hat{F}}, \eta) = (i_{ Y_{\hat{F}}} \Omega)(\eta)  \label{KillDerive}
\end{eqnarray}
after using (\ref{GOmega}).  This implies that the Schr\"odinger vector field (\ref{SchrVField}) determined by the observable $\hat{F}$ is exactly the Hamiltonian vector field $X_F$ generated by the expectation value of $\hat{F}$ \cite{Schilling}.  In index notation this is
\begin{equation}
Y_{\hat{F}}^a = \Omega^{a b}\partial_b F  \label{SigmaVec}
\end{equation}
which clearly relates the expectation value and a corresponding vector field on the manifold. \\ \\
This relation between the (normalised) expectation values of operators and the corresponding vector fields on the manifold will be crucial to the dictionary we develop from chapter \ref{GravChap} onwards.  We will thus provide the proof relevant to the construction (\ref{Prov2}) explicitly.  We will make one generalisation namely that we will calculate the vector field associated with a, in general, non-hermitian operator.  We start with the inner product
\begin{equation}
I_s(|a\rangle, |b\rangle) = \langle a | b \rangle - \langle a | s\rangle \langle s | b\rangle
\end{equation} 
which is the inner product of the components orthogonal to $|s \rangle $ of two tangent vectors at $|s \rangle $.  What is important to note is that the metric and anti-symmetric two-form of (\ref{Prov2}) can be written as
\begin{equation}
I_s( \partial_i |s\rangle, \partial_j |s\rangle   ) = g_{ij}(s) + i \sigma_{ij}(s) \equiv G_s( \partial_i |s\rangle, \partial_j |s\rangle   ) + i \Omega_s( \partial_i |s\rangle, \partial_j |s\rangle   ).  
\end{equation}
We have defined $G_s$ and $\Omega_s$ in accordance with (\ref{AshDef}).  A number of properties of $G_s$ and $\Omega_s$ are important for our purposes.  First we have that
\begin{equation}
I_s(N(s)|s\rangle, |a\rangle) = 0 = G_s(N(s)|s\rangle, |a\rangle) = \Omega_s(N(s)|s\rangle, |a\rangle)  \label{Prop1}
\end{equation}
where $N(s) |s\rangle$ is proportional to $|s\rangle$.  Secondly we can show that
\begin{equation}
G_s(A |s\rangle, \partial_j |s\rangle) = \frac{1}{2} \partial_j \left( \frac{(s|A|s)}{(s|s)} \right) = \frac{1}{2}\partial_j \langle A \rangle \label{Prop2}
\end{equation}
if $A$ is an hermitian operator.  Finally we have the property (\ref{GOmega}) which means
\begin{equation}
\Omega_s( -i |a\rangle , |b\rangle ) = G_s(|a\rangle, |b\rangle).  \label{Prop3}
\end{equation}
We express this, as discussed, as a linear mapping $J$ that acts as
\begin{equation}
i \partial_k |s) = J^j_{\ k} \partial_j |s)  \label{Prop4}
\end{equation}
which implies that $J^i_{\ k} J^{k}_{\ j} = -1$ and $\sigma_{ij} = J^{k}_{\ i} \ g_{kj}$.  With the properties (\ref{Prop1})-(\ref{Prop4}) we can now relate the expectation value of an operator to a vector field.  \\ \\
Consider an operator $A$ that has the effect on the family of states of
\begin{equation}
A|s\rangle = f(s) |s\rangle - i \chi_A |s\rangle \equiv f(s) |s\rangle - i \chi^\mu_A \partial_\mu |s\rangle   \label{ATrans}
\end{equation} 
of which an example are the generators of dynamical symmetries (\ref{DynamicSymm}).  However, we do not assume that $A$ is hermitian and the transformation may be defined only locally.  We would like to relate the vector field $\chi^\mu_A \partial_\mu$ or $\chi_A$, to the expectation value of $A$.  We first split $A$ up into its hermitian and anti-hermitian parts i.e. $A = A_1 + i A_2$ where $A_1$ and $A_2$ are hermitian.  From (\ref{ATrans}) and (\ref{Prop1}) we have that
\begin{equation}
\Omega_s(i A|s\rangle, \partial_j|s\rangle ) = \Omega_s(\chi_A^i \partial_i |s\rangle, \partial_j|s\rangle ) = \chi_a^i \sigma_{ij}.   \label{VecField}
\end{equation}
We can also expand the left-hand side as
\begin{eqnarray}
\Omega_s(i A|s\rangle, \partial_j|s\rangle ) & = & \Omega_s(i A_1 |s\rangle, \partial_j|s\rangle) + \Omega_s(-A_2|s\rangle, \partial_j|s\rangle ) \nonumber \\
& = & \Omega_s(i A_1 |s\rangle, \partial_j|s\rangle) + J^k_{\ j} \Omega_s(-i A_2|s\rangle, \partial_k|s\rangle ) \nonumber \\
& = & -G_s(A_1 |s\rangle, \partial_j |s\rangle ) + J^k_{\ j} G_s( A_2|s\rangle, \partial_k|s\rangle) \nonumber \\
& = & -\frac{1}{2} \partial_j \langle A_1 \rangle + \frac{1}{2} J^k_{\ j}\partial_j \langle A_2 \rangle.   \label{ExpToVec}
\end{eqnarray}
In the first line we used the linearity of $\Omega_s$, in the second line property (\ref{Prop4}), in the third line property (\ref{Prop3}) and in the final line property (\ref{Prop2}).  Combining (\ref{VecField}) and (\ref{ExpToVec}) we now find
\begin{equation}
\chi_A^i = -\frac{1}{2} \left( \sigma^{ij} \partial_{j} \langle A_{1} \rangle + g^{ij}\partial_j \langle A_{2} \rangle   \right) \label{AshOpVec}
\end{equation}
which relates the vector field to the (normalised) expectation value explicitly.  The equation (\ref{AshOpVec}) will be indispensable in the development of the dictionary from chapter \ref{GravChap} onwards.  

\section{A Brief Look at Other Possible Constructions}

Before we proceed to construct the metrics and anti-symmetric two-forms for specific quantum mechanical models we take note of other possible metrics that may be used.  The discussion will also provide some insight into why we have chosen our particular construction.

\subsection{Left- and Right Multiplication Symmetries}

To lead into such a discussion we introduce the idea of left- and right multiplication symmetries.  The dynamical symmetries of quantum states are so-called left multiplication symmetries, as can be seen in (\ref{alphaStates}).  The transformations follow from the left multiplication by a unitary operator on a group element (which then acts on some reference state from the Hilbert space).  \\ \\
One can also define right multiplication symmetries where the transformation is induced by right multiplication by a unitary operator.   Given some group element $G(x_1, ... , x_n)$ left multiplication symmetries are defined via left multiplication by some unitary operator $U_g G(x_1, ... , x_n) \propto G(g_l(\vec{x}))$ and right multiplication symmetries by right multiplication of that group element $G(x_1, ... , x_n) U_g \propto G(g_r(\vec{x}))$.  The transformations induced by left and right multiplication symmetries are usually different, except in special circumstances, and thus represent different symmetries.  \\ \\
With the dynamical symmetries, due to the presence of the reference state $| \phi_0)$, (at least some and in the cases we will investigate all of) the right multiplication symmetries are removed.  Group elements, upon right multiplication, that have the reference state  $| \phi_0)$ as an eigenvector perform only trivial transformations and are factored out.  Of the remaining right multiplication transformations only the ones that cannot be reproduced by some left multiplication are considered.  For the cases we will consider that set is empty and the quantum states do not possess right multiplication symmetries.  \\ \\
We may wish, for some reason, to construct a metric that encodes both left and right multiplication symmetries.  These metrics are called bi-invariant \cite{Gibbons} since they are both left- and right-invariant.  The geometry would encode the full set of symmetries of the underlying group.  Intuitively it is hard to imagine that there is then enough freedom to encode the dynamics of a given quantum mechanical model and we thus will not use these for the purposes of building duals.  Nonetheless, these metrics provide us with some instructive examples and we will indicate how they may be augmented to encode information of the dynamics.  

\subsubsection{Metrics on Group Manifolds}

Suppose we have a group $G$ with a finite number of generators in some representation $S(g)$ where $g \in G$ which may or may not be a unitary representation. The metric on the group manifold is defined as \cite{Gibbons}
\begin{equation}
g_{\mu\nu} \equiv tr\left( \left\{ S^{-1} \partial_{(\alpha_\mu)} S, S^{-1} \partial_{(\alpha_\nu)} S  \right\} \right) \label{groupMetric}
\end{equation}
where $\partial_{\alpha_\mu}$ indicates a derivative with respect to $\alpha_\mu$.  This metric contains both left multiplication isometries and right multiplication isometries.  The left multiplication isometries follow almost immediately since if $S \rightarrow U S$ then
\begin{equation}
S^{-1}\partial_{(\alpha_\mu)} S \rightarrow S^{-1}(\vec{\alpha}) U^{-1} \partial_{(\alpha_\mu)} U S(\vec{\alpha}) = S^{-1}(\vec{\alpha})\partial_{(\alpha_\mu)} S(\vec{\alpha}),
\end{equation}
which implies that
\begin{eqnarray}
S^{-1}(\vec{\alpha})\partial_{\alpha_\mu} S(\vec{\alpha}) & = & S^{-1}(\vec{\alpha}) U^{-1} \partial_{(\alpha_\mu)} U S(\vec{\alpha}) \nonumber \\
 & = & S^{-1}(\beta(\alpha) ) \partial_{(\alpha_\mu)} S(\beta(\alpha)) \nonumber \\ 
 & = & S^{-1}(\beta(\alpha) ) \partial_{(\beta_\gamma)} S(\beta(\alpha)) \frac{\partial \beta^\gamma}{\partial \alpha^\mu}.  
\end{eqnarray}
This implies that indeed
\begin{equation}
g_{\mu\nu}(\beta) = g_{\gamma\delta}(\alpha) \frac{\partial \alpha^\gamma}{\partial \beta^\mu} \frac{\partial \alpha^\delta}{\partial \beta^\nu}.
\end{equation}
The proof for the right multiplication isometries follow similarly once the cyclic property of the trace is employed.  It is important to note that all the coordinates that are transformed by the group multiplication have to be included on the geometric manifold in order for the transformations to be encoded as isometries.  If only a subset of the coordinates is used then all symmetries that transform the fixed coordinates are not included as isometries.  An example of this will follow in the next section.  \\ \\
Instead of referring to ``a" metric on the group manifold we referred to the metric (\ref{groupMetric}) as ``the" metric on the group manifold.  This is because all metrics that have the dimension of the number of generators and possesses all left and right isometries must be proportional.  The number of isometries pins down the form of the metric up to a constant factor and coordinate transformation.  \\ \\
As two examples, instructive for later purposes, we consider the metrics of $SU(2)$ and $SU(1,1)$ in a $2\times 2$ matrix representation.  A general $SU(2)$ group element can be written as  
\begin{equation}
S_{SU(2)} = \left( \begin{array}{cc} e^{-i t} \cos(r) & -e^{-i\theta}\sin{(r)} \\ e^{i\theta}\sin{(r)} & e^{it} \cos{(r)} \end{array}\right).
\end{equation}
By now appying the definition of the metric (\ref{groupMetric}) and ensuring a positive signature we find the metric
\begin{equation}
ds^2 = 4\cos^2(r) dt^2 + 4 dr^2 + 4 \sin^2(r) d\theta^2  \label{SU2GroupMetric}
\end{equation}
which is a parametrisation of a $2+1$-dimensional de Sitter metric with scalar curvature $R = \frac{3}{2}$.  A similar calculation for a general $SU(1,1)$ group element
\begin{equation}
S_{SU(1,1)} = \left(\begin{array}{cc} e^{-i t}\sqrt{1 + r^2} & e^{-i \theta} r \\ e^{i\theta} r & e^{i t}\sqrt{1+r^2} \end{array} \right)
\end{equation}
yields the metric
\begin{equation}
ds^2 = -4( 1+r^2 ) dt^2 + \frac{4}{1 + r^2} dr^2 + 4 r^2 d\theta^2  
\end{equation}
which is the $AdS_{2+1}$ metric with scalar curvature $R = -\frac{3}{2}$.  These results are as one would anticipate.  These are three-parameter groups which thus yields six symmetries, three left-multiplication and three right multiplication symmetries.  The resulting three-dimensional metric will thus be maximally symmetric.  The compact group yields a de Sitter metric and the non-compact group an anti-de Sitter metric.   \\ \\
For these examples we have considered the groups in $2 \times 2$ matrix representation.  As a final note we discuss the role of the representation of groups on the level of the metrics.  The objects 
\begin{equation}
S^{-1} \partial_\mu S = f^i_\mu K_i
\end{equation}
where $K_i$ are group generators and $f^i_\mu$ are scalar functions that depend on the coordinates.  By inserting this result into the definition of the metric (\ref{groupMetric}) we find
\begin{equation}
g_{\mu\nu} = f^{i}_\mu f^{j}_\nu \  tr\left( \left\{ K_i, K_j \right\} \right) \equiv f^{\alpha}_\mu \ f^{\beta}_\nu \  P_{\alpha \beta}.
\end{equation}
Now clearly the coordinate dependency is completely contained in the tensors $f^{i}_\mu$ and they are representation independent.  The tensor $P_{\alpha \beta}$, on the other hand, can depend on the specific representation.  In most cases these only differ by a constant factor (and they simply multiply geometric quantities like the scalar curvature by some constant) but they may be different when reducible and irreducible representations are considered. 

\subsubsection{Density Matrix as a Symmetry Filter}

\label{DMatSymmFilter}

It is possible to augment the group metric (\ref{groupMetric}) so that we construct metrics that possess a more restricted symmetry.  We introduce the density matrix $\rho = e^{-\beta H}$ and write
\begin{equation}
g^{F}_{\mu\nu} = tr\left( \rho \left\{ S^{-1} \partial_{(\alpha_\mu)} S, S^{-1} \partial_{(\alpha_\nu)} S  \right\} \right).
\end{equation}
Clearly the above metric possesses left multiplication symmetry by default.  The right multiplication symmetries, however, are only encoded as isometries if 
\begin{equation}
U_g\rho(\beta) U_g^\dag = \rho(\beta) .\label{rhoSymmetry}
\end{equation}  
The density matrix thus acts as a filter for right multiplication symmetries.  It is also possible to write down definitions where it is rather the left multiplication symmetry (or both) which is broken. \\ \\ 
If the parameter $\beta$ is included as a coordinate on the manifold then a transformation of $\beta$ is allowed for a symmetry transformation i.e. the transformation will represent a symmetry if
\begin{equation}
U_g\rho(\beta) U_g^\dag = \rho(\beta')
\end{equation}  
Explicitly we then consider the metric
\begin{equation}
ds^2 = c_1 \partial^2_\beta ( tr(\rho) ) d\beta^2 + g^F_{\mu\nu}d\alpha^\mu d\alpha^\nu
\end{equation}
where $c_1$ is some real constant.  On the level of the operators, an operator $A$ with the property
\begin{equation}
[H, A] = q H \ \ \ ; \ \ \ q \in \mathbb{R}  \label{CommSymm}
\end{equation}
 with the Hamiltonian will be included as symmetries if $\beta$ is a coordinate on the manifold.  If it is not included as a coordinate then only the operators $A$ that truly commute with $H$ will induce symmetry transformations.   \\ \\
This approach to constructing metrics is not ideal, since we need to include a large number of coordinates on the manifold.  By inserting the density matrix filter we are breaking all the right multiplication symmetries except those that satisfy (\ref{rhoSymmetry}).  These metrics will thus always possess a rather small number of symmetries so that we do not expect to find simple duals for the simple quantum models.   This is why the construction (\ref{Prov2}) is better suited than these metrics constructed by applying symmetry filters to group manifolds.  Nonetheless, some of the examples we have pointed out are instructive for our later purposes.    

\subsection{Bures Metric}

The density matrix itself permits the definition of a metric.  Though we will not be using this metric in our subsequent construction, the Bures metric \cite{Bures}, \cite{Bures2}, is stated here for the sake of completeness and possible future study.  Since the geometry is computed directly from physical information of the quantum mechanical model it is an appealing construction.  \\ \\
The Bures metric considers how the density matrix changes with an infinitesimal shift.  Suppose we have $\rho_0(s) = e^{H(s)}$ where $H(s)$ is some Hamiltonian and we make an infinitesimal shift
\begin{equation}
\rho_0(s+ds) = e^{H(s) + dH} = \rho_0(s) + \rho_0(s) dF_0 \equiv \rho_0 + d \rho_0
\end{equation}
where 
\begin{equation}
dF_0 = \int_0^1 e^{-\lambda H} dH e^{\lambda H} d\lambda = -\int_0^1 e^{\lambda ad_H} dH d\lambda =  \left( \frac{1 - e^{ad_H}}{ad_H} \right) dH. \label{dF0}
\end{equation}
Here $ad_H$ refers to the adjoint representation mapping for the algebra $ad_H X = [X, H]$.  The change in the density matrix is then $\rho_0 dF_0$.  In the definition of the Bures metric we will not be interested in $dF_0$ but rather the operator $dG_0$ which is defined as
\begin{eqnarray}
d\rho_0 & = & \rho_0 dG_0 + dG_0 \rho_0 \nonumber \\
\Rightarrow dF_0 & = & dG_0 + \rho_0^{-1} dG_0 \rho_0 \nonumber \\
                 & = & \left(1 + e^{ad_H} \right) dG_0.   \label{dG0Def}
\end{eqnarray}
Combining (\ref{dF0}) and (\ref{dG0Def}) yields a formal expression for the operator $dG_0$
\begin{equation}
dG_0 = \frac{\tanh\left(\frac{1}{2} ad_H  \right)}{ad_H} dH.
\end{equation}
The Bures metric is then defined as 
\begin{equation}
ds^2 = tr\left( \rho dG_0 dG_0 \right) - tr\left(\rho dG_0 \right)^2 \label{Bures}
\end{equation}
where $\rho = \frac{\rho_0(s)}{tr(\rho_0(s))}$ is the normalised density matrix.  \\ \\
The explicit calculation of the metric (\ref{Bures}) is not a trivial task and, though the Bures metric is appealing for its physical content, the construction we employ will encode the desired symmetries with significantly less calculational hazard.

\section{Summary}

In chapter \ref{BackChap} we noted the central importance of symmetries in the formulation of the gauge/gravity duality.  Our goal in this thesis is to build duals to quantum mechanical models in a systematic way and we chose the global symmetries as our point of departure.  In this chapter we presented the construction (\ref{Prov2}) that we will be employing throughout the thesis, highlighted some of its most important features and argued why it is a more sensible starting point than other ways to construct metrics from quantum states.  The primary reason is that it is computationally simple but also preserves the symmetry.  \\ \\
The metric is constituted of the expectation values of the products of algebra elements with respect to the reference state and, in the case of a non-normalisable reference state, the states have to be appropriately regularised.  This regularisation naturally gives rise to the idea of a boundary, where the unregularised states live, and the bulk where the states are states are normalisable.  We mentioned, but have not yet shown, that the regularisation scheme can be chosen in such a way that the boundary symmetries of this manifold of states are encoded as bulk isometries.  This will be shown in the next chapter when we consider specific examples.  \\ \\
We showed furthermore that states parametrised by complex coordinates give rise to a metric and anti-symmetric two-form that is equivalent to those used in the geometric formulation of quantum mechanics.  This duality between the expectation values of operators and vectors on a manifold will be of great use to us in chapter \ref{GravChap}.

\chapter{Geometry of Time-Evolved States}

\label{GeomChap}

Having introduced the construction that encodes the dynamical symmetries of quantum states as isometries of a metric, we will now proceed to apply this to physical models.  The construction will produce for us a geometry which we will later, in chapter \ref{GravChap}, link in some way to a dual theory of gravity.  \\ \\
We will start with the simplest possible class of physical models for our purposes - quantum mechanics in $0+1$ dimensions.  However, there is possibly some care that needs to be taken.  The one- and two-dimensional conformal group is of a very different character from its counterparts in higher dimensions and needs to be treated carefully in the $AdS/CFT$ dictionary.  One would thus also expect that the treatment of the lower dimensional examples to be different in our construction.  \\ \\
Fortunately our approach here will generalise rather simply to the higher dimensional cases since the geometry is only sensitive to the global symmetries.  The infinite dimensional conformal symmetry, a novelty of the one- and two-dimensional conformal group, will not feature as metric isometries.  

\section{Regularised States}

Consider a family of quantum states generated by the time translation of some reference state, $|\phi_0)$, in the Hilbert space
\begin{equation}
| t  ) \equiv \left \{ \begin{array}{cc}  U^\dag(t, 0) | \phi_0 ) & t \geq 0  \\ U(0, t) | \phi_0 ) & t < 0 \end{array}     \right. \label{tStates}
\end{equation}  
with some time evolution operator.  As a shorthand notation we will refer to the time evolution of these states simply as 
\begin{equation}
|t ) \equiv U(t) |\phi_0).  \
\end{equation}
The symmetry transformations of (\ref{tStates}), if there are any, are unitary actions on the state as
\begin{equation}
U_g(\alpha) |t) = [f_g(t)]^{-1} |g(t))
\end{equation}
and have the effect of parametrising time and changing the normalisation of the state (\ref{DynamicSymm}).  Suppose now the reference state $\left| \phi_0 \right)$ is non-normalisable (a position eigenstate for example).  As discussed in section \ref{GroupFamily}, in order to define a sensible metric we have to regularise the reference state by some means.  We therefore consider the states
\begin{equation}
| t, \beta) \equiv U(t) e^{-\beta H_0}| \phi_0 ) \ \ ; \ \ \beta > 0 \label{States}
\end{equation} 
where $H_0$ is some operator bounded from below that does not have $| \phi_0 )$ as an eigenstate.  We will make the assumption that the states (\ref{States}) are now normalisable.  For all the examples we will consider in this thesis this assumption will hold.  \\ \\
We will interpret the parameter $\beta$ to be a coordinate on the now two-dimensional manifold of states.  The reason for this is simply that, for the examples we will consider and appropriate choices of $H_0$, this will allow us to retain the symmetries of the original states $|t)$ (\ref{tStates}) as isometries in the bulk ($\beta >0$). \\ \\ 
Note that the quantum states of our model (\ref{tStates}) are defined on the $\beta\rightarrow 0$ boundary of this two-dimensional manifold of states (\ref{States}) and that the added dimension, $\beta$, has the interpretation of an energy scale.  This coincides well with the conventional $AdS/CFT$ picture where the $AdS$ radial coordinate is associated with an energy scale \cite{AdSUserguide} and the quantum theory lives on the boundary of the dual theory.  

\section{The Physical Content of the Metrics}

For the purposes of this discussion we will be adding the additional dimension, $\beta$, regardless of whether the reference state is normalisable or not.  We will show in section (\ref{CohSection}) why this is a beneficial choice (and, of course, essential for non-normalisable reference states).  For the purposes of our construction we thus define the states as in eq. (\ref{States}), where we may now specify the reference state, the time evolution and the regularisation.  \\ \\
We will assume that the time-evolution operator satisfies
\begin{equation}
\partial_t U(t) \equiv i G(t)U(t) \ \ \; \ \ \ G(t) = H_1 + \gamma(t)H_2  \label{TimeEvolveDef}
\end{equation}
i.e. that $G(t)$ splits into a time-independent part and time-dependent part where $\gamma(t)$ is a time-dependent coupling.  Note that the time evolution (\ref{TimeEvolveDef}) is a kind of interaction picture where the states evolve backwards in time.  Applying the metric formula (\ref{Prov2}) now yields
\begin{equation}
ds^2 = \langle \beta | (\delta G(t))^2    | \beta \rangle dt^2 + i \langle \beta | [G(t), H_0] | \beta \rangle d\beta dt + \langle \beta | (\delta H_0)^2    | \beta \rangle d\beta^2 \label{ConstructMetric}
\end{equation}
where $\delta A = A - \langle \beta | A | \beta \rangle$ and $|\beta\rangle \propto e^{-\beta H_0}|\phi_0)$ is the normalised regularised reference state.  As was also shown more generally in equation (\ref{GroupMetric}), the information contained in the metric is essentially variances of the different generators.      \\ \\
An interesting case is when $G(t) = H_0$ i.e. the Hamiltonian is time-independent and the regularisation is just the Hamiltonian itself.  We will refer to these cases as unsourced and the resulting metrics as unsourced metrics.  Inserting these choices into the more general metric (\ref{ConstructMetric}) then yields
\begin{eqnarray}
ds^2 & =&  \langle \beta | (\delta H_0)^2    | \beta \rangle \left( d\beta^2 + dt^2\right) \nonumber \\
& \equiv & C_{H_0}(\beta)\left( d\beta^2 + dt^2\right) \label{aAdSmetric}
\end{eqnarray}   
where we have defined $C_{H_0}$ as the conformal factor of the metric.  The scalar curvature can be expressed exclusively in terms of this conformal factor
\begin{equation}
R = -\frac{\partial^2_\beta \log(C_{H_0} )}{C_{H_0}}.  \label{scaleCurve}
\end{equation}
This is true for any choice of $H_0$.  Note that since the states used are of the form 
\begin{equation}
e^{i(t + i\beta)H_0}|\phi_0) \equiv e^{i\tau H_0}|\phi_0) \equiv |\tau).   \label{tauStates}
\end{equation}
In the coordinate $\tau = t + i\beta$ it is clear that the manifold of states is K\"ahler \cite{Kahler1}, \cite{Kahler2}.  The K\"ahler structure gives rise to many useful features.  For our future purposes two are worthwhile to note.  First, there exists a quantity called the K\"ahler potential and explicit expressions for the curvature tensors can be given in terms of the K\"ahler potential.  The K\"ahler potential is given by
\begin{equation}
K(\overline{\tau}, \tau) = \log(\overline{\tau}|\tau).  
\end{equation}
Second, there exists a compatible symplectic structure.  The metric may be calculated by taking derivatives of the K\"ahler potential
\begin{equation}
ds^2 = \partial_\tau \partial_{\overline{\tau}} \log(\overline{\tau}|\tau)   d\tau d\overline{\tau} = \frac{1}{4}\partial_\beta^2 \log (\beta | \beta)(d\beta^2 + dt^2).  \label{KahlerProvost}
\end{equation}
Note that the kets (\ref{tauStates}) are holomorphic in $\tau$ while the bras are anti-holomorphic.  This feature implies that not necessary to impose the condition that the coordinates of the bra's and kets be set equal (like in (\ref{Prov2})).  This is because the complex coordinates relate the primed coordinates to $\overline{\tau}$ and the unprimed coordinates to $\tau$ while $\partial_\tau \overline{\tau} = \partial_{\overline{\tau}} \tau = 0$.  It follows that the conformal factor in eq. (\ref{aAdSmetric}) can be expressed in terms of $(\overline{\tau}|\tau)$.  

\subsection{Non-normalisable Reference State and $aAdS$}

If, in addition to our choice of $G(t) = H_0$, the reference state $|\phi_0)$ is also non-normalisable the overlap of the states $(\beta, t| \beta, t)$ is singular at $\beta = 0$.  For the cases we will be considering later this is not an essential singularity and there exists some exponent $m>0$ so that we may expand the overlap of (\ref{States}) in the form
\begin{equation}
(t, \beta| t, \beta) = (\beta|\beta)  = \sum_{n=0}^{\infty} c_n \beta^{-m + n}
\end{equation}
where the $c_n$'s are arbitrary expansion coefficients.  The metric can then be calculated to be
\begin{equation}
ds^2 = \frac{m}{4 \beta^2}\left(1 +  \sum_{n=2}^{\infty} k_n \beta^n \right) \left( d\beta^2 + dt^2\right)  \label{aAdSmetricDef}
\end{equation}
where the $k_n$'s are $c_n$-dependent expansion coefficients.  The metric can be recognised as an asymptotically $AdS_2$ ($aAdS_2$) metric \cite{Skenderis}.  This is because, near the $\beta \rightarrow 0$ boundary, the metric looks like $AdS_2$ 
\begin{equation}
ds^2 = \left(\frac{m}{4 \beta^2} + O(1) \right) d\beta^2 + \left(\frac{m}{4 \beta^2} + O(1)\right) dt^2.
\end{equation}
For a non-normalisable reference state $\left| \phi_0 \right)$ an $aAdS_{2}$-geometry thus arises quite generally from a family of states of the form (\ref{States}).  This is even true for a large number of cases where $U(t) \neq e^{i t H_0}$ and depends on the regularisation chosen relative to the time evolution operator.  Examples of this will follow later.  This very natural emergence of an $aAdS$ geometry is a comforting and insightful result for the purpose of our toy model of $AdS/CFT$.  \\ \\
If, however, the reference state $|\phi_0)$ is normalisable it appears that not much can be said in general.  The possible geometry is not restricted to asymptotically $AdS_2$ and may be asymptotically $dS_2$ or asymptotically flat.  Some examples will feature in section \ref{CohSection}. 

\section{Unsourced Metrics for $H_0 \in su(1,1)$ }

\label{UnsourcedSect}

Having examined some general properties we now turn our attention to specific examples of interest.  We start with arguably the simplest quantum mechanical model, the free particle (\ref{Schr11Coord}).   We will work with a more general class of Hamiltonians that contain the free particle as a special case, namely where the Hamiltonian is an algebra element of $su(1,1) $.  The algebra is given in eq. (\ref{Schr11Comm}), in terms of position and momentum operators, and is restated here, with a slight generalisation
\begin{eqnarray}
H = \frac{1}{2}P^2 + \frac{g}{2 X^2} \ \ \ ; \ \ \  D = -\frac{1}{4}(XP + PX) \ \ \ ; \ \ \  K =  \frac{1}{2}X^2 \nonumber \\
\left[H, D \right] = i H \ \ \ ; \ \ \  \left[K, D \right] = -i K \ \ \ ; \ \ \ \left[H, K \right] = 2 i D.   \label{kIrrep} 
\end{eqnarray}
We have generalised the free particle Hamiltonian $\frac{1}{2}P^2$ to the scale symmetric Hamiltonian $H$ by including the term proportional to $\frac{1}{X^2}$.  The purpose of these generalisations will reveal itself shortly.  The representation (\ref{kIrrep}) is the $k = \frac{1}{2}\sqrt{g + \frac{1}{4} }$ irrep of $SU(1,1) $ and the Casimir is given by $C = -D^2 + \frac{1}{2}(H K + KH) = k(k-1)$.  A second basis for this algebra will also be relevant, the Cartan-Weyl basis
\begin{eqnarray}
& & K_0 = \frac{1}{2}(K + H) \ \ \ ; \ \ \ K_\pm = \frac{1}{2}(K - H) \pm i D \nonumber \\
& & [K_{0}, K_{\pm}] = \pm K_{\pm} \ \ \ ; \ \ \ [K_{-}, K_{+}] = 2 K_{0}.   \label{LadderBase}
\end{eqnarray}
The $su(1,1)$ algebra was prominent in section (\ref{DynSymFreeP}) where we identified it as the dynamical symmetry generators of the free Schr\"odinger equation that transform time (\ref{Schr11Coord}).   \\ \\
We will consider the family of states
\begin{equation}
|t) = e^{i t H_0} |\phi_0) \ \ \ ; \ \ \ H_0 = u H + v D + w K \ \ \ ; \ \ \  |\phi_0) = ``|x = 0)" \equiv e^{-K_+}|0\rangle \label{su2Operators}
\end{equation}
where $|0\rangle$ is the eigenstate of $K_0$ that is also annilated by $K_{-}$ (\ref{LadderBase}).  See Appendix \ref{HolsteinApp}.  We call the reference state $|x=0)$ because it holds that 
\begin{equation}
K |\phi_0) = \frac{1}{2}X^2 |\phi_0) = 0.  
\end{equation}
Since both the reference state $|\phi_0)$ and $|x=0)$ are non-normalisable we will not claim that these states are, in fact, the same. \\ \\
The free particle Hamiltonian can be recovered from (\ref{su2Operators}) when $v,w = 0$ and the coupling constant, $g = 0$.  The choices for the parameters $u, v, w$ order the $su(1,1)$ operators (\ref{su2Operators}) into three different categories \cite{Alfaro}, \cite{Hyperbolic}.  For this purpose we define the parameter 
\begin{equation}
d^2(H_0) = 4uw - v^2.  \label{d2Expression}
\end{equation}
The $d^2 > 0$ case corresponds to compact operators with discrete spectra and normalisable eigenvalues.  These transformations are classified as elliptic.  \\ \\
The $d^2 = 0$ operators generate parabolic transformations, with continuous spectra and non-normalisable eigenstates.  Finally, if $d^2 < 0$ the operators generate hyperbolic transformations and the spectrum is not bounded from below.  This makes the regularisation ($H_0$ with $d^2 < 0$) unsuitable in (\ref{States}) and we will thus not be considering Hamiltonians, and thus unsourced metrics, of this type.  For all intents and purposes the parabolic transformations can be thought of as the scale symmetric Hamiltonian, $H_0 = H $, and the elliptic transformations as harmonic oscillators, $H_0 = H + \omega^2 K $ with $d^2 = 4\omega^2 $. This is because Hamiltonians with the same value for $d^2$ can be shown to be unitarily equivalent i.e. we can find a unitary transformation that transforms any Hamiltonian with positive $d^2$ or zero $d^2$ into the harmonic oscillator and free particle respectively.  \\ \\ 
It is thus sufficient to only work with the harmonic oscillator and $H$.  One may proceed to calculate the overlap for the harmonic oscillator states (\ref{tauStates}) which is done in \ref{HOLapCalc} using the $SU(1,1)$ BCH formulas.  Note that the scale symmetric Hamiltonian case is recoverable as the $\omega \rightarrow 0$ limit.  We find
\begin{equation}
(\overline{\tau}| \tau ) = \left(\frac{\omega}{4 \sinh(2 \omega \beta)}\right)^{2k}  \label{HOrealLap}
\end{equation}
where $k$ is the representation label given by $K_0 |0\rangle = k |0\rangle$ (see Appendix \ref{HolsteinApp}).  From the overlap (\ref{HOrealLap}) one may derive the metric and anti-symmetric two-form
\begin{equation}
ds^2 = \frac{ 2 k \omega^2}{\sinh^2(2 \omega \beta)}\left(d\beta^2 + dt^2 \right) \ \ \ ; \ \ \  \sigma_{\beta t} = \frac{ 2 k \omega^2}{\sinh^2(2 \omega \beta)} = -\sigma_{t \beta} \label{FullMetric}
\end{equation}
which we will refer to as the harmonic oscillator metric.  The free particle metric, recovered in the $\omega\rightarrow 0 $ limit, is precisely the $AdS_2$ metric in Euclidean signature in the Poincare patch coordinates
\begin{equation}
ds^2_{H_0 = H} = \frac{k}{2\beta^2}(d\beta^2 + dt^2)  \label{freeMet}
\end{equation}
with scalar curvature $R = -\frac{4}{k}$.  This result is comforting since the intuition is that the free particle should produce (something as close as possible to) a purely $AdS$ metric.  The result is also to be expected - the non-normalisable nature of the reference state $|x=0)$ already demands an $aAdS$ metric (\ref{aAdSmetricDef}) but we know further that the free particle states contain three dynamical symmetries, namely time translation, scaling and special conformal transformations.  The two dimensional metric (\ref{freeMet}) is thus maximally symmetric and therefore $AdS_2$.  Note that the scalar curvature is determined by the representation of $su(1,1)$ labeled by $k$.  We note that this is in accordance with the prediction of \cite{xpModel}  though their metric is constructed differently.  \\ \\ 
Considering now the harmonic oscillator metric (\ref{FullMetric}), one may proceed to analyse it and one finds that it too is $AdS_2$ with scalar curvature $R = -\frac{4}{k}$.  This may be verified by noting that the metric is maximally symmetric.  For a two dimensional this means that there should exist three Killing vectors which satisfy Killing's equation
\begin{equation}
\nabla_\nu \chi_\mu + \nabla_\mu \chi_\nu = 0  \label{Killing}
\end{equation}
Explicitly, the Killing vectors are
\begin{eqnarray}
	\chi^\mu_H \partial_\mu &=&2^{-1}\left[(1+\cosh[2\beta\omega]\cos[2t\omega])\partial_t-\sin[2t\omega]\sinh[2\beta\omega]\partial_\beta\right] \nonumber \\
	\chi^\mu_D \partial_\mu&=&(2\omega)^{-1}\left[\cosh[2\beta\omega]\sin[2t\omega]\partial_t+\cos[2t\omega]\sinh[2\beta\omega]\partial_\beta\right] \nonumber \\
	\chi^\mu_K \partial_\mu&=&(2\omega^2)^{-1}\left[(1-\cosh[2\beta\omega]\cos[2t\omega])\partial_t+\sin[2t\omega]\sinh[2\beta\omega]\partial_\beta\right] \label{KillingSU11}
\end{eqnarray}
so that the metric possesses three isometries and is maximally symmetric.  Note that $(\chi^\mu_H + \omega^2 \chi^\mu_K)\partial_\mu = \partial_t$ generates time translation as it should.  The family of metrics (\ref{FullMetric}), parametrised by $\omega$,  are  equivalent (related by coordinate transformation).  Thus, the metrics we produce from the free particle states, which possess scale symmetry, and the harmonic oscillator states, which don't, are both $AdS_2$.    

\section{The Meaning of Time Translation}

The fact that we produce an $AdS_2$ geometry from states that do not possess scale symmetry seems to suggest that our construction is somehow badly defined.  It must be remembered, however, that the isometries of the metric are the dynamical symmetries which are of physical significance.  The equivalence of the two metrics is a result of the fact that the generators of dynamical symmetry for the free particle and harmonic oscillator are the same, namely $su(1,1)$.  This result was proved by Niederer \cite{NiedererFree}, \cite{NiedererHO}.  In fact, the generators of dynamical symmetry are identical (namely $schr_{d+1}$) even when position is included as a degree of freedom.  The metrics will thus be the same up to coordinate change even if we include position.  The geometries are thus inescapably equivalent.  \\ \\
The question one may rightly ask is, if these (clearly) different quantum mechanical models possess the same generators of dynamical symmetry, in what way are they different?  The answer lies in what exactly is meant by time translation in the two models.  Consider the states of the form (\ref{su2Operators})
\begin{equation}
e^{i t H_0}|x = 0) \ \ \ ; \ \ \ \textnormal{where} \ \ \ d^2(H_0) \geq 0. \label{H0states} 
\end{equation}
The generators of dynamical symmetry is $su(1,1)$ for valid (elliptic or parabolic) choices of $H_0$.  These states differ by the choice of the operator $H_0 \in su(1,1)$ that is the generator of time translation.  The states (\ref{H0states})  have the same set of dynamical symmetry generators, but they differ in the sense that a specific symmetry (time translation) is associated with a different element of the $su(1,1)$ algebra.  \\ \\
This choice of the generator of time translation can only constitute a reparametrisation of the group i.e. a different choice of coordinates.  Indeed, the transformation
\begin{equation}
t_f + i\beta_f = \frac{1}{\omega}\tan(\omega t_{ho} + i \omega \beta_{ho}) \label{freeToHO} % = \frac{1}{\omega}\frac{\sin(2\omega t_{h0}) - i \ 2\cosh(\omega \beta_{ho})\sinh(\omega \beta_{h0})}{\cos(2 \omega t_{h0}) + \cosh(2 \omega \beta_{h0})} \label{freeToHO}
\end{equation}
maps the free particle propagator (with $t_f$ and $\beta_f$) onto the harmonic oscillator propagator (with $t_{ho}$ and $\beta_{ho}$), up to a change in phase.  From (\ref{freeToHO}) it is clear that, though $\omega$ does not introduce a scale into the curvature it does introduce a scale into the coordinate patch related to the periodicity in time.

\section{Conformal Quantum Mechanics}

\label{CQMSection}

The $SU(1,1)$ Hamiltonian states we have discussed thus far possess $SU(1,1)$ symmetry i.e. the global symmetries of the one-dimensional conformal group discussed in section \ref{VirasSection}.  In this section we will now introduce the model of conformal quantum mechanics (CQM) \cite{Alfaro} where we can give physical meaning to the full conformal group.  We will show the relation between this model of conformal quantum mechanics and the $SU(1,1)$ Hamiltonian states explicitly.  In this discussion we will once again see that the free particle and harmonic oscillator are related by a transformation of the time-coordinate.  We will, in fact, see that the free particle is related to models with (even a) time-dependent quadratic source in the same way.

\subsection{The Global Symmetries}

The action we will be considering is that of the one-dimensional field theory
\begin{equation}
S[\phi] = \int dt \left(\frac{1}{2}(\dot{\phi})^2 - \frac{g}{2 \phi^2}\right) \ \ \ ; \ \ \ g>0 \label{CQMAction}
\end{equation}
where the dot indicates a time derivative.  This is the most general scale-invariant theory in one dimension with at most double derivatives.  It is studied in detail in \cite{Alfaro}.  Note that the Lagrangian corresponds precisely to that of the scale symmetric Hamiltonian (\ref{kIrrep}).  The model has three global symmetries compactly stated as
\begin{equation}
t' = \frac{\alpha t + \beta}{\gamma t + \delta} \ \ \ ; \ \ \ \phi'(t') = (\gamma t + \delta)^{-1} \phi(t) = \sqrt{\frac{\partial t'}{\partial t}} \phi(t)
\end{equation}
where $\alpha \delta - \beta \gamma = 1$.  These transformations can be split into time translation, where $\alpha = \delta = 1 \ ; \ \gamma = 0$, dilatation where $\beta = \gamma = 0 \ ; \delta = \frac{1}{\alpha}$ and a special conformal transformation where $\alpha = \delta = 1 \ ; \ \beta = 0$.  These transformations are generated by
\begin{equation}
V_1 = i\partial_t \ \ \ ; \ \ \ V_0 = i t\partial_t \ \ \ ; \ \ \ V_{-1} = i t^2 \partial_t
\end{equation} 
which close on the $su(1,1)$ algebra since
\begin{equation}
[V_1, V_0] = i V_1 \ \ \ ; \ \ \ [V_1, V_{-1} ] = 2i V_0 \ \ \ ; \ \ \ [V_{-1}, V_0] = -i V_{-1} .  \label{CQMGens}
\end{equation}
We can recover the $SU(1,1)$ model of CQM by varying the action (\ref{CQMAction}) and quantizing the on-shell conserved quantities associated with the symmetries.  The field equation of the action (\ref{CQMAction}) is
\begin{equation}
\frac{\delta S}{\delta \phi} \equiv F[\ddot{\phi}, \phi ] = \ddot{\phi} - \frac{g}{\phi^3} = 0. \label{CQMFieldEq}
\end{equation}
One can find the on-shell energy by finding a function which has as derivative a function proportional to the field equation (\ref{CQMFieldEq}).  One finds that the quantity $D_1 = \frac{1}{2}(\dot{\phi})^2 + \frac{g}{2 \phi^2}$ is such that $\dot{D_1} = F \dot{\phi} =0$ so that
\begin{equation}
\int_{t_1}^{t_2} dt \ \ \partial_t\left( D_1 \right) = = D_1(t_2) - D_1(t_1) = 0.
\end{equation}
This is nothing other than the statement that the total energy be conserved and is a consequence of time translation symmetry.  The other conserved quantities can be derived as in \cite{Alfaro}
\begin{equation}
D_1 = \frac{1}{2}(\dot{\phi})^2 + \frac{g}{2 \phi^2} \ \ \ ; \ \ \  D_0  = D_1 t - \frac{1}{4}( \dot{\phi}\phi + \phi \dot{\phi}) \ \ \ ; \ \ \  D_{-1} = D_1 t^2 + 2 (D_0 - D_1 t) + \frac{1}{2} \phi^2.
\end{equation}
Quantising these on-shell conserved quantities via $[\dot{\phi}(t), \phi(t)] \equiv [P, X] = -i$ i.e. associating $\dot{\phi}$ with the operator $P$ and $\phi$ with $X$ one finds the quantum mechanical operators
\begin{eqnarray}
D_1 & \rightarrow & \frac{1}{2}P^2 + \frac{g}{2 X^2}, \nonumber \\
D_0 & \rightarrow & \frac{1}{4} e^{-i t H} (XP + PX) e^{i t H}, \nonumber \\
D_{-1} & \rightarrow & \frac{1}{2} e^{-i t H}  X^2 e^{i t H}.  
\end{eqnarray}
These are the $su(1,1)$ generators with some time dependence.  The time-dependence is such that the operators are time-independent in the Heisenberg picture.  The time-evolution is determined, of course, by the generator of time translational symmetry, $H$.  The quantum states are 
\begin{equation}
|t) = e^{i t H} |x=0) \ \rightarrow \ |t, \beta) = e^{i t H} e^{-\beta H_0} |x=0)
\end{equation}  
after regularisation by $H_0 \in su(1,1)$.  We have already considered this $SU(1,1)$ model of CQM in section \ref{UnsourcedSect}.

\subsection{The Effect of a General Coordinate Transformation}

Consider a general transformation $t' = f(t)$ and \hbox{$\phi'(t') = \sqrt{\dot{f}(t)}\phi(t)$} which transforms the action (\ref{CQMAction}) to
\begin{eqnarray}
S' & = & \ \ \ \frac{1}{2}\int dt' \left( (\partial_{t'} \phi'(t'))^2 - \frac{g}{\phi'^2(t')} \right)  \nonumber  \\ 
	&   & -\frac{1}{4} \int dt' \left(\frac{d f(t)}{d t}\right)^{-2} \left(\partial_t h(t) -\frac{1}{2} h(t)^2 \right) \phi'^2(t') \nonumber \\ & &  \ \ + \frac{1}{2}\int dt' \partial_t' \left( (\partial_t f(t))^{-1} h(t) \phi'(t')^2 \right)
\end{eqnarray}
where $h \equiv \frac{\ddot{f}}{\dot{f}}$.  The final integral only contributes a surface term.  A general coordinate transformation thus maps the action (\ref{CQMAction}) onto one of the same form plus a time-dependent quadratic source term i.e.
\begin{eqnarray}
S' & = & \frac{1}{2}\int dt' \left( (\partial_{t'} \phi')^2 - \frac{g}{\phi'^2} + \gamma(t') \phi'^2 \right) \label{gammaAction} \\
 \textnormal{where} \ \ \gamma(t') & = & -\frac{1}{2}\left(\frac{d f(t)}{d t}\right)^{-2} \left(\partial_t h(t) -\frac{1}{2} h(t)^2 \right)  \label{gammaDefin}
\end{eqnarray}
The global symmetry transformations satisfy the condition $\dot{h} - \frac{1}{2}h^2 = 0$ so that no quadratic source term emerges and the action remains invariant.  All transformations that are not global symmetries do give rise to this quadratic source term.  A special example of this is the transformation that maps, up to a phase, the free particle propagator to the harmonic oscillator propagator (eq. (\ref{freeToHO}) with $\beta_f = \beta_{ho} = 0$).  This gives
\begin{equation}
t' = \frac{1}{\omega} \tan^{-1}(\omega t) \ \ \ ; \ \ \ \gamma(t') = -\frac{1}{2}\left(\frac{d f(t)}{d t}\right)^{-2} \left(\partial_t h(t) -\frac{1}{2} h(t)^2 \right) = \omega^2
\end{equation}
so that the quadratic coupling term is constant.  The Lagrangian thus corresponds to that of the harmonic oscillator.  By now varying the action one finds the field equation   \begin{equation}
F'[\ddot{\phi'}, \phi' ] \equiv \ddot{\phi}' - \frac{g}{(\phi')^3} + \omega^2\phi' = 0.  \label{fFieldEqHO}
\end{equation}
and the conserved energy $D_1'+ \omega^2 D_{-1}' = \frac{1}{2}(\dot{\phi}')^2 + \frac{g}{(2 \phi')^2} + \frac{1}{2}\omega^2 (\phi')^2$.  Quantising the field now will clearly yield the harmonic oscillator Hamiltonian.  The conserved charges can be calculated as 
\begin{eqnarray}
D_1' & = & \frac{1}{2} \cos^2(\omega t)\left( (\dot{\phi})^2 + \frac{g}{\phi^2} \right) + \frac{1}{4}\omega \sin(2 \omega t) \left(\dot{\phi} \phi + \phi \dot{\phi} \right) + \frac{1}{2}\omega^2 \sin^2(\omega t) \left( \phi^2 \right) \nonumber \\
D_0' & = & \frac{\sin(2\omega t)}{2\omega} \left( (\dot{\phi})^2 + \frac{g}{\phi^2} \right) + \frac{1}{4}\cos(2\omega t)\left(\dot{\phi} \phi + \phi \dot{\phi} \right) -\frac{\omega}{2}\sin(2\omega t) \\
D_{-1}' & = & \frac{\sin^2(\omega t)}{2\omega^2}\left( (\dot{\phi})^2 + \frac{g}{\phi^2} \right) + \frac{\sin( \omega t)}{2 \omega}\left(\dot{\phi} \phi + \phi \dot{\phi} \right) + \frac{1}{2} \cos^2(\omega t) \left( \phi^2 \right) \nonumber
\end{eqnarray}
so that one again finds that these are the time-dependent generators of $su(1,1)$.  The time-evolution is now the harmonic oscillator so that the quantum states are
\begin{equation}
|t) = e^{i t (H+\omega^2 K)} |x=0) \ \rightarrow \ |t, \beta) = e^{i t (H+\omega^2 K)} e^{-\beta H_0} |x=0).
\end{equation}
For a general transformation $t \rightarrow t' = f(t)$ the conserved charges are again these that yield the $su(1,1)$ generators after cononical quantisation, but now with time evolution operator $\partial_t U(t) = i(H + \gamma(t) K )$.  Quantising this yields nothing else than a sourced Hamiltonian put together from the $su(1,1)$ algebra elements, $H + \gamma(t)K$ i.e. the states
\begin{equation}
|t) = U(t)|x=0) \ \rightarrow \ |t, \beta) = U(t) e^{-\beta H_0}|x=0) \ \ \ \textnormal{where} \ \ \ \partial_t U(t) = i(H + \gamma(t) K ).  
\end{equation}
This model is related to the free particle and harmonic oscillator by a reparametrisation of time and thus there are three conserved charges.  On the level of the quantum states this means that the quantum states will again possess three symmetries and the resulting geometry should be $AdS_2$.

\subsection{The Conformal Symmetry of CQM}

One may recover the Witt algebra (\ref{WittAlgebra}) from the general infinitesimal transformation $t' = t + \epsilon(t)$.  We did this explicitly for the two-dimensional conformal group in section \ref{VirasSection} and the argument here is identical.  One identifies the generators
\begin{equation}
V_n = -t^{1-n} \partial_t
\end{equation} 
which close on the centerless Virasoro algebra (\ref{Viras})
\begin{equation}
[V_n, V_m] = (n - m)V_{n+m}.  
\end{equation}
The generators $V_n$ where $|n| > 1$ do not induce symmetries of the action (\ref{CQMAction}) and instead induce a time-dependent coupling to a quadratic source.  The model (\ref{CQMAction}) will possess full conformal symmetry if it has the freedom to change the time-dependent coupling $\gamma(t)$ (\ref{gammaAction})\footnote{Note that this is reminiscent of a gauge degree of freedom.  The global symmetries of the model affect the states only while there are additional local symmetries of the model that not only affect the states but the gauge as well.  We do not claim, however, that this time-dependent coupling is in fact a gauge degree freedom though the similarities are intriguing}.  \\ \\
We note lastly the transformation of the time-dependent coupling $\gamma(t)$ under a general coordinate transformation of the action (\ref{gammaAction}), $t' = f(t)$.  The coupling will get two contributions.  The first comes from the term
\begin{equation}
\int dt \gamma(t) \phi^2 = \int dt' \gamma(t) \left(\dot{f} \right)^{-1} \left( \left(\dot{f}\right)^{-\frac{1}{2}}\phi'\right)^2 = \int dt' \left(\frac{dt'}{dt} \right)^{-2} \gamma(t) (\phi')^2.  
\end{equation}
The second part comes from the transformation of the kinetic term which will add a quadratic source term defined in (\ref{gammaDefin}).  Putting these two contributions together yield
\begin{equation}
\gamma'(t') = \left( \frac{d t'}{d t} \right)^{-2} \left(\gamma(t) - \frac{1}{2}\partial_t \left( \frac{\partial_t^2 t'}{\partial_t t'} \right) -\frac{1}{4}\left( \frac{\partial_t^2 t'}{\partial_t t'} \right)^2  \right).  \label{gammaFinTrans}
\end{equation}
This is precisely the transformation property of an energy momentum tensor (\ref{EMTensorTrans}) with central charge $c=6$.

\section{Coherent States}

\label{CohSection}

Another useful way of thinking about the $SU(1,1)$ family of states is simply as coherent states \cite{Perel1}, \cite{Perel2}.  Indeed, the procedure of extending the boundary $SU(1,1)$ Hamiltonian states into the bulk gives us simply a different parametrisation of $SU(1,1)$ coherent states.  This is precisely why all the $SU(1,1)$ models end up with equivalent geometries.  \\ \\ 
We presented a general procedure for constructing coherent states in section \ref{GroupFamily}.  The result is that given a Hilbert space and a group $G$, with generators $\{ A_i \}$, acting on this Hilbert space one chooses a reference state in the Hilbert space, $|\phi_0)$, and calculates the stationary subgroup $H$ as defined in (\ref{HDefine}).  The coherent state is then
\begin{equation}
U|\phi_0) = e^{\sum_i \alpha_i A_i}| \phi_0 ) \equiv |\vec{\alpha}) \ \ \ ; \ \ \ U \in G/H .    \label{CohDef}
\end{equation}
The states $|\vec{\alpha})$ are generalised coherent states and allow a resolution of the identity operator as \cite{Perel1}
\begin{equation}
I \propto \int d \mu(\vec{\alpha}) | \vec{\alpha} )(\vec{\alpha}   | \label{CohIdentity}
\end{equation}
where $d\mu$ is the invariant measure.  This is because the operator (\ref{CohIdentity}) can be shown to commute with all elements of $G$.  The rough outline for the proof is as follows.  Acting with a group element on the left of the operator (\ref{CohIdentity}) and on the right with its conjugate leads to a different parametrisation of the states (\ref{CohDef}).  This different parametrisation can be absorbed in the integral ($\vec{\alpha} \rightarrow \vec{\alpha}'$) and consequently the operator (\ref{CohIdentity}) commutes with all group elements. If working with an irreducible representation then by Schur's lemma \cite{Perel1} the operator (\ref{CohIdentity}) must thus be proportional to the identity.  The coherent states thus form an overcomplete basis for the Hilbert space.  \\ \\
We mentioned in section \ref{GroupFamily} that a reference state with a larger stationary subgroup produces coherent states that are parametrised by fewer coordinates.  This in turn leads to a metric of fewer dimensions.  We will illustrate these beneficial choices of reference states by example.  

\subsection{$SU(1,1)$ Coherent States}

For the $SU(1,1)$ group such a convenient choice of reference state is the state $|0\rangle$ such that (see Appendix \ref{HolsteinApp})
\begin{equation}
K_0|0\rangle = k |0\rangle \ \ \ ; \ \ \ K_{-}|0\rangle = 0
\end{equation} 
where $k$ labels the representation.  A general $SU(1,1)$ group element can be written as 
\begin{equation}
U(z, d, \overline{z}) = e^{z K_+} e^{i d K_0} e^{\overline{z}K_{-}}  \ \ \ ; \ \ \ d \in \mathbb{R} \ ; \ |z| < 1
\end{equation}
by using the BCH formula \cite{Lie1}, \cite{Lie2}. Thus, using the properties of the reference state $|0\rangle$, we have the coherent states
\begin{equation}  
| z ) \equiv U(z, d, \overline{z})|0\rangle \propto e^{z K_+}|0\rangle \ \ \ ; \ \ \ z\in \mathbb{C} \ \ \ ; \ \ \ |z| < 1.    \label{SU11CohStates}
\end{equation}
It can be shown that the regularised time-evolved states (\ref{States}) for the $SU(1,1)$-Hamiltonians are nothing other than a different parametrisation of the states (\ref{SU11CohStates}).  Specifically we can reparametrise the coherent states (\ref{SU11CohStates}) as
\begin{equation}
| z') \equiv e^{-\frac{-i + z'}{i + z' }K_{+}}|0\rangle = (1 - i z')^{2k} e^{i z' H} e^{-K_+}|0\rangle = (1 - i z')^{2k} e^{i z' H} |x=0) \ \ \ ; \ \ \ Im(z') > 0   \label{PhysicsPar}
\end{equation}
The condition that $Im(z') >0$ is equivalent to the condition that $|z| < 1 $.  From this vantage point of coherent states one can clearly see why Hamiltonians linear in $su(1,1)$ algebra elements will produce equivalent geometries.  For any $SU(1,1)$ group element the BCH formulas imply that
\begin{equation}
U_{SU(1,1)}|x=0) = e^{i f_H H} e^{i f_D D} e^{i f_K K}|x = 0) \propto e^{i f_H H}|x=0).
\end{equation}
The geometry one will thus generate is exactly the geometry of the $SU(1,1)$ coherent state up to the coordinate transformation $f_H \rightarrow f_H'$.  

\subsection{$SU(2)$ and Glauber Coherent States}

Examples of interesting geometries other than $AdS_2$ can be recovered by simply considering different coherent states.  In particular it is also possible to produce $dS_2$ and flat space metrics from the appropriate coherent states.  The $SU(2)$ coherent states produce a de Sitter metric.  The group manifold of $SU(2)$ was also de Sitter (\ref{SU2GroupMetric}).  Note, however, that the group manifold (\ref{groupMetric}) had both left- and right multiplication symmetries while the dynamical symmetries are only the left multiplication symmetries.  The $su(2)$ algebra is given by (\ref{su2Algebra}), and the coherent states (which may be derived by a similar argument as in the previous section) by
\begin{equation}
|z) = e^{z J_{+}} | -j \rangle   \label{SU2Coh}
\end{equation}
where $|-j\rangle$ is the lowest eigenvector of $J_z$ i.e. $J_z |-j\rangle = -j|-j\rangle$ and $J_{-}|-j\rangle = 0$.  We are considering this specific irreducible representation of $SU(2)$ labelled by $j$.  One can find an operator $H_{su(2)}\in su(2)$ such that the coherent states can then be parametrised as
\begin{equation}
|z) = e^{z(\beta, t) J_{+}}|-j\rangle \ = \ e^{i(t + i\beta)H_{su(2)}}e^{-J_{+}}|-j\rangle \ \ \ ; \ \ \ z\in \mathbb{C}
\end{equation}
It is thus possible to recast the $SU(2)$ coherent states in a similar form to the $SU(1,1)$ coherent states i.e. as generated by a Hamiltonian with complexified time acting on some reference state.  The situation in the case of $SU(2)$ is different since the reference states $|-j\rangle$ and $e^{-J_{+}}|-j\rangle$ are normalisable.  The states thus don't require a regularisation parameter (such as complex time) in order to produce sensible metrics.  Nonetheless it is a beneficial choice to work with complex coordinates anyway since this allows us to view the states as coherent states and exploit their many useful properties \cite{Perel1}.  \\ \\
One may proceed to calculate the metric and anti-symmetric two-form from the $SU(2)$ coherent states (\ref{SU2Coh}) and one finds
\begin{equation}
ds^2 = \frac{2j}{(1 + z\overline{z})^2} dz d\overline{z} \ \ \ ; \ \ \ \sigma_{z\overline{z}} = \frac{2 i j}{(1 + z\overline{z})^2} = -\sigma_{\overline{z} z} \label{SU2Metric} 
\end{equation}
which is the metric for the 2-sphere.  It is a maximally symmetric manifold with Killing vectors
\begin{eqnarray}
\chi_1^\mu \partial_\mu & = & i z\partial_z - i \overline{z}\partial_{\overline{z} } \nonumber \\
\chi_2^\mu \partial_\mu & = & i(z^2 - 1)\partial_z - i(\overline{z}^2 - 1)\partial_{\overline{z}} \nonumber \\
\chi_3^\mu \partial_\mu & = & (z^2 + 1)\partial_z + (\overline{z}^2 +1)\partial_{\overline{z}}  \label{su2Killing}
\end{eqnarray} positive scalar curvature $R = \frac{4}{j}$ and indeed the Euclidean signature version of $dS_2$.  \\ \\
The scalar curvature scales like $\frac{1}{j}$ and approaches zero in the $j\rightarrow \infty$ limit.  \\ \\
The coherent states (\ref{SU2Coh}) tend to the Glauber coherent states in this limit, as can be seen from a Holstein-Primakoff expansion (\ref{HP1}).  The Glauber coherent states are expressed in terms of bosonic creation and annihilation operators (\ref{HeisenbergAlg2}) as
\begin{equation}
| z ) = e^{z a^\dag}|0\rangle.   \label{Glauber}
\end{equation}
The metric of these Glauber coherent states is the flat space metric in two dimensions
\begin{equation}
ds^2 = dz d\overline{z} \ \ \ ; \ \ \ \sigma_{z \overline{z}} = i = -\sigma_{\overline{z} z}.  \label{GlauberMet}
\end{equation}
Considering the flat space metric the Killing vectors are
\begin{eqnarray}
\chi_1^\mu\partial_\mu & = & z \partial_z - \overline{z}\partial_{\overline{z}} \nonumber \\ 
\chi_2^\mu \partial_\mu & = & \partial_z + \partial_{\overline{z}}  \nonumber \\
\chi_3^\mu \partial_\mu & = & i \partial_{z} - i\partial_{\overline{z}}.  
\end{eqnarray}

\section{A Comment on a Result From the $AdS_2/CFT_1$ Literature}

\label{JackiwSection}

Having concluded our discussion of the geometries of the two-dimensional manifold of states we are in a position to comment on and extend a result from the $AdS_2/CFT_1$ literature.  In \cite{Jackiw} the authors show that the appropriate form for the $2$- and $3$-point functions (\ref{PrimaryCorrelators}) can be found by considering the quantum mechanics of a particle on a half-line subject to an inverse square potential term.   \\ \\
Their analysis begins on the $AdS_2$ side of the duality where the isometries of $AdS_2$, generated by the $so(2,1)$ algebra elements, are identified.  States $|t)$ are introduced, defined in terms of their transformations with respect to these generators
\begin{equation}
(t| H = i \partial_t (t| \ \ \ ; \ \ \ (t| D =  i(t \partial_t + r_0)(t| \ \ \ ; \ \ \ (t| K = i(t^2 \partial_t + 2 r_0 t) (t| \label{tJackStates}
\end{equation}
where $r_0$ is the lowest eigenvalue of $K_0$ (\ref{kIrrep}) - we call this $k$ in Appendix \ref{HolsteinApp}.  Using symmetry arguments the authors then show that the states (\ref{tJackStates}) produce the appropriate forms for the $2$- and $3$-point functions 
\begin{eqnarray}
(t_1|t_2) & \propto & |t_1 - t_2|^{-2 r_0}  \nonumber \\
(t_1|B(t)|t_2) & \propto & |t - t_1|^{-\delta}|t - t_2|^{-\delta} |t_1 - t_2|^{\delta - 2 r_0}  \label{CQMCorrFuncs}
\end{eqnarray}
where $B(t)$ is a primary of dimension $\delta$.  This is despite two puzzles pointed out by the authors.  Firstly, the states $|t)$ are not normalizable and secondly there exists no conformally invariant state in the Hilbert space.  They conclude that the $K_0$ lowest eigenstate, $|0\rangle$ is the averaging state and that there exists an operator, $A(t)$ such that $A(t)|0\rangle$ produces the appropriate states $|t)$.  Neither the averaging state $|0\rangle$ nor the operator $A(t)$ transforms appropriately, however.  By this it is meant that the averaging states does not transform like the $CFT$ vacuum and the operator $A(t)$ does not transform like a primary operator.  Remarkably their ``defects" (as the authors call them) seem to cancel in order to produce the appropriate correlation functions (\ref{CQMCorrFuncs}).  \\ \\
Their construction is noteworthy in that a quantum model is constructed from $AdS_2$ in a systematic way and that this model exhibits some of the desired properties of $CFT_1$.  Their discussion also points out very important differences between quantum mechanics and field theory that we need to be mindful of as we progress with our construction.  Their analysis can be unpacked from our current construction in a natural way that also addresses their puzzles directly and we are, in fact, in a position to extend some of their results.  \\ \\
The states $|t)$ that they define are nothing other than the free particle states we have discussed in section \ref{UnsourcedSect}.  Indeed, the transformation properties of (\ref{tJackStates}) can be shown explicitly using the states and $su(1,1)$ symmetry generators.  We know, by construction, that $AdS_2$ is the appropriate geometry to associate with these states.  The absence of gravity in their analysis is also clear.  As we have pointed out, our construction in this chapter is, for the most part, simply a different view on the work of \cite{Ashtekar} so that all the information of the quantum states can be extracted from the geometry alone.  It is possible to repackage this further as a theory of gravity but when one is interested in the expectation values the symmetries and geometry is sufficient.  \\ \\
The non-normalisable nature of the states is something that we have already addressed.  Indeed, it is precisely this that gives rise to an anti-de Sitter geometry and the bulk coordinate of $AdS_2$ is a regularisation parameter.  In the bulk these states are normalisable and the regularisation scheme is such that the boundary symmetries are retained as bulk isometries.  \\ \\
The key observation to understand why one still produces the appropriate two- and three-point functions despite these differences is that the symmetry transformations for operators and quantum states come about differently.  For vacuum to vacuum correlators in field theory this distinction does not arise since under a unitary transformation the field theory vacuum transforms trivially so that it is only the operators that transform.  The quantum state and operators transform as
\begin{equation}
|s) \rightarrow U |s) \ \ \ \textnormal{and} \ \ \ A(s) \rightarrow U A(s) U^\dag 
\end{equation}
respectively.  This ensures that we can induce symmetry transformations by inserting the identity in the form of the corresponding unitary operators.  This, along with sufficient symmetry to restrict their form, is precisely why the appropriate form for the two- and three-point functions are reproduced.  \\ \\
One matter that needs to be taken note of is that the correlators of this quantum mechanical problem (\ref{CQMCorrFuncs}) can only be resolved up to a normalisation.  This is, in fact, a direct consequence of the fact that symmetry transformations may change the normalisation of the state.  Once the condition of normalisability has been relaxed the form of the correlation functions (\ref{CQMCorrFuncs}) are completely determined by the dynamical symmetry.   \\ \\
Our systematic construction adds to the analysis \cite{Jackiw} in several ways.  Firstly, it provides the explicit mapping from quantum mechanical state to geometry.  This gives, for instance, the physical interpretation of the $AdS_2$ radius as the representation label.  The mapping used to go from the geometry to the quantum symmetry generators is to identify the algebra of the Killing vectors.  This is used in \cite{Jackiw}, but we will extend this in the next chapter by showing this property for general states of complex coordinates.  Secondly, our construction has illuminated the physical significance of coordinates and that, geometrically, the $SU(1,1)$ states are all equivalent.  This implies that the analysis of \cite{Jackiw} can be applied also, after coordinate transformation, to the harmonic oscillator (and indeed to any of the $SU(1,1)$ Hamiltonians).  Lastly, and possibly most importantly, our construction can be generalised to, not only other quantum models, but to a description that includes the CQM local symmetries as well.  The role and existence of gravitational duals can be investigated systematically.    

\section{Summary}

This concludes our discussion of the two-dimensional metrics.   We take particular note of the K\"ahler structure that arises for the unsourced metrics considered in section \ref{UnsourcedSect}.  This will allow us to utilise the link between the expectation values of operators and vector fields on the manifold, see section (\ref{GeoReform}).  \\ \\
The fact that we can decipher the results and puzzles of existing work in the $AdS_2/CFT_1$ literature \cite{Jackiw} in such a natural way indicates that our construction, at least the procedure that goes from quantum state to geometry, is sensible.  We will, in chapter \ref{GravChap}, add another step to the procedure that takes the quantum states and this metric and matches these to a dual gravitational theory.  The main advantage is that our approach is systematic so that no guesswork is required to put together the appropriate dual.  \\ \\
Our task will now be to repackage the geometric information, by some means, into a gravitational dual description of these quantum mechanical models.

\chapter{Gravitational Duals in Two Dimensions}

\label{GravChap}

In this chapter we will now, finally, present a systematic way of constructing gravitational duals of quantum mechanical models.  Our work up to this point has identified a construction (\ref{Prov2}) that maps a given set of quantum states to a metric.  Importantly the dynamical symmetries are encoded as isometries of the metric.  This property, along with its relatively simple straightforward computation, was the single most important property that set this construction apart from others that were discussed in chapter \ref{ConstrChap}.  \\ \\
Our next task is to incorporate this metric in some way as an ingredient in a theory of gravity. The theory should be chosen, of course, in such a way that it represents a dual description of the original quantum mechanical model.  As was the case with the possible ways to construct metrics from quantum mechanical models, it is unavoidable that many possible choices for a dual theory exist.  We will again provide some intuitive arguments for why the choice we will make in this chapter is a sensible one and show, when we consider the two-dimensional manifold of states, that this choice is in agreement with existing works in the $AdS_2/CFT_1$ literature.  Our work here will recover these existing results in a natural way and the systematic approach will provide direct access to the quantum mechanics / gravity dictionary.  This will allow us to extend these existing results in several ways. 

\section{Proceeding to a Gravitational Dual}

\label{ProcGravDual}

Let us begin this chapter with a summary of what we have discussed thus far.  The construction (\ref{Prov2}) takes as input a family of quantum states (possessing some symmetries) and gives as output a metric (with corresponding isometries), $g_{\mu\nu}$ and an anti-symmetric two-form $\sigma_{\mu\nu}$ (\ref{ProvSym}) from this family of states.  \\ \\
A case that is of special importance is when the family of states are parametrised by $n$ complex coordinates $|z_1, ..., z_n )$.  This is typical of coherent states but can be more general.  In this special case the metric and anti-symmetric two-form are closely related, most easily seen by considering them with one co-variant and one contra-variant index.  In general for these states one finds
\begin{equation}
g_{z_k}^{\ z_j} = \delta_k^j = g_{\overline{z}_k}^{\ \overline{z}_j} \ \ \ ; \ \ \ \sigma_{z_k}^{\ z_j} = i\delta_k^j = -\sigma_{\overline{z}_k}^{\ \overline{z}_j}
\end{equation}
and, consequently \cite{Kahler1}, that
\begin{equation}
\nabla_\alpha \sigma_{\mu\nu} = 0.    \label{Compat}
\end{equation}
By $\nabla_\alpha$ we mean the covariant derivative, see Appendix \ref{AppGeo}.  Note that the indices of $\sigma$ transform like a tensor (the indices are raised and lowered by the metric).  However, $\sigma^{\mu\alpha}$ is not the inverse of $\sigma_{\alpha \nu}$ but rather
\begin{equation}
\sigma^{\mu\alpha} \sigma_{\alpha \nu} = -\delta^{\mu}_{\ \nu}.  \label{sigmaInverse}
\end{equation}
Note that in these complex coordinates the anti-symmetric two-form is now a symplectic form.  \\ \\
In section \ref{GeoReform} we showed explicitly how vector fields can be related to operators.  If an operator $A$ is hermitian then the vector fields are
\begin{equation}
\chi_{A} \equiv \chi^\mu_{A} \partial_\mu \equiv -\frac{1}{2}\sigma^{\mu \alpha} \ \nabla_{\alpha} \phi_{A} \partial_\mu \ \ \ ; \ \ \ \chi_{i A} \equiv \chi^\mu_{iA} \partial_\mu \equiv -\frac{1}{2}g^{\mu \alpha} \ \nabla_{\alpha} \phi_{A} \partial_\mu.  \label{OperatorVector}
\end{equation}
where $\chi_{A}$ is the vector field and $\phi_{A} =\langle \vec{z} | A | \vec{z} \rangle$, the normalised expectation value of the operator $A$.  In the present context the dynamical symmetries are generated by hermitian operators so that the first vector field in (\ref{OperatorVector}) will be applicable.   Note that the vector field associated with the operator $A$ and $A + c$ where $c$ is a constant is the same.  This freedom will prove useful later.  As a specific example of (\ref{OperatorVector}) consider the Killing vectors which satisfy Killing's equations (\ref{Killing}) and are related, via (\ref{OperatorVector}), to the normalised expectation values of symmetry generators.  In general, the vector fields (\ref{OperatorVector}) satisfy the relation
\begin{equation}
\nabla_\nu \chi_\mu + \nabla_\mu \chi_\nu = \tau_{\mu\nu}   \label{FlowKill}
\end{equation} 
where we will refer to the tensor $\tau_{\mu\nu}$ as the flow parameters.  The flow parameters can be viewed as an indication of how far the vector field deviates from a Killing vector.  For conformal Killing vectors, which transform the metric up to a conformal factor, the flow parameters are given by
\begin{equation}
\tau_{\mu\nu} = \kappa(\vec{\overline{z} }, \vec{z}) g_{\mu\nu}  \label{kappaDef}
\end{equation}
where $\kappa(\vec{\overline{z} }, \vec{z})$ is an arbitrary function of the coordinates.  \\ \\
The mapping (\ref{Prov2}) from quantum states to metric encodes all the symmetries of the quantum states as isometries of the metric.  The Killing vectors of the metric are thus related to the normalised expectation values of symmetry generators of the original quantum states in a simple way.  This relation is of critical importance.  The Killing vectors, vector fields on the manifold - a ``geometric concept" - is directly related to the normalised expectation value of a generator of symmetry - an operator, a quantum observable.  Similarly the expectation value of an arbitrary operator on the quantum side may be related to a corresponding vector field on the geometric side of the construction (\ref{FlowKill}).  In this way there exists a duality between the quantum mechanical operators and the vector fields.  \\ \\
Note that we have not even begun to bring any sort of gravitational theory into the discussion.  The relation between the quantum mechanical expectation values and vector fields on the manifold relate the quantum mechanics to purely geometric quantities.  As is clear from the work of \cite{Ashtekar}, a dual geometric description of quantum mechanical models can be constructed along these lines.  \\ \\
Our interest, due to the $AdS/CFT$ correspondence, is to find a dual gravitational theory to a given quantum mechanical model.  We will, for the purposes of this thesis, think of gravitational theories in terms of an action formulation.  Consider, then, an action that is a functional of a metric and some set of fields
$S_{grav}[g_{\mu\nu}, \phi_1, \phi_2, ..., \phi_n]$.  A simple way in which our metric can be linked to a classical theory of gravity is to interpret it as the metric that yields a stationary value for the action i.e.
\begin{eqnarray}
\left. \frac{\delta S_{grav}}{\delta g^{\mu\nu}} \right|_{ \left\{ g_{\mu\nu} \ ; \ \vec{\phi} = \vec{\phi}_{cl} \right\} } & = & 0 \nonumber \\
 \left. \frac{\delta S_{grav}}{\delta \phi_i} \right|_{ \left\{ g_{\mu\nu} \ ; \ \vec{\phi} = \vec{\phi}_{cl} \right\} } & = & 0 \label{ActionField}
\end{eqnarray}
where $\vec{\phi}_{cl}$ are the values for the fields that, along with our metric, solve the equations of motion of the classical gravitational model.  Since the metric is tied to the gravitational theory by means of the stationary action we can expect that our dual theories will likely be built around the semi-classical approximations of gravity models.  As a first attempt to constructing duals this is a workable simplification.  If this first attempt gives sensible results then we can ask the question how one may interpret the fluctuations around the stationary action.  We leave this topic for future work. \\ \\
Of course, the simplest subset of these gravitational theories is where no fields are present so that the action is a functional of only the metric.  We will not be considering actions of this type for two reasons.  Firstly, in the $AdS/CFT$ correspondence the fields that feature in the theory of gravity play a vital role since their boundary values are the sources of the generating functional (\ref{PathInt}).  Any gravitational dual that would hope to be comparable to existing literature must thus contain fields.  Secondly, for the two-dimensional manifolds - our starting point - the Einstein equations are trivial.  We will speak more to this in section \ref{JTSection}.  \\ \\
We will be looking to construct a dual theory where the gravitational action is a functional of both the metric and fields.  The immediate question is then - what interpretation can we attach to the fields?  As mentioned in section \ref{CorrGenF}, the conventional interpretation of the fields is that their boundary values are the sources of the generating functional.  However, our metric (\ref{Prov2}) is constructed from the overlap of states which carries no knowledge of the sources.  In order to proceed with this interpretation and analyse it systematically we will probably have to develop a rather sophisticated algorithm.  It should be possible to investigate this -  a topic for future research.  The interpretation we will propose for this thesis is a much more direct and simple one and it will allow us to generate some interesting results.  \\ \\
It is worthwhile to recall that the quantities of interest are the correlation functions (\ref{CorrFuncsDef}) i.e. the expectation values of arbitrary strings of observables.  The simplest, most direct way that we can extract these expectation values is to interpret the fields that solve the equations of motion (\ref{ActionField}) as these same expectation values.  This may be a beneficial choice as well.  We know already from our discussion in section \ref{GeoReform} that the expectation values of operators are related to vector fields on the manifold.  The equations of motions that expectation values of the generators of symmetry should solve are thus likely to be of a simple form, since these should be related to the Killing equation (\ref{Killing}).  \\ \\
We will show in the chapters ahead how, despite being at odds with the conventional interpretation in $AdS/CFT$, this interpretation of the fields in the gravitational model will allow us to calculate the expectation values of arbitrary string of symmetry generators, at least for the simplest quantum models.   Also, intriguingly, we will reproduce several existing results of the $AdS_2/CFT_1$ literature which seemingly indicate that these are sensible choices.  \\ \\
Now that we have selected the criteria for our dual gravitational model (\ref{ActionField}) we can proceed to find one that fits them.  Before we do this we will first discuss a general way in which the symmetries of the quantum states can be used to calculate the expectation values of arbitrary strings of symmetry generators.  This procedure will again highlight the key role played by symmetries.  

\subsection{A Quick Example of the Expectation Value / Killing Vector Relation}

The relation between the expectation values of symmetry generators and Killing vectors may be verified for the metrics (\ref{FullMetric}) of the $SU(1,1)$ family of states and the corresponding Killing vectors (\ref{KillingSU11}) by using the normalised expectation values
\begin{eqnarray}
\langle \beta, t| H | \beta, t\rangle & = & \frac{ k\omega( \cosh(2 \omega \beta) + \cos(2 \omega t))}{\sinh[2 \beta \omega]} \nonumber \\
\langle \beta, t| D | \beta, t\rangle & = & \frac{ k \sin(2 \omega t) }{\sinh(2 \omega \beta)} \nonumber \\
\langle \beta, t| K | \beta, t\rangle & = & \frac{ k( \cosh(2 \omega \beta) - \cos(2 \omega t))}{\omega \sinh[2 \beta \omega]} \label{HOExpect}
\end{eqnarray}
for the harmonic oscillator states.  Applying eq. (\ref{OperatorVector}) to the expectation values (\ref{HOExpect}) yields the Killing vectors (\ref{KillingSU11}).  Note that, as it should, the vector fields are related to the expectation values even though they are not written in complex coordinates.  The choice of coordinates does not affect these relations.  

\subsection{Lemmas Pertaining to Dynamical Symmetries}
\label{SymmCalc}

We will now show how many physical quantities can be calculated using only the expectation values of symmetry generators and symmetry transformations.  First we will derive results for an arbitrary set of real coordinates, $\left\{s_1, s_2, ..., s_n \right\}$ and then show some noteworthy features particular to the complex coordinates. \\ \\
We note the following general feature.  Given a generator of symmetry $K'$ and its corresponding Killing vector $\chi_{K'}^\mu$ we note that
\begin{equation}
e^{\alpha \chi^\mu_{K'} \partial_\mu} \langle \psi|\vec{s} \rangle = \langle \psi|g_{K'}(\vec{s} ) \rangle = (f_{K'}(\vec{s}))^{-1} \langle \psi|e^{i\alpha {K}'} |\vec{s} \rangle \label{KillOp}
\end{equation}
where $\langle \psi|$ is an arbitrary bra independent of the coordinates and $f_{K'}$ a normalisation factor associated with the dynamical symmetry (\ref{DynamicSymm}).\\ \\
The consequence of (\ref{KillOp}) is that
\begin{eqnarray}
e^{\alpha \chi_{K'}^\mu \partial_\mu} \frac{ (\vec{s} | A |\vec{s})   }{(\vec{s}|\vec{s})} & = & \frac{ (g_{K'}(\vec{s})| A |g_{K'}(\vec{s}))   }{(g_{K'}(\vec{s})|g_{K'}(\vec{s}))} \nonumber \\
& = & \frac{ (g_{K'}(\vec{s})| A |g_{K'}(\vec{s}))   }{(g_{K'}(\vec{s})|g_{K'}(\vec{s}))} \frac{N^{*}_{K'} N_{K'} }{N^{*}_{K'} N_{K'}} \nonumber \\
& = & \frac{ (\vec{s}|e^{-i\alpha K'} A e^{i\alpha K'} |\vec{s})   }{(\vec{s}|e^{-i\alpha K} e^{i\alpha K}|\vec{s})}
\end{eqnarray}
where $A$ is an arbitrary operator.  From this we derive (from the leading order in $\alpha$) that
\begin{equation}
\chi_K' \langle \vec{s}| \hat{A}| \vec{s} \rangle = \chi_{K'}^\mu \partial_\mu \langle \vec{s}| \hat{A}| \vec{s} \rangle = i \langle \vec{s}| \ [\hat{A}, K'] \ | \vec{s} \rangle.   \label{KillComm}
\end{equation}
In the complex coordinates $\left\{ \overline{\vec{z}}, \vec{z}\right\}$ we can exploit another property.  It can be shown that
\begin{equation}
\chi_{K'}^{z_\mu} \partial_{z_\mu} \langle \vec{z} | \hat{A}| \vec{z} \rangle = i \langle \vec{z}| \ \hat{A} \ (\delta K') \ | \vec{z} \rangle .  \label{KillGen}
\end{equation}
where the derivatives $\partial_{z_\mu}$ are with respect to the holomorphic coordinates only and $\delta K' = K' - \langle \vec{z}| K' | \vec{z} \rangle$.  This procedure thus inserts an operator into the existing expectation value.  Note that we require the expectation value of an operator as a starting point for this insertion procedure.  The derivation of the result is almost identical to (\ref{KillComm}).  From the overlap (\ref{KillGen}) one may calculate the overlap $\langle \vec{z} | \hat{O} K'| \vec{z} \rangle$ by algebraic means if the overlap $\langle \vec{z} | K'| \vec{z} \rangle$ is also known.  A more general expression than eq. (\ref{KillGen}) is given by
\begin{equation}
\chi_{i G} \langle \vec{s}| A | \vec{s} \rangle =  - \langle \vec{s} | \delta G A + A \delta G |\vec{s} \rangle  \label{antiComm}
\end{equation}
 where $G$ is hermitian and $A$ is an arbitrary operator.  The two lemmas (\ref{KillComm}) and (\ref{KillGen}) now allow one to generate an arbitrary string of symmetry generators $\langle \vec{z}| K'_1 K'_2 ... K'_n |\vec{z} \rangle$ inside the expectation value.  It is crucial, for this procedure to work that the states can be parametrised by complex coordinates and that the normalised expectation values of the symmetry generators are known.    

\subsection{Calculations Utilising Symmetries}

\label{CalcUseSymm}

The results of (\ref{KillComm}) and (\ref{KillGen}) indicate that, by using only the expectation values of symmetry generators and complex coordinates as a starting point, the expectation value of an arbitrary string of symmetry generators can be calculated.  The symmetry transformations can be calculated by solving the Killing equations (\ref{Killing}) and the normalised expectation values can be calculated from their relation to the Killing vectors (\ref{OperatorVector}).  Of course, the larger the symmetry group the more expectation values can be calculated in this way.  For coherent states of some group we can calculate any string of operators from the enveloping algebra using only these quantities.  \\ \\
Two comments are in order here.  Firstly, it is clear that if our interest is limited to the expectation values of symmetry generators there is no need to consider a gravitational dual since the geometry will suffice.  The Killing vectors capture the symmetries and the expectation values of the corresponding symmetry generators may be calculated from these (\ref{OperatorVector}).  As shown in section \ref{SymmCalc}, these quantities are all that are required to calculate an arbitrary string of symmetry generators, when dealing with complex coordinate states.  In terms of the goal of constructing a gravitational dual this is a promising feature since it guarantees that these same quantities of interest can be extracted from the gravitational dual.  However, it also muddles the motivation for seeking a gravitational dual in the first place - why bother if the geometry is already sufficient to calculate the quantities of interest? \\ \\
It needs to be remembered that the results we have derived thus far depend on the fact that we have states parametrised by complex coordinates.  Certainly, there are many interesting choices of normalisable states that do not have this feature and, indeed, we will encounter examples in chapter \ref{McGChap}.  The choice of complex coordinates is beneficial since it has very useful properties and will allow us to construct gravitational duals more simply.  It thus serves as a useful benchmark before less favourable examples are considered.  In addition it is likely that there may be quantities of interest that are not expectation values e.g. the central charge.  

\section{Equations of Motion}

\label{EOMSection}

Our aim is to proceed from the relation between the expectation values of operators and vector fields on the manifold (\ref{OperatorVector}) to a gravitational dual.  As mentioned, our strategy is to identify an action where the metric and expectation values of operators leave the action integral stationary.  In order to gain some insight into how such an action may be chosen we will now proceed to write down equations of motion for the expectation values of operators.  Once we have these equations of motion our second step will be to match these to the field equations of a theory of gravity.  The discussion presented here will make no assumptions of the dimension of the manifold but we will specialise to the two-dimensional case in section \ref{TwoDimEOM} and proceed with that in the remainder of this chapter.  We will return to the higher-dimensional case in chapters \ref{HighDim} and \ref{McGChap}.  Again, we emphasise that we assume that the quantum states may be parametrised by a set of complex coordinates.  \\ \\
A general vector field on the manifold satisfies (\ref{FlowKill}), restated here for convenience,
\begin{equation}
\nabla_\mu \chi_\nu + \nabla_\nu \chi_\mu = \tau_{\mu\nu},  \label{FlowEq}
\end{equation}  
where $\tau_{\mu\nu}$ are what we refer to as the flow parameters.  Two notable examples are the Killing vectors which have $\tau_{\mu\nu} = 0$ and conformal Killing vectors with $\tau_{\mu\nu} = \kappa(\vec{z}, \vec{z}) g_{\mu\nu}$ (\ref{kappaDef}).  Furthermore, like any vector field, they also satisfy
\begin{equation}
(\nabla_\mu \nabla_\nu - \nabla_\nu \nabla_\mu) \chi_\gamma = R_{\mu\nu\gamma}^{\ \ \ \ \delta} \chi_\delta \label{RiemDef}
\end{equation}
where $R_{\mu\nu\gamma}^\delta$ is the Riemann curvature tensor.  Combining cyclic permutations of (\ref{RiemDef}) and using (\ref{FlowEq}) leads to \cite{Wald}
\begin{equation}
\nabla_\alpha \nabla_\beta \chi_\gamma = R_{\gamma \beta \alpha}^{\ \ \ \  \delta} \chi_\delta - \frac{1}{2}\left( \nabla_\alpha \tau_{\beta \gamma} + \nabla_{\beta} \tau_{\alpha \gamma} - \nabla_{\gamma} \tau_{\alpha \beta}    \right) \equiv R_{\gamma \beta \alpha}^{\ \ \ \  \delta} \chi_\delta -\frac{1}{2}\tau_{\alpha\beta\gamma}
\end{equation}
where the tensor $\tau_{\alpha\beta\gamma}$ is defined in the second expression.  The vector fields, as discussed, are related to the normalised expectation values of operators (\ref{OperatorVector}).  We label the arbitrary expectation value by $\Phi$, which then satisfies
\begin{equation}
\sigma_{\gamma}^{\ \epsilon} \nabla_\alpha \nabla_\beta \nabla_\epsilon \Phi = R_{\gamma \beta \alpha}^{\ \ \ \  \delta}\sigma_{\delta}^{\ \epsilon} \nabla_{\epsilon} \Phi - \frac{1}{2}\tau_{\alpha \beta \gamma}   \label{EOM}
\end{equation}
after using the property that $g_{\mu\nu}$ and $\sigma_{\mu\nu}$ have zero covariant derivative (\ref{Compat}).  Eq. (\ref{EOM}) can be thought of as an equation of motion for the scalar $\Phi$.  On the left hand side we have a third order differential operator while on the right hand side we have a first order differential operator. \\ \\
In order to link these equations of motion to the equations of motion of a theory of gravity we would like them to have at most second order derivatives.  We should then be able to find a theory of gravity that yield, as the field equation solution, the expectation value of some operator.  It is worthwhile to note an alternative, namely that we may consider the equation (\ref{EOM}) as a second order differential equation for the vector $\nabla_\epsilon \Phi$.  It should be possible to find a gravitational action, which is a functional of a metric and vector fields, that matches these equations of motion as the field equations of the action.  This is a possible avenue of future study.  For our purposes a second order differential equation for the scalar fields $\Phi$ will be convenient especially in two dimensions.  This will allow us to give elegant interpretations for the expectation values of symmetry generators and conformal symmetry generators in the context of a gravitational dual.  

\subsection{Two Dimensional Equations of Motion}
\label{TwoDimEOM}

Extracting simple and elegant equations of motion with at most second order derivative from (\ref{EOM}) is not always possible.  In order to simplify matters we will focus our attention on the two-dimensional manifolds.  In two dimensions we can find exact second order expressions for these equations of motion owing to the special form for the Riemann curvature tensor (see Appendix \ref{AppGeo})
\begin{equation}
R_{\alpha \beta \gamma \delta} = \frac{R}{2}(g_{\alpha \gamma} g_{\beta \delta} - g_{\alpha \delta} g_{\beta \gamma}) \label{2DRiem}
\end{equation}
where $R$ is the scalar curvature.  One may use the identity (\ref{2DRiem}) in (\ref{EOM}) to find
\begin{equation}
\sigma_{\mu}^{\ \gamma} R_{\gamma \beta \alpha}^{\ \ \ \ \delta}\sigma_{\delta}^{\ \epsilon} \nabla_{\epsilon} \Phi = \frac{R}{2}g_{\mu\beta} \nabla_{\alpha} \Phi.
\end{equation}
We may further use the property (\ref{sigmaInverse}) to rewrite (\ref{EOM}) as
\begin{equation}
\nabla_\alpha \nabla_\beta \nabla_{\gamma} \Phi = -\frac{R}{2} g_{\beta \gamma} \nabla_\alpha \Phi + \frac{1}{2}\tau_{\alpha \beta \delta} \  \sigma^{\delta}_{\ \gamma}.  \label{EOM2}
\end{equation}
This then implies that
\begin{equation}
\nabla_\alpha \left( \nabla_\beta \nabla_\gamma \Phi + \frac{R}{2}g_{\beta \gamma} \Phi \right) = -\frac{1}{2}g_{\beta \gamma} \Phi \nabla_\alpha R + \frac{1}{2}\tau_{\alpha \beta \delta} \  \sigma^{\delta}_{\ \gamma} \equiv \nabla_\alpha \tilde{T}_{\beta \gamma}, \label{EnergyMom}
\end{equation}
which in turn implies that
\begin{equation}
\nabla_\beta \nabla_\gamma \Phi + \frac{R}{2}g_{\beta \gamma} \Phi = \lambda g_{\beta \gamma} + \tilde{T}_{\beta\gamma}   \label{EOMFinal}
\end{equation}
where $\lambda$ is a constant.  Note that the tensor $\tilde{T}_{\beta \gamma}$ has been defined in the last step of (\ref{EnergyMom}).  The undetermined term proportional to the metric is a consequence of the fact that the metric has zero covariant derivative and that the expression is symmetric (so that a term proportional to the symplectic form is not applicable).  We will show later how the tensor $\tilde{T}_{\beta\gamma}$ may be interpreted as an energy momentum tensor in a gravitational model.  We will, however, not be calculating it explicitly here for any scalar curvature and flow parameter.  Note that if the scalar curvature is constant and $\Phi$ is the expectation value of a generator of symmetry then $\tilde{T}_{\beta\gamma} = 0$.  \\ \\
Not all of the three equations in (\ref{EOMFinal}) are independent \cite{Mann}.  We may extract two independent equations as follows.  The first comes from a contraction of indices in (\ref{EOMFinal}).  This yields
\begin{equation}
\nabla^2 \Phi + R\Phi = 2 \lambda + g^{\alpha \beta} \tilde{T}_{\alpha \beta}.   \label{ScalarfieldEq}
\end{equation}
The second can be extracted by multiplying (\ref{EOMFinal}) by $\nabla_\delta \Phi$ and contracting the index $\delta$ with one of the existing indices
\begin{eqnarray}
g^{\beta \delta } \left( \lambda g_{\beta \gamma} + \tilde{T}_{\beta\gamma} \right) \nabla_\delta \Phi & = & g^{\beta\delta}\left( \nabla_\beta \nabla_\gamma \Phi + \frac{R}{2}g_{\beta \gamma} \Phi \right) \nabla_\delta \Phi \nonumber \\  & = & \frac{1}{2}g^{\beta\delta}\nabla_\gamma \nabla_\beta \Phi \nabla_\delta \Phi + \frac{1}{2}g^{\delta\beta}\nabla_\gamma \nabla_\delta \Phi \nabla_\beta \Phi + \frac{R}{2} \Phi \nabla_\gamma \Phi \nonumber \\
& = & \nabla_\gamma \left( \frac{1}{2}(\nabla\Phi)^2 + \frac{R}{4}\Phi^2   \right) \equiv \nabla_\gamma M.  \label{MassField}
\end{eqnarray}
The potential $M$ is defined in the last line.  To be more specific, the quantity $M$ is defined as
\begin{equation}
M = \frac{1}{2}(\nabla\Phi)^2 + \frac{R}{4}\Phi^2 \label{MassDefine}
\end{equation}
while the field equation is
\begin{equation}
\nabla_\gamma M = g^{\beta \delta } \left( \lambda g_{\beta \gamma} + \tilde{T}_{\beta\gamma} \right) \nabla_\delta \Phi.  
\end{equation}
Note that $M + c$ where $c$ is a constant will satisfy the same field equation.  In the discussions of the gravity models from section \ref{JTSection} onwards we will give a physical interpretation to the potential $M$.  It is customary to normalise the potential appropriately (see e.g. \cite{Mann}) i.e. $M \rightarrow \frac{M}{M_0}$ but we will only do this later.  The discussion up to that point will be independent of this normalisation.

\subsection{Equations of Motion for Constant Scalar Curvature}

We may simplify the equations of motion even further in the case of a constant, non-zero scalar curvature since we have the freedom to add a constant to the scalar $\Phi$, as mentioned.  By now adjusting $\Phi \rightarrow \Phi + \frac{2 \lambda}{R}$ we can get rid of the constant $\lambda$ in (\ref{EOMFinal}) i.e.
\begin{equation}
\nabla_\beta \nabla_\gamma \Phi + \frac{R}{2}g_{\beta \gamma} \Phi = \tilde{T}_{\beta\gamma}.  \label{DilEOM}
\end{equation}
Furthermore, for the constant scalar curvature case, if $\Phi$ is the expectation value of a conformal symmetry generator (\ref{FlowEq}), we can again derive simple equations of motion.  We start from (\ref{EOM2}) and contract indices so that
\begin{eqnarray}
\nabla^2 \nabla_\gamma \Phi + \frac{R}{2}\nabla_\gamma \Phi & = & \frac{2-d}{2}\sigma^\delta_{\ \gamma} \nabla_\delta \kappa = 0 \ \ \ \textnormal{for} \ \ \ d=2 \nonumber \\
\Rightarrow \nabla^2 \Phi + R \Phi & = & 0.  \label{ConfField}
\end{eqnarray}
Here we have again shifted the scalar $\Phi$ to get rid of the arbitrary constant $\lambda$. The function $\kappa$ that appears in (\ref{ConfField}) is defined in (\ref{kappaDef}) and features here because we are considering conformal transformations.  \\ \\
We note finally that the field equations are a condition on the second order covariant derivative of the expectation values $\Phi$.  The $\Phi$'s will be the solutions of the equations of motion of the gravitational model.  In the gravity model we have to give as input the appropriate energy momentum tensor $\tilde{T}_{\beta\gamma}$ in order to ensure that we will get the desired $\Phi$'s as the solutions.  The discussion here shows that, equivalently, we can specify the values of $M$ (\ref{MassField}) and $g^{\alpha \beta} \tilde{T}_{\alpha \beta}$ (\ref{ScalarfieldEq}) instead of the value for $\tilde{T}_{\beta\gamma}$.  

\subsection{What Then is the Procedure?}

Before we proceed to specific examples of the duality we will summarise our steps from a quantum model to gravitational dual carefully.  We start from a family of quantum states, which for simplicity we assume to be parametrisable in terms of complex coordinates.  The quantities we are interested in on this quantum mechanical side are the expectation values of arbitrary strings of observables.  On the quantum mechanical side these can be calculated using, for instance, commutation relations of the operators. \\ \\
Our focus for this thesis is rather on how these quantities can be calculated in a dual gravitational description.  In this regard we showed that the expectation values of operators can be thought of as the solutions to field equations characterised by a metric and an energy momentum tensor.  In other words, you give as input an energy momentum tensor and some boundary conditions and the output is the desired expectation value.  The components of the energy momentum tensor, in two dimensions, cannot all be chosen arbitrary.  Indeed, we only have the freedom to specify two of them.  From these we extracted two scalar quantities from two different contractions of the energy momentum tensor - the trace of the energy momentum tensor and the potential $M$.  If you give as input the trace and $M$, one can write down the field equations which have as solution the desired expectation value.  \\ \\
The expectation value of an operator can thus be calculated either directly from the commutation relations (on the quantum side) or by solving the appropriate field equations (on the gravity side).  The natural question to ask is how the energy-momentum tensor is to be chosen?  To this we can only provide an answer in the simplest cases e.g. where the scalar curvature is constant and one is interested in the expectation value of a generator of symmetry.  For this reason the method we show here should not be thought of as a calculation tool but a proof of concept - that such a repackaging of a quantum mechanical into a gravitational theory is possible.  If we have the quantum expectation values we can calculate the appropriate energy momentum tensor and vice versa.  \\ \\
What we have not shown, up to this point, is that the equations of motion (\ref{EOMFinal}) belong to a proper theory of gravity.  We will address this point now.

\section{Dilaton Gravity}

\label{JTSection}

When we discussed (qualitatively) possible gravitational duals  in section \ref{ProcGravDual} we mentioned briefly that Einstein gravity (general relativity) in two dimensions is not a sensible choice.  This is due to the fact that any metric in two dimensions is Einstein (see Appendix \ref{AppGeo}) so that the field equations are always trivial and physical quantities such as the energy momentum tensor cannot be imprinted on the metric.  We have to consider an alternative model of gravity as a gravitational dual for two dimensional quantum states\footnote{Even in higher dimensions an action that is a functional of both the metric and fields remains useful since we can attach a direct quantum mechanical interpretation to these fields.}.   \\ \\
Such an alternative model of gravity in two dimensions is dilaton gravity.  The dilaton is a field in these models that has the physical interpretation of the volume of compactified dimensions (which is allowed to fluctuate).  The presence of a dilaton field is a quite general consequence when a higher dimensional model of gravity is compactified to give an equivalent lower dimensional one \cite{CoordChange}, \cite{DilatonReview}.  \\ \\
This choice of gravitational model is made for two main reasons.  Firstly, it is a relatively simple choice of a gravitational model and indeed, the Jackiw-Teitelboim (JT) model which is only linear in the dilaton field, is arguably the simplest model of gravity one can consider beyond Einstein gravity.  Secondly, it is precisely this model of Jackiw-Teitelboim gravity that has been studied in the context of the $AdS_2/CFT_1$ correspondence \cite{Cadoni1}-\cite{CadoniCQM}.  With this choice of gravitational model we can hopefully relate our construction to these existing results.  For this reason we will discuss the main results of these works before we proceed to analyse the dilaton gravity duals ourselves.  \\ \\
We will first define the model of dilaton gravity and show that its field equations may be matched to the equations of motion we have derived (\ref{EOMFinal}).  It is a well-studied fact that \cite{Gegenberg3} a generic dilaton model with at most second order derivatives can be transformed by appropriate redefinitions of the fields and metric into the form
\begin{equation}
S_{JT}[\eta, g_{\mu\nu}, L_M] = \int{d^2 x \sqrt{|g|}\left( R\eta + V(\eta, L_M) \right)} \label{JTAction}
\end{equation}
where $\eta = e^{-\phi}$ and $\phi$ is the dilaton.  We will, however, be referring to the field $\eta$ as the dilaton for conciseness.  $V$ is a functional of both the dilaton and matter content $L_M$.  The dilaton can be coupled to both the metric and the matter content, (see e.g. \cite{Cosmology}).  \\ \\ 
Several sets of field equations can be derived from this model depending on which degrees of freedom are considered dynamic and which static.  See Appendix \ref{FieldEqApp}  for some of the detailed derivations.  For instance, varying with respect to the dilaton, $\eta$, yields
\begin{equation}
\frac{\delta S_{JT}}{\delta \eta} = 0  \ \Rightarrow \ R = -\frac{\partial V}{\partial \eta} \equiv V_{\eta}.  \label{DilField}
\end{equation}
Varying the action (\ref{JTAction}) with respect to the inverse metric yields
\begin{equation}
0 = \frac{1}{\sqrt{-g}}\frac{\delta S_{JT}}{\delta g^{\mu\nu} } \Rightarrow  -\nabla_\mu\nabla_\nu \eta + g_{\mu\nu} \nabla^2\eta = V_{\mu\nu}.     \label{JTdilmetric}
\end{equation}
We single out the quantity 
\begin{equation}
V_{\mu\nu} \equiv -\frac{\partial \sqrt{g} V}{\partial g^{\mu\nu} },
\end{equation}
which is an energy momentum tensor.  The field equations can be rewritten after multiplying by $g^{\mu\nu}$ and summing over the indices
\begin{eqnarray}
\nabla^2 \eta & = & g^{\mu\nu} V_{\mu\nu} \nonumber \\
\Rightarrow \nabla_\mu\nabla_\nu \eta & = & g_{\mu\nu} g^{\alpha\beta}V_{\alpha\beta} - V_{\mu\nu}. \nonumber \\
\Rightarrow \nabla_\mu\nabla_\nu \eta + \frac{R}{2} g_{\mu\nu}\eta & = & \left(\frac{R}{2} g_{\mu\nu}\eta +  g_{\mu\nu} g^{\alpha\beta}V_{\alpha\beta} - V_{\mu\nu} \right) \label{DilatonEq}
\end{eqnarray}
The equation (\ref{DilatonEq}) is written in this form so that it may be compared easily with the left hand side of (\ref{EOMFinal}).  By comparing (\ref{DilatonEq}) and (\ref{EOMFinal}) it is clear that we could make a choice for the function $V(\eta, L_M)$ that will ensure that the dilaton gravity field equations match the equations of motion of the expectation values of operators. Specifically, we can make a choice for $V$ linear in the dilaton i.e.
\begin{equation}
V = -R_s \eta + T(L_M),  \label{VForm}
\end{equation}
which determines the geometry independent of the dilaton (\ref{DilField}) i.e. $R = R_s$. We use the symbol $R_s$ to emphasise that $\frac{\delta R_s}{\delta g^{\mu\nu}} = 0$.  $T$ is some functional of the matter content.  Note that, as before, of the three field equations (\ref{DilatonEq}) only two are independent.  The energy momentum tensor $V_{\mu\nu}$ therefore only contains two independent entries.  For this purpose we use the trace of the energy momentum tensor and the potential (\ref{MassDefine}) $M = \frac{1}{2}(\nabla \eta)^2 + \frac{R}{4}\eta^2$.  The energy momentum tensor can be reconstructed from these two quantities.\\ \\
Having now identified dilaton gravity as fitting all our requirements we can proceed with building the quantum mechanics / dilaton gravity dual in two dimensions.  Before we proceed with this we take note of some important works from the $AdS_2/CFT_1$ literature.  These will guide our discussions and place our findings into context.  

\section{Some Relevant $AdS_2/CFT_1$ Results from the Literature}

\label{RelResults}

Dilaton gravity, and specifically the Jackiw-Teiltelboim model \cite{JTOrig}, has been studied in a series of works (\cite{Cadoni1}, \cite{Cadoni2}, \cite{Cadoni3}, \cite{Cadoni99}, \cite{CadoniCQM}) centered around the $AdS_2/CFT_1$ correspondence.  One of the motivations behind these works is to calculate the entropy of a two-dimensional dilaton gravity black hole by using its $CFT_1$ dual.  Though this calculation can be carried out without the $AdS_2/CFT_1$ dictionary it serves to show that quantities in $2d$ dilaton gravity can be calculated from the dual quantum theory.  Since the conformal group in one and two dimensions is infinite dimensional the $CFT_1$ and $CFT_2$ e- and higher-point correlation functions are more restricted \cite{Francesco}.  These low dimensional examples, where the field theory is comparatively simple, thus provide a good opportunity to study low-dimensional gravity by using its dual theory.  \\ \\
The development of these works is a progressively more refined calculation of the conformal algebra central charge culminating in \cite{CadoniCQM} with the correct value for the black hole entropy.  In this final work it was identified that the on-shell equations for the asymptotic fluctuations of the dilaton and metric can be matched with the equations of motion for a scale invariant system coupled to a time-dependent quadratic source.  This is the model of CQM discussed in section \ref{CQMSection}. \\ \\
We sketch their calculations and arguments here for the sake of completeness and to later relate it to our framework.  To relate it more explicitly we use a choice of coordinates more akin to those used in this thesis.  The model of gravity under consideration is the JT-model of dilaton gravity
\begin{equation}
S_{JT}[\eta, g_{\mu\nu}, L_M] = \int{d^2 x \sqrt{|\det(g_{\mu\nu})|}\left( R\eta + \frac{4}{k} \eta \right)}.   \label{SU11JT}
\end{equation}
Varying with respect to the dilaton and inverse metric yields the field equations which are of the form (\ref{DilField}) and (\ref{JTdilmetric}).  The on-shell geometry is thus locally $AdS_2$.  However, the dilaton must also be taken into account in the geometry.  In particular the dilaton allows the definition of the potential $M$ (\ref{MassField}).  The presence of the dilaton now gives rise to three inequivalent solutions of the field equations.  These are labelled $AdS_{+}$, $AdS_{0}$ and $AdS_{-}$ where the subscript is the sign of $M$.  The reason these solutions are inequivalent is because, firstly, $M$ is unaffected by coordinate transformations.  Secondly, one may rescale the dilaton $\eta \rightarrow N \eta$ which yields $M \rightarrow N^2 M$ so that a (real) rescaling can change the magnitude but not the sign of $M$.  Complex rescalings are not allowed since the dilaton is a real scalar function.  There is thus no legitimate transformation that can map between these three different solutions.  \\ \\
Dilaton gravity in two dimensions permits the definition of a two-dimensional black hole, if $M$ is positive \cite{Gegenberg2}.  The black hole does not give rise to a curvature singularity but has the same causal structure as a conventional black hole \cite{JTOrig}, \cite{Gegenberg2}.  The potential $M$ we have defined is a normalisation factor away from the black hole mass $M'$.  For our current analysis this normalisation is not important but we will show how it may be calculated in section \ref{DilNorm}.  \\ \\
The thermodynamic black hole entropy is given by \cite{CadoniCQM}
\begin{equation}
S = 4 \pi \sqrt{-\frac{M}{R}}
\end{equation}
and the authors of \cite{Cadoni1}-\cite{CadoniCQM} set out to calculate this entropy from the boundary $CFT_1$ of the JT-model.  This is done by using the Cardy-Verlinde formula \cite{Cardy} which requires the central charge of the $CFT_1$.  This calculation is a two-dimensional analogue of the counting of quantum states used to calculate the entropy of the Ba\~nados-Teitelboim-Zanelli (BTZ) black hole in the $AdS_3/CFT_2$ correspondence \cite{BTZAds3}.  The BTZ black hole is a solution to the Einstein-Hilbert equations in three dimensions that may be interpreted as a black hole.  Though there is no curvature singularity the causal structure of spacetime is like that of a black hole.  Associated with this black hole is a temperature and entropy.  \\ \\
With the two-dimensional correspondence there is a novelty namely that $AdS_2$ has two boundaries that needs to be taken into account and not just the asymptotic boundary.  Both of these boundaries give a contribution to the central charge and are handled separately.  

\subsection{The Central Charge Calculation}

To extract the central charge of the boundary $CFT_1$ from the dilaton gravity one requires two ingredients.  First, one requires a quantity from the gravitational theory that can be related to the energy momentum tensor of the $CFT_1$.  Second, one requires a set of allowed transformations of time on the boundary.  The central charge may then be calculated by varying the energy momentum tensor with these transformations, like in (\ref{EMCoordTrans}). 

\subsubsection{The Positivity Condition of the Dilaton}

\label{DilNorm}

One technical detail that needs adressing before we do this is the requirement of the positivity of the dilaton - the dilaton is by definition a strictly positive function.  We have built up the dictionary by choosing the interpretation for the field equation solution, the dilaton, as the expectation values of an operator.  This can seemingly cause issues.   Even if the operators are hermitian their expectation values are not strictly positive.  This is not a major detriment to our calculations since we can calculate the expectation value in the region where it is strictly positive and simply extend the solution to the other regions afterwards. \\ \\
In terms of interpreting the expectation values as dilaton gravity objects their positivity is an important requirement, however.  One of the main features of this positivity condition is that the potential $M$ (\ref{massDef}) can be normalised in such a way that it has the interpretation of the mass of the black hole.  The normalisation constant can be found by applying Birkhoff's theorem \cite{Cadoni1} which states that there always exists a choice of coordinates $\left\{t, x \right\}$ such that the metric and dilaton can be written as
\begin{equation}
ds^2 = -(\lambda^2 x^2 - a^2)dt^2 + (\lambda^2 x^2 - a^2)^{-1}dx^2 \ \ \ ; \ \ \ \eta = \eta_0 \lambda x \label{Birkhoff}
\end{equation}
where the manifold is restricted to $x > 0$.  The normalisation constant $\eta_0 \lambda$ is positive owing to the positivity of the dilaton.   This normalisation of the dilaton is used to normalise the potential appropriately and we define
\begin{equation}
M' = \frac{M}{\lambda \eta_0}. \label{massDef}
\end{equation}
The quantity $M'$ (\ref{massDef}) now has the interpretation of an energy \cite{Gegenberg2} which is the mass of a black hole \cite{Cadoni1}-\cite{CadoniCQM}.  If the dilaton is not strictly positive then we may induce a sign flip in (\ref{massDef}).\\ \\
We classified the three inequivalent solutions of the vacuum $JT$-model as $AdS_{\pm}$, $AdS_0$ where the subscript indicates the sign of the potential $M$.  The convention is rather to use the mass $M'$ to do this classification.  $M'$ transforms linearly with the dilaton normalisation $M' \rightarrow N M'$ but again a sign change cannot be made since the dilaton normalisation must be positive.  

\subsubsection{The Transformations and Energy-Momentum Tensor}

\label{AsympCentral}

In order to calculate the central charge we require two ingredients - a set of allowed transformations and an energy momentum tensor.  We first identify the transformations.  These are identified by examining the asymptotic symmetries of the model.  The asymptotic form of the $AdS_2$ metric is defined as
\begin{equation}
ds^2 = \left[\frac{k}{2 \beta^2} + \gamma_{\beta\beta}(t) + O(\beta)\right] d\beta^2 + \left[\gamma_{t\beta}(t)\beta + O(\beta^2) \right]  d\beta dt + \left[ \frac{k}{2 \beta^2} + \gamma_{tt}(t)+ O(\beta)\right] dt^2. \label{AsympMetric}
\end{equation}
The asymptotic form of the dilaton is given by
\begin{equation}
\eta = \eta_0\frac{\rho(t)}{\beta} + \eta_0\gamma_\eta(t) \beta + O(\beta^2). \label{AsympDil} 
\end{equation}
Note that the asymptotic form of the metric and dilaton contain five independent functions of time, three for the metric and two for the dilaton.  The on-shell equations of motion provide two restrictions for the five arbitrary functions in the asymptotic limit, namely
\begin{eqnarray}
0 & = & \gamma_{tt}(t) \dot{\rho}(t) - \frac{\rho(t)}{2}\dot{\gamma}_{\beta\beta}(t) - k \dot{\gamma}_\eta (t) \label{Field1} \\
\ddot{\rho}(t) & = & \frac{2}{k}\left( \gamma_{\beta\beta}(t) \rho(t) + \gamma_{tt}(t) \rho(t) + k\gamma_\eta(t) \right)  \label{Field2}.
\end{eqnarray}
In addition the function $\gamma_{\beta t}(t)$ does not affect the analysis.  This can be understood intuitively since a coordinate change can always eliminate the off-diagonal entry of the two-dimensional metric.  Due to this and the two restrictions (\ref{Field1}), (\ref{Field2}) there are only two independent functions of time (out of the original five) that prove relevant. \\ \\
The allowed transformations are now the ones that preserve the asymptotic forms of the metric (\ref{AsympMetric}) and dilaton (\ref{AsympDil}) namely
\begin{equation}
X^t = \epsilon(t) - \frac{1}{2}\ddot{\epsilon}(t)\beta^2 + \alpha^t (t) \beta^4 + O(\beta^5) \ \ \ ; \ \ \ X^\beta = \dot{\epsilon}(t)\beta + \alpha^\beta(t) \beta^3 + O(\beta^4)   \label{AsympTransform}
\end{equation}
where the $\alpha^t(t)$ and $\alpha^\beta(t)$ are arbitrary functions of $t$ and represent the so-called ``pure gauge" diffeomorphisms \cite{Cadoni1}.  The explicit transformation of the dilaton induced by these transformations are
\begin{eqnarray}
\delta \eta = X^\mu \partial_\mu \eta & = &  \ \ \frac{\epsilon(t) \dot{\rho}(t) - \dot{\epsilon}(t) \rho(t)}{\beta} \nonumber \\
& & + (\epsilon(t) \dot{\gamma}_{\eta}(t) + \gamma_{\eta} \dot{\epsilon}(t) + \frac{1}{2} \dot{\rho}(t) \ddot{\epsilon}(t) - \alpha^\beta(t)\rho(t) )\beta + O(\beta^2) \nonumber \\
& \equiv & \frac{1}{\beta}\delta_{\epsilon(t)} \rho + \delta_{\epsilon(t)} \gamma_{\eta} \beta + O(\beta^2).   \label{DilVar}
\end{eqnarray}
By $\delta_{\epsilon(t)}$ we mean the change in the function when we change $t \rightarrow t + \epsilon(t)$.  The diagonal entries of the metric are transformed (using $\delta g_{\mu\nu} = \frac{1}{2}(\nabla_\mu X_\nu + \nabla_\nu X_\mu)$) as
\begin{eqnarray}
\delta_{\epsilon(t)} \gamma_{\beta\beta}[\epsilon(t)] & = & \epsilon(t) \dot{\gamma}_{\beta\beta}(t) + 2 \gamma_{\beta\beta}(t) \dot{\epsilon}(t) + 2k \alpha^\beta(t), \label{MetVar2} \\
\delta_{\epsilon(t)} \gamma_{tt}[\epsilon(t)] & = & \epsilon(t) \dot{\gamma}_{tt}(t) + 2 \gamma_{tt} \dot{\epsilon}(t) - \frac{k}{2}\partial_t^3 \epsilon(t) - k\alpha^\beta(t) \epsilon(t). \label{MetVar1}
\end{eqnarray}
Note that the diffeomorphism associated with $\alpha^t(t)$ is absent from these near-boundary variations.  This is indicative of the fact that only two independent functions of time are relevant (out of the original three in (\ref{AsympTransform})).  \\ \\
The quantity in dilaton gravity that is related to the $CFT_1$ energy momentum tensor is found as follows.  In the Hamilton analysis of dilaton gravity one needs to add a boundary term to the action in order to find consistent field equations \cite{Cadoni1}.  The variation of the boundary term $J$ is calculated first and is given in terms of the variation of the dilaton (\ref{DilVar}) and the metric (\ref{MetVar2}) and (\ref{MetVar1}).  From this the authors calculate the boundary term to be
\begin{equation}
J[\epsilon(t)] =  -2 \eta_0 \epsilon \ddot{\bar{\rho}} \equiv \epsilon T_{tt}  \label{BoundaryTerm}
\end{equation}
where $\overline{\rho}(t) +1 = \rho(t)$.  The function $\rho(t)$ is the leading order of the dilaton (\ref{AsympDil}).  Note from eq. (\ref{BoundaryTerm}) that $T_{tt} = -2\eta_0 \ddot{\bar{\rho}}$.  The quantity $T_{tt}$ transforms as
\begin{equation}
\delta_{\omega(t)} T_{tt} = \omega \dot{T}_{tt} + 2\dot{\omega} T_{tt} + 2 \eta_0 \partial_t^3 \omega(t)   \label{ccEMT}
\end{equation}
 i.e. like the energy momentum tensor of a $CFT_1$ (\ref{EMCoordTrans}) with central charge $c = 24 \eta_0$.  This is the asymptotic boundary contribution to the total central charge. \\ \\
For our purposes the details of this calculation are not crucial.  Indeed, in this thesis we will not be recalculating the central charge at all.  What is important is that there are two ingredients to the analysis.  Firstly, a set of allowed transformations (that retain the asymptotic form of the dilaton and metric) and secondly, a specific quantity, the boundary term (\ref{ccEMT}), from which the central charge is extracted via the expression (\ref{ccEMT}). We will show, using our construction, that one arrives at the transformations (\ref{AsympTransform}) in a very natural way.  In addition our construction will set up the problem in such a way that the boundary term is of a very simple form.  

\subsection{The Inner Boundary Contribution to the Central Charge}
 
The contribution of the inner boundary, $\beta \rightarrow \infty$ to the central charge \cite{CadoniCQM} is calculated differently than that of the outer boundary.  We borrow the notation \cite{CadoniCQM} directly since we will provide the explicit transformations and motivations later.  In \cite{CadoniCQM} its contribution is attributed to entanglement of states on the $\beta \rightarrow \infty$ boundary and comes about due to the fact that the vacuum and black hole solutions, on the boundary, are related simply by a reparametrisation of time $t \rightarrow \tau$.  The contribution of the time-reparametrisation to the central charge needs to be subtracted from the full contribution.  The finite form of the transformation (\ref{ccEMT}) is given by
\begin{equation}
t = t(\tau) \ \ \ ; \ \ \ T_{\tau\tau} = \left( \frac{d t}{d\tau} \right)^2 T_{tt} - \frac{c_{ent}}{12}\left( \frac{d t}{d \tau}  \right)^2 \left\{ \tau, t \right\}  \label{FiniteTrans}
\end{equation}
where $\left\{ \tau, t \right\}$ is the Schwarzian derivative.  The explicit transformation $t = t(\tau)$ between the vacuum and black hole solution is substituted in (\ref{FiniteTrans}).  The authors of \cite{CadoniCQM} now state that by ``fixing the diffeomorphisms on shell" one can always have $T_{\tau\tau} = \frac{\lambda M}{\eta_0}$.  For the vacuum $M$ is zero while for the black hole $M$ is a constant.  Substituting everything yields $c_{ent} = -12 \eta_0$.  \\ \\
Taking both the inner and outer boundary contributions into account yields the value for the central charge compatible with the black hole entropy \cite{CadoniCQM}, $c + c_{ent} = 12 \eta_0$.  \\ \\
The observation that the diffeomorphisms can always be fixed so that $T_{\tau\tau} = \lambda\frac{M}{\eta_0}$ is an important one for our analysis in section \ref{OurCCharge}.  Our construction will set up the problem in such a way that the boundary term is always proportional to $M$ i.e. in the words of \cite{CadoniCQM} the ``diffeomorphisms on shell always remain fixed".  We contend that this not only allows for a less technical calculation of the central charge but also keeps the quantum mechanics / dilaton gravity dictionary neat and simple.  

\subsection{An Issue of Normalisation}

\label{NormSec}

In our ensuing construction the reader will note the absence of an explicit dilaton normalisation, $\eta_0 \lambda$ as in (\ref{Birkhoff}).  This is certainly not a trivial omission - the central charge of the boundary $CFT$ depends on these constants.  Our analysis in section \ref{OurCCharge} should thus be viewed as a qualitative one.  We again emphasise that we do not aim to reproduce the calculations of \cite{Cadoni1}-\cite{CadoniCQM} here.  Rather, we aim to show that, by following our construction, all the required ingredients to perform the calculation come about in a very natural way.  \\ \\
We hope to have this technical issue addressed when we present the analysis in a future publication.  

\section{The Dilaton Gravity Dual for the $SU(1,1)$ Model of CQM}

After this brief detour in the $AdS_2/CFT_1$ literature we may now finally analyse the quantum mechanics / dilaton gravity dictionary in our construction.  It is worthwhile to recall exactly how we got to this point.  A family of states are used to form a manifold of states and a metric is defined (\ref{Prov2}) on this manifold.  When the states are parametrised by complex coordinates the metric, and more specifically the symmetries of the metric, can be used to calculate the expectation value of an arbitrary string of symmetry generators.  Thus, in principle, a theory of gravity that only produced the metric would be sufficient.  However, the fields in the $AdS/CFT$ correspondence play a vital role and, furthermore, in two dimensions the Einstein tensor is trivial so that one is forced to consider alternate models of gravity.  The crucial choice we had to make was what interpretation we attach to the fields.  Motivated by the work of \cite{Ashtekar} we made a very direct choice - the fields are the expectation values of operators.  Though this choice is simple, it will be sufficient and match several existing works from the $AdS_2/CFT_1$ literature.  By examining the equations of motion of the expectation values of operators we proposed that a model of dilaton gravity (\ref{JTAction}) can be used as a gravitational dual in two dimensions.  For simplicity we consider the dilaton models where the functional $V$ is of the form (\ref{VForm}).  To complete the dictionary we need to find the appropriate matter content to include in the action to source a specific expectation value.  The calculation of a desired expectation value can then be completed on the gravity side of the duality.  As input one gives the appropriate matter content and the field equation solution is then the expectation value.  \\ \\
As a first example we will now examine the case where the family of quantum states used to calculate the metric are the $SU(1,1)$ Hamiltonian states.  As we have already shown in section \ref{UnsourcedSect}, the scalar curvature for the $SU(1,1)$ family of states is $AdS_2$ with scalar curvature $R = -\frac{4}{k}$ where $k$ is the representation label.  The dual dilaton gravity action is thus
\begin{equation}
S = \int{d^2 x \sqrt{|g|}\left( R\eta + \frac{4}{k} \eta + T(L_M) \right)} \label{SU11Action}
\end{equation}
where $T$ must be chosen so as to produce the appropriate energy momentum tensor and is only a functional of matter.  Our task is now to find the matter content that would source a desired expectation value.  

\subsection{The $SU(1,1)$ Symmetry Generators}

In the section \ref{EOMSection} we have shown that the field equations are significantly simpler when one considers the generators of symmetry.  Indeed, since the scalar curvature is also constant for the $SU(1,1)$ Hamiltonians the energy momentum tensor that sources the expectation values of the symmetry generators $H, D, K$ is zero.  We thus set $T=0$ in (\ref{SU11Action}).  The field equations correspond exactly to those studied in \cite{Cadoni1}-\cite{CadoniCQM} namely
\begin{eqnarray}
R & = & -\frac{4}{k} \\
0 & = & -\nabla_\mu \nabla_\nu \eta + g_{\mu\nu} \nabla^2 \eta - \frac{1}{2}g_{\mu\nu}\left(\frac{4}{k}\right) \eta \ \ \ \ \ \ \ \ \ \ \ \ \ 
\end{eqnarray} 
or equivalently  
\begin{eqnarray}
R & = & -\frac{4}{k} \label{One}\\
\nabla^2 \eta & = & -R\eta \label{Two}\\
\partial_\mu M & = & 0 \ \ \ \textnormal{where} \ \ \ M = \frac{1}{2}(\nabla \eta)^2 + \frac{R}{4}\eta^2 \label{Three}
\end{eqnarray}
There are three linearly independent solutions of the field equations corresponding to the expectation values of $H, D$ and $K$.  This is a coordinate independent result but can be verified for the free particle metric (\ref{freeMet}) and the expectation values (\ref{HOExpect}), which we restate here for convenience
\begin{equation}
ds^2_{H_0 = H} = \frac{k}{2\beta^2}(d\beta^2 + dt^2)
\end{equation}
and
\begin{eqnarray}
\langle \beta, t| H | \beta, t\rangle & = & \frac{k}{\beta},\nonumber \\
\langle \beta, t| D | \beta, t\rangle & = & \frac{k t}{\beta}, \nonumber \\
\langle \beta, t| K | \beta, t\rangle & = & \frac{k(t^2 + \beta^2)}{\beta}.  
\end{eqnarray}
As discussed in section \ref{RelResults}, the $JT$-model gives three globally distinct solutions characterised by the sign of the potential $M$ (\ref{MassField}), $AdS_{\pm}$ and $AdS_0$. The natural question to ask is whether there is a relation between the three linearly independent dilaton solutions and the sign of the potential.  To gain clarity on this we examine the general $SU(1,1)$ symmetry generator, $u H + v D + w K$, (the expectation values are found in (\ref{HOExpect})) and substitute this into (\ref{MassField}) to find
\begin{equation}
M_{u H + v D + w K} = \frac{1}{R}(u w - \frac{1}{4}v^2).  
\end{equation}
The potential is constant, as it should be since the symmetry generator solves the on-shell field equations.  Note that it is proportional to the value of $d^2$ that classifies the $SU(1,1)$ as elliptic, parabolic or hyperbolic (\ref{d2Expression}).  The quantity $M$ thus indicates which $SU(1,1)$ element expectation value, up to unitary transformation, is the solution of the field equations.  The expectation values of the elliptic operators are thus the dilaton solutions of $AdS_{+}$, the parabolic operators of $AdS_0$ and the hyperbolic operators the dilaton solutions of $AdS_{-}$.  Again, this is a coordinate independent result but may be verified for the harmonic oscillator metric and expectation values.  \\ \\
Note furthermore that $M$ will no longer be constant if $\eta \rightarrow \eta +c$.  This implies that the potential $M$ contains information of both the fact that the dilaton solves the field equation ($\nabla_\nu M = 0$) and a boundary condition for the dilaton.  The fact that the on-shell field equations (with $V_{\mu\nu} = 0$) are satisfied tells us that the solution must be a linear combination of the symmetry generators, $u H + v D + w K$.  The value for the potential then tells us the value of $d^2$ that this linear combination must satisfy i.e. the operator must be of the form $u H \pm \frac{1}{2}\sqrt{uw - d^2}D + wK$.  This specifies the $su(1,1)$ operator up to a unitary transformation.   In order to select the expectation value of a specific operator with this value of $d^2$ we must give boundary conditions.   \\ \\
The dictionary for the $SU(1,1)$ symmetry generators can be summarised as follows
  \begin{center}
  \begin{tabular}{ | c | c | }
    \hline
    SU(1,1) model of CQM & JT-model of dilaton gravity  \\ \hline \hline
		Quantum states related by phase (ray) & Point of the manifold of states \\ \hline
		Symmetries of states & Isometries of the metric \\ \hline
    Hamiltonian & Choice of time coordinate  \\ \hline
    Representation label, $k$ & Scalar curvature $R = -\frac{4}{k}$ \\  \hline
		Operator expectation value & Solution of Dilaton e.o.m. \\ \hline
		Operator $\in su(1,1)$ & $V = 0$ \ i.e. \ $\nabla_\mu M = 0 \ , \ g^{\alpha\beta}V_{\alpha\beta} = 0$ \\ \hline
		$d^2$ & Potential, $M$ \\ \hline
		Specific unitary operator from $d^2$ family & Boundary condition on the dilaton \\ \hline
  \end{tabular}
\end{center}
The first four entries of this dictionary are simply due to how we construct the metric from a family of quantum states.  The fifth is a choice that we made - the field equation solutions are precisely the expectation values of operators.  The last three entries tell us what matter content (and which boundary conditions) must be put into the dilaton gravity model in order to source a desired generator of symmetry.  \\ \\
We have already begun to extend the work of \cite{Cadoni1}-\cite{CadoniCQM} by firstly providing the explicit dictionary between dilaton gravity and quantum mechanics.  This allows us direct access to the dictionary which we have now started to fill out.  The systematic nature of our construction allows us to go beyond just the generators of symmetry and we can give a dual interpretation to the dilatons in the case where the matter content is non-zero.  This will, of course, be the expectation values of operators that are not generators of symmetry.

\subsection{Extending Beyond the Generators of Symmetry of $SU(1,1)$}

\label{CasimirSection}

We will now extend the $SU(1,1)$ dictionary by investigating the explicit energy momentum tensors that source the expectation values of operators from the enveloping algebra of $su(1,1)$.  Before that we make a number of remarks that will prove useful.  \\ \\
The Killing vectors close on the same commutation relations as their corresponding operators and, indeed, are nothing other than a differential operator representation of these operators.  These differential operators act on functions of the coordinates and in particular they act on the expectation values of operators.  It can be verified that this is a representation since, by using (\ref{KillComm}), it can be shown that
\begin{equation}
[-i \chi_{A_1},  -i \chi_{A_2} ] \langle s| A | s\rangle = -i(-i \chi_{i[A_1, A_2]}) \langle s| A | s\rangle
\end{equation}
for symmetry generators $A_1$ and $A_2$ and an arbitrary operator $A$.  As an example, the Killing vectors of $su(1,1)$ satisfy the commutation relations
\begin{eqnarray}
\left[-i\chi_D, -i \chi_H\right] & = & -i (-i \chi_H) \nonumber \\
 \left[-i\chi_D, -i \chi_K\right] & =& i (-i \chi_K) \\ 
\left[-i\chi_H, -i \chi_K\right] & = & -i (-i \chi_{(-2D)}) = 2i (-i\chi_{D}) \nonumber 
\end{eqnarray}
which may be compared to (\ref{kIrrep}). The Laplacian is a differential operator, quadratic in derivative, that commutes with all the Killing vectors and must thus be proportional to a differential operator representation of the Casimir.  In our present case this is
\begin{equation}
\nabla^2 = -\frac{R}{2}\left( \chi_D \chi_D - \frac{1}{2}( \chi_H \chi_K - \chi_K \chi_H)\right)   \label{LapCasimir}
\end{equation}
which is proportional to the $su(1,1)$ Casimir.  \\ \\
We are interested in finding the trace of the energy momentum tensor and the potential (\ref{MassDefine}).  The trace of the energy momentum tensor features when one applies the Laplace operator to the dilaton (\ref{DilatonEq}).  Thus the properties of the Casimir can prove useful in determining the appropriate trace of the energy momentum tensor.  \\ \\
For determining the appropriate value of the potential we do not currently possess an algorithm.  The appropriate value for the potential will thus have to be calculated (for operators that are not generators of symmetry) by direct substitution. The trace of the energy momentum tensor lends itself to symmetry arguments and may be calculated as follows. \\ \\
A basis for the enveloping algebra of $su(1,1)$ are all hermitian operators that are the product of $n$ $su(1,1)$ elements $H, D, K$, where $n$ runs from $1$ upwards.  For instance, when $n=2$ we have $9$ operators which we may order as
\begin{eqnarray}
H^2  &     &   \nonumber \\
 HD + DH & \ \ \ \ \  i(HD - DH) & \nonumber \\
 D^2 + \frac{1}{4} (HK + KH)& \ \ \ \ \ i(HK - K H) & \ \ \ \frac{1}{2}(HK + KH) - D^2 \label{TensOps2}   \\
KD + DK & \ \ \ \ \ i(KD - DK) & \nonumber \\
K^2 & & \nonumber 
\end{eqnarray}
Note that the operators in (\ref{TensOps2}) are tabled according to scaling dimension i.e.
\begin{equation}
[D, A] = i q A \ \ \ i.e. \ \ \ \chi_D \langle A \rangle = q \langle A \rangle
\end{equation}
where $q$ is the scaling dimension.  All operators on the same row have the same scaling dimension.  Note further that one may go up in a column (up to normalisation) by a commutator with $H$ and down by a commutator with $K$.  This implies that we only need the topmost (or equivalently the bottommost) operator of every column.  We label the operators of (\ref{TensOps2}) by $L_{j, q}^n$ and find for the topmost operator 
\begin{equation}
L_{j=2, -2}^2 = H^2 \ \ \ ; \ \ \ L_{j=1, -1}^2 = i(HD - DH) \ \ \ ; \ \ \ L_{j=0, 0}^2 = \frac{1}{2}(HK + KH) - D^2
\end{equation}
and define the rest up to normalisation by $L_{j, q+1}^n \propto [L_{j, q}^n, K]$ and $L_{j, q-1}^n \propto [L_{j, q}^n, H]$.  We have defined the operators in such a way that the value for $|q|$ can never exceed $j$.  In terms of the expectation values of $L_{j, q-1}$ i.e. $\langle L_{j, q-1} \rangle$ this implies that $\chi_K \langle L_{j, q} \rangle \propto \langle L_{j, q+1} \rangle$ and $\chi_K \langle L_{j, q} \rangle \propto \langle L_{j, q-1} \rangle$  \\ \\
Another instructive example is $n=3$.  For these we find that the topmost operators are
\begin{eqnarray}
& & L^3_{j=3, -3} = H^3  \nonumber \\
& & \left._{(1)}L^3_{j=2, -2} \right. = i (H^2 D - D H^2) \ \ \ ; \ \ \ \left._{(2)}L^3_{j=2, -2} = H D H - \frac{1}{2} (H^2 D + D H^2) \right. \nonumber \\ 
& &  \left._{(1)}L^3_{j=1, -1} \right. = H D^2 + D^2 H -\frac{1}{2}( H^2 K + 2HKH + KH^2) \nonumber \\ & & \left._{(2)}L^3_{j=1, -1} \right. = D^2 H + H D^2 - D H D - H K H \nonumber \\ & & \left._{(3)}L^3_{j=1, -1} \right. = i(H D^2 - D^2 H - \frac{1}{2}H^2 K + \frac{1}{2} K H^2) \nonumber \\
& & \left._{(1)}L^3_{j=1, -1} \right. = i(K D H - H D K + 2 H K D -2 D K H).  
\end{eqnarray}
Note that we had to include another index.  The set of operators $\left._{(\alpha)}L^n_{j, q} \right.$ constitute a basis for the enveloping algebra of $su(1,1)$.  The operators can be simplified furtherby using the commutation relations and indeed one finds that, for instance $\left._{(1)}L^3_{j=2, -2} \right. \propto H^2$ and $\left._{(0)}L^3_{j=2, -2} \right. = 0$.  For the purpose of our further discussion we will only consider the operators in the chain where $j$ assumes the maximal value and we define
\begin{equation}
L^n_{q}  = L^n_{j=n, q} \ \ \ ; \ \ \ \eta^n_q = \langle L^n_{q}  \rangle 
\end{equation}
where we essentially drop the index $j$.  This classification scheme for the expectation values of operators amounts to the construction of spherical tensor operators.  The expectation values of the operators are eigenfunctions of $\chi_D$ (associated with $J_z$ while the role played by the vector fields $\chi_H$ and $\chi_K$ are that of ladder operators (associated with $J_-$ and $J_+$).  \\ \\
The expectation values of the operators will thus be simultaneous eigenfunctions of the Laplacian (\ref{LapCasimir}) (associated with $J^2$) and we find that
\begin{equation}
\nabla^2 \eta^n_q = -\frac{R n(n+1)}{2} \eta_q^n.   \label{weightEq}
\end{equation} 
The relation (\ref{weightEq}) is most easily verified using the free particle metric (\ref{freeMet}) and the expectation value
\begin{equation}
\langle \beta | H^n | \beta \rangle \propto \frac{1}{\beta^n}.  
\end{equation}
The two derivatives with respect to $\beta$ yield the factor $n(n+1)$ and the multiplication with the inverse metric restores the appropriate order in $\beta$ and yields the factor $-\frac{R}{2}$.  Note also that the normalisation of the states used is irrelevant and, as per usual, the result is coordinate independent.  \\ \\
The form of the appropriate potential is more complicated than the trace of the energy momentum tensor.  For this reason we provide only the appropriate potential to source the expectation values of $L_{j=n, -n}^n$ and $L_{j=n, n}^n$.  The appropriate potential to source the other expectation values can be achieved by applying the ladder operators (\ref{KillComm}).  For these expectation values we find
\begin{equation}
M^n_{\pm n}[\widetilde{V}_{\mu\nu}] = -\frac{R(n+1)(n-1)}{4}(\eta^n_{\pm n}  )^2.   \label{orderEq}
\end{equation}
The expressions (\ref{weightEq}) and (\ref{orderEq}), along with boundary conditions for the dilaton, now allow one to calculate a desired expectation value purely on the gravitational side of the duality.  An arbitrary operator from the $su(1,1)$ enveloping algebra is broken up in terms of the basis $\left._{(\alpha)}L^n_{j, q} \right.$ and the appropriate trace and potential put together from (\ref{weightEq}) and (\ref{orderEq}).  The energy momentum tensor is reconstructed from these and the field equations solved.  The dilaton solution of these field equations is then the desired expectation value.  
\\ \\
The dictionary for the $su(1,1)$ enveloping algebra operators reads
  \begin{center}
  \begin{tabular}{ | c | c | }
    \hline
    SU(1,1) model of CQM & JT-model of dilaton gravity  \\ \hline \hline
		Quantum states related by phase (ray) & Point of the manifold of states \\ \hline
		Symmetries of states & Isometries of the metric \\ \hline
    Hamiltonian & Choice of time coordinate  \\ \hline
    Representation label, $k$ & Scalar curvature $R = -\frac{4}{k}$ \\  \hline
		Operator expectation value & Solution of Dilaton e.o.m. \\ \hline
		Operator $L_q^{n}$ & $g^{\alpha\beta}V_{\alpha\beta} = -\frac{R n(n+1)}{2} \eta_q^n$ \\ \hline
		Operator $L_{\pm n}^n$ & Potential, $M = -\frac{R(n+1)(n-1)}{4}(  \eta^n_{\pm n}  )^2$ \\ \hline
		Specific unitary operator from solution set & Boundary condition on the dilaton \\ \hline
  \end{tabular}.  
\end{center}

\subsection{Extending to the Conformal Transformations}

\label{CadoniConf}

Another interesting set of transformation are the conformal transformations.  For two dimensions we derived the equation of motion (\ref{ConfField}) for the generators of conformal transformations of the metric.  Note that this is the trace of the dilaton field equations i.e. eq. (\ref{Two}) of the set (\ref{One}), (\ref{Two}), (\ref{Three}).  We are thus requiring that the fields solve only one of the field equations, not the full set.  This is known as an unconstrained variation (as opposed to a constrained variation where the fields need to solve the full set of field equations).  

\subsubsection{The Unconstrained Solutions}

We are now interested in finding fields that solve (\ref{Two}).  \\ \\
For simplicity of notation we now consider the free particle model and work with complex coordinates where $z = t+i\beta$ and $\overline{z} = t-i\beta$.  As before the result is coordinate independent and .  The solutions of (\ref{Two}) are any linear combination of the functions
\begin{equation}
\Phi_n \equiv \langle z| i^{1 - n}V_n |z \rangle = -i k \frac{(n+1)z^{1-n} + (1-n)z^{-n}\overline{z}}{z - \overline{z}} \label{ConfDilatons}
\end{equation}
as well as their complex conjugates.  These are the expectation values of the operators
\begin{equation}
V_n  = (V_0 - kn) \frac{\Gamma(V_0 +1 - k - n)}{\Gamma(V_0 + 1 - k )} V_1^n \ \ \ ; \ \ \ V_{-n} = (V_0 + kn) \frac{\Gamma(V_0 + k )}{\Gamma(V_0 + k + n )} V_{-1}^n \ \ \ ; \ \ \ n \geq 0.  \label{FairlieViras}
\end{equation}
The factor of $i^{1 - n}$ in (\ref{ConfDilatons}) ensures that the hermitian part of the operator in the expectation value is given by $\frac{1}{2}(A + A^\dag)$\footnote{This is most easily verified for $V_{-1}, V_{0}$ and $V_1$}.  The operator function $\Gamma(A + 1) \equiv A\ \Gamma(A)$ is the gamma function and we have defined $V_1 = H$, $V_0 = i D$ and $V_{-1} = K$.  The functions (\ref{ConfDilatons}) can be derived from the unnormalised overlap $(z|z) = (i z - i \overline{z})^{-2k}$ by acting on it with the differential operator representation
\begin{eqnarray}
V_n & = & i(1-n)k z^{-n} + iz^{1-n} \partial_z \nonumber \\
\overline{V}_n & = & -i(1-n) k \overline{z}^{-n} -i\overline{z}^{1-n}\partial_{\overline{z}} \label{HoloViras2}
\end{eqnarray} 
It can be verified that the operators $V_n$ close on the centerless Virasoro algebra (\ref{Viras}) (i.e. $c=0$) and indeed, the operators (\ref{FairlieViras}) are a representation of the centerless Virasoro algebra in terms of $su(1,1)$ algebra elements \cite{Fairlie}. \\ \\
Near the $\beta \rightarrow 0$ boundary we may expand the expectation values (\ref{ConfDilatons}) to find
\begin{equation}
\Phi_n = -\frac{k t^{1-n}}{\beta} - \frac{1}{2} k (n^2 - n) t^{-1-n}\beta + \frac{i}{3} k (n^3 - n)t^{-2-n}\beta^2 +  O(\beta^3).  \label{ConfSField}
\end{equation}
The operators (\ref{FairlieViras}) are not all hermitian and we need to calculate the vector fields with a bit more care than previously.  We write these in terms of the expectation values of hermitian and anti-hermitian operators (\ref{ConfDilatons}).  This can be done by 
\begin{eqnarray}
\Phi_{1, n} & = & \frac{1}{2}(\Phi_n + \Phi_n^*) \\ 
\Phi_{2, n} & = & \frac{1}{2}(\Phi^*_n - \Phi_n).  
\end{eqnarray}
The vector fields for hermitian and anti-hermitian operators are calculated as in eq. (\ref{OperatorVector}) and we thus have
\begin{eqnarray}
\chi_{1, n} & \equiv & -\frac{1}{2}\sigma^{\mu\nu}\nabla_\nu (\Phi_{1,n}) \partial_\mu \nonumber \\
& = & \ \  \left(-(n-1)t^{-n}\beta - \frac{1}{2}(n^3 - n)\beta^3 + O(\beta^4) \right) \partial_\beta \nonumber \\
&   & + \left(t^{1-n} - \frac{1}{2}(n^2 - n)t^{-1-n}\beta^2 + O(\beta^4) \right)\partial_t   \label{FlowSField1} \\  
\chi_{2, n} & \equiv & -\frac{1}{2} g^{\mu\nu}\nabla_\nu (\Phi_{2,n}) \partial_\mu \nonumber \\
& = & \ \ \left( \frac{2}{3}(n^3 - n)t^{-2-n}\beta^3 + O(\beta^4) \right) \partial_\beta \nonumber \\ 
&   & + \left(-\frac{1}{3}(n^3 - n^2)(n + 2) t^{-3-n}\beta^4 + O(\beta^5 ) \right) \partial_t \label{FlowSField2}
\end{eqnarray}
The fields (\ref{ConfSField}) and vector fields (\ref{FlowSField1}), (\ref{FlowSField2}) clearly form a power series expansion.  The general form for the expectation values of the hermitian operators is any linear combination of (\ref{ConfSField}) and is thus given by
\begin{equation}
\sum_n a_n (\Phi_n + \Phi_n^*) + i(\Phi_n - \Phi_n^*) =  \frac{\rho(t)}{\beta} + \frac{1}{2} \ddot{\rho}(t) \beta + O(\beta^2)  \label{DilSol}
\end{equation}
where $\rho(t) = \sum_n k a_n t^{1-n}$, see eq. (\ref{ConfSField}).  Similarly the most general vector fields are given by
\begin{equation}
\sum_{n} b_n \chi_{1,n} = [ \dot{\epsilon}(t) \beta + O(\beta^3) ] \partial_\beta + [ \epsilon(t) - \frac{1}{2}\ddot{\epsilon}(t) \beta^2 + O(\beta^3) ] \partial_t.  \label{VirasFlows}
\end{equation}
and
\begin{equation}
\sum_{n} c_n \chi_{2,n}\partial_\mu = [\alpha(t)\beta^3 + O(\beta^4)] \partial_\beta + [\frac{1}{2}\dot{\alpha}(t)\beta^4 + O(\beta^5) ]\partial_t  \label{MetrFlows}
\end{equation}
where $\epsilon(t) = \sum_n b_n t^{1-n}$ and $\alpha(t) = \sum_n \frac{2c_n}{3} (n^3 -n)t^{-2-n}$, see eqs. (\ref{FlowSField1}), (\ref{FlowSField2}).  The $a_n$'s, $b_n$'s and $c_n$'s are arbitrary expansion coefficients.  \\ \\
Comparing these with the transformations and field solutions of (\ref{AsympTransform}) and (\ref{AsympDil}) reveal that the vector fields comprise the $\epsilon(t)$ and $\alpha_\beta(t)$ part of the transformations.  These are precisely the transformations required to perform the analysis of section \ref{AsympCentral}.  This is the first of two ingredients required to perform that analysis.  What remains for us is to find the second ingredient namely the energy momentum tensor on the gravity side. 

\subsubsection{The Asymptotic (Outer Boundary) Contribution to the Central Charge}
\label{OurCCharge}

There is one thing, however, we do need to specify further.  Note that between the field solutions (\ref{DilSol}) and vector fields (\ref{VirasFlows}), (\ref{MetrFlows}) we have three functions of time.  In the analysis of section \ref{AsympCentral} there are only two independent functions of time that prove relevant.  \\ \\
This is due to the fact that, in that analysis, an additional constraint is imposed.  Here we have only required that the fields satisfy (\ref{Two}) while the analysis of section \ref{AsympCentral} also requires them to satisfy (\ref{Three}).  At first glance this seems fundamentally different to what we have done.  In the analysis of section \ref{AsympCentral} the constrained field equations are imposed while our very starting point was that the conformal transformations satisfy the unconstrained field equations\footnote{We note that this imposing of the constrained field equations is discussed as a weakness in \cite{NBRef}.}.  Up to this ``missing" constraint we recover exactly the transformations used in the analysis.  It thus seems that one may recover these transformations in at least two ways - either allow the metric and dilaton to fluctuate asymptotically and impose the constrained variation or work with an identically $AdS_2$ metric but impose only the unconstrained variation.  \\ \\
What to do then about the extra function of time that we find in our analysis?  The main insight is that in requiring that the fields solve the equation (\ref{Two}) we are requiring that the trace of the energy momentum tensor is fixed.  We do not place any restriction on the potential $M$ in the bulk.  What we may specify is a boundary condition for the potential.  In \cite{Cadoni2} two possible boundary conditions are discussed and we will discuss the effects of both.  \\ \\
The first boundary condition is the one that will reproduce the analysis of section \ref{AsympCentral} namely $ \lim_{\beta \rightarrow 0} \nabla M  = 0$.  This means that, though we do not require (\ref{Three}) to hold in the bulk we do require it to hold near the boundary.  Since the analysis is carried out near the boundary we thus have the same allowed transformations and constraints as in section \ref{AsympCentral}.  \\ \\
Importantly we derived these transformations without changing the metric from $AdS_2$ to asymptotically $AdS_2$.  We are thus working in an $AdS_2$ background.  For the $AdS_2$ metric the boundary term $J$ is simply given by \cite{Gegenberg2}
\begin{equation}
J \propto M.   \label{ccMass}
\end{equation}
By expanding around the zero mass solution $\rho = 1 - \overline{\rho}(t)$ we find
\begin{equation}
M \propto \dot{\rho}(t)^2 - 2 \rho (t) \ddot{\rho}(t) = -2 \ddot{\overline{\rho}}(t) + O(\overline{\rho}^2).  
\end{equation}
The potential is thus of the appropriate form to be interpreted as an energy momentum tensor \cite{NBRef}.  As mentioned in section \ref{NormSec} the precise normalisation still needs to be clarified.  We have merely recast the arguments of section \ref{AsympCentral} in a different set of coordinates so that we should get exactly the same contribution to the central charge from this calculation. 

\subsubsection{The Inner Boundary Contribution to the Central Charge}

The insight that the implication of the unconstrained variation is a fluctuating potential term is of great value to clarify why the contribution of the inner boundary needs to be taken into account along with the contribution of the outer boundary.  We can simply argue this in terms of the set of transformations that we allow.  \\ \\
The key is the boundary condition.  By imposing the condition $\lim_{\beta \rightarrow 0} \nabla M  = 0$ we are not excluding the possibility that a transformation may be included that changes the potential from positive (indicative of a black hole) to zero (which is indicative of the vacuum).  We know from our discussion of conformal quantum mechanics (\ref{CQMAction}) that such transformations do form part of the conformal group.  Indeed, it is the transformation that maps the scale-invariant Hamiltonian to the harmonic oscillator.  \\ \\
This situation is clearly unacceptable since the vacuum solutions and black hole solutions are mixed.  We must thus factor out the transformation that maps the black hole solutions onto the vacuum solutions.  This is the step (\ref{FiniteTrans}).  \\ \\
We could have achieved the same feat if we started off by working with the second boundary condition proposed in \cite{Cadoni2}, namely that $M = \textnormal{fixed} > 0$.  Indeed, in \cite{Cadoni2} they work with this boundary condition and find the correct expression for the central charge without taking the inner and outer boundary into account.  The authors argue that the gravitational content with these two boundary conditions are different.  For our purposes this is not a significant issue - we are merely interested in understanding how the CQM central charge may be described in the dual.  Indeed, the unconstrained variation does not yield the metric as part of the field equation solutions (\ref{One}) and is thus not, in a strict sense, a theory of gravity.  Our aim is to show that the central charge can be incorporated in a dual description - a feat that can be achieved using both boundary conditions.  
 \\ \\
We conclude by again emphasising that what we have presented in this section are not new results in the $AdS_2/CFT_1$ context but rather show that the construction we have made allow these existing results to be derived in a simple and elegant way.  The explicit realisation of the dictionary also serves to add insights to the work of \cite{Cadoni1}-\cite{CadoniCQM} and allows for future generalisations of these works in a very natural way.  

\subsubsection{Massive Scalar Field Action}

\label{ScalarFieldSection}

We will now show how the unconstrained variation can be contained neatly as the variation of an action.  \\ \\ 
We showed in section \ref{JTSection} that the constrained variation can be achieved by varying the dilaton action (\ref{JTAction}) with respect to the fields and the inverse metric.  \\ \\
Consider first the quadratic dilaton action
\begin{equation}
S_{\eta^2}[\eta, g_{\mu\nu}] = \int d^2 x \sqrt{g} ((\nabla \eta)^2 + \frac{1}{2}R \eta^2 - \frac{3}{2}R_s \eta^2 + L_M)  \label{eta2Action}
\end{equation}
where $L_M$ is some minimally coupled matter (independent of the metric).  It can be chosen so that the constrained field equations of (\ref{eta2Action}) are equivalent to those of the JT-model.  We will only be interested in the unconstrained variation so that $L_m$ plays no role.  \\ \\
When we consider the metric as fixed as $AdS_2$ with scalar curvature $R = -R_s$ in (\ref{eta2Action}) we find
\begin{equation}
S_{\eta^2}[\eta] = \int d^2 x \sqrt{g} ((\nabla \eta)^2 - R_s \eta^2 + L_M).   
\end{equation}
This can be identified as the scalar field action.  The variation with respect to the dilaton now yields
\begin{equation}
\nabla^2 \eta =  - 2 R_s \eta  \label{eta2DilEq}
\end{equation}
which is precisely the unconstrained variation.  For the action quadratic in the fields we can thus give a very clear physical distinction between the constrained and unconstrained variation.  The constrained variation comes from a model of dilaton gravity where we may vary the action with respect to the metric and dilaton.  The unconstrained variation is where we fix the background metric and only vary with respect to the field.  The solutions of this unconstrained variation yields the expectation values of the Virasoro elements related to the conformal transformations.  \\ \\
We note that the conformal symmetry of the massive scalar field in $AdS_2$ was studied in \cite{Cruz}.  

\subsubsection{A More Direct Approach from CQM}

\label{DirectCQM}

In \cite{CadoniCQM} it is pointed out that the field equations of the asymptotic metric and dilaton corresponds to that of a scale-invariant quantum mechanical model coupled to a time-dependent quadratic source.  This is nothing other than the model of conformal quantum mechanics.  In this section we will tackle our central charge analysis from another angle that will make the connection between conformal quantum mechanics and dilaton gravity even more explicit. \\ \\
To find the quantum states from the CQM action we quantised the fields from which we found the generator of time translations.  The time-evolved states were then defined accordingly.  After regularising the time-evolved states we may construct the metric from (\ref{Prov2}).  The symmetries of the states are imprinted as isometries of the metric.  The conformal quantum mechanics action possesses full conformal symmetry if one allows the time-dependent coupling to change.  On the level of the time-evolved states this is equivalent to the statement that the Hamiltonian should be allowed the freedom to change.  \\ \\
Thus consider the quantum states
\begin{equation}
|\beta, t) \equiv U(t) e^{-\beta H} |x = 0) \ \ \ ; \ \ \ \partial_t U(t) = i (H + \gamma(t) K ) U(t).    \label{boundaryDeform}
\end{equation}
The symmetries of these states are generated by the $su(1,1)$ elements $H, D$ and $K$.  The form of these states can be identified as that of the sourced quantum states we defined in (\ref{States}).  \\ \\
The metric may be calculated from (\ref{Prov2}) to be
\begin{equation}
ds^2 = \frac{k}{2 \beta^2} d\beta^2 + \frac{k}{2 \beta^2} (1 - \gamma(t) \beta^2)^2 dt^2 \label{CQMSourceMet}
\end{equation}
which has constant scalar curvature $R = -\frac{4}{k}$ and is another parametrisation of $AdS_2$.  The dynamics of the quantum system is imprinted very clearly on the metric.  The manifold is still K\"ahler, but now written in a form where the transformation to the holomorphic and anti-holomorphic coordinates are not clear.  The dictionary we have developed in this chapter thus still holds.  The form of (\ref{CQMSourceMet}) is reminiscent of the form of the asymptotic $AdS_2$ metrics (\ref{AsympMetric}) but with more limited freedom.  This freedom will be all we require, however.  \\ \\
The general solution of the field equation is
\begin{equation}
\eta(t, \beta) = \frac{\eta_1(t)}{\beta} + \eta_2(t) \beta.  
\end{equation}
The unconstrained variation yields the constraint
\begin{equation}
\ddot{\eta}_1(t) = -2 \gamma(t) \eta_1(t) + 2 \eta_2(t) \label{CQMField2}
\end{equation}
while the constrained variation yields the additional constraint
\begin{equation}
0 = \gamma(t) \dot{\eta}_1(t) + \dot{\eta}_2(t).    \label{CQMField1}
\end{equation}
The potential, $M$, if the dilaton solves the constrained field equations is given by
\begin{equation}
M = \frac{\dot{\eta}_1(t)^2 - 4 \eta_1(t) \eta_2(t)}{k}
\end{equation}
which is constant.  The constrained field equations yield the expectation values of the three generators of symmetry as solutions.  
\\ \\
The potential, when only the unconstrained field equation (\ref{CQMField2}) is imposed, is given by (in an expansion around $\beta = 0$)
\begin{equation}
M = \frac{\dot{\eta}_1(t)^2 - 4 \eta_1(t) \eta_2(t)}{k} + \frac{2(\gamma(t)\dot{\eta}_1(t) + \dot{\eta}_2(t))\dot{\eta}_1(t) \beta^2}{k} + O(\beta^3).   \label{MCorrect}
\end{equation}
In (\ref{MCorrect}) the difference between the constrained and unconstrained variation is clear.  The constrained variation yields a constant value for $M$ while the unconstrained variation allows for asymptotic correction terms.  In \cite{CadoniCQM} the authors identify the asymptotic corrections with the kinetic energy of the conformal quantum mechanical model.  We showed in section (\ref{CQMSection}), our discussion of conformal quantum mechanics, that the time-dependent coupling (after conformal transformation) is due to the kinetic term.   \\ \\
The transformation properties of the dilaton gravity potential $M$ can thus clearly be used to extract the central charge.  The interpretation of a changing coupling term, $\gamma(t)$ in $CQM$ is thus that the potential of dilaton gravity, related to the black hole mass, is allowed to fluctuate asymptotically.  \\ \\
We thus present the following additions to the dictionary for conformal quantum mechanics and dilaton gravity
  \begin{center}
  \begin{tabular}{ | c | c | }
    \hline
    Conformal Quantum Mechanics & JT-model of dilaton gravity  \\ \hline \hline
		Local Symmetries & Conformal transformations of the metric  \\ \hline 
		Expectation values of & Dilaton solutions \\
						conformal generators				 &	in unconstraind variation							\\ \hline
		Changing $\gamma(t)$ & Asymptotic fluctuations in $M$ \\ \hline
  \end{tabular}
\end{center}

\subsection{The Duals of the $SU(2)$ Quantum Models}

We next examine the $SU(2)$ Hamiltonian systems.   As we showed (\ref{SU2Metric}) we find that the scalar curvature is also constant as it was for the $SU(1,1)$ models.  The only adjustment we need to make is the sign of the scalar curvature so that we now consider a $dS_2$ background as opposed to an $AdS_2$ background.  Two dimensional de Sitter geometries have been considered in the context of dualities before, most notably in the $dS_2/CFT_1$ correspondence \cite{CadonidS2}, \cite{dS2Second}.  The field equation (\ref{ConfField}) holds for the conformal transformations of any two-dimensional metric so that the conformal transformations of $dS_2$ can be handled on similar footing to $AdS_2$.   \\ \\
The $su(2)$ and $su(1,1)$ algebras are related by an appropriate complexification of the generators.  It should thus be possible to perform an analysis that is very similar to the $AdS_2/CFT_1$ analysis we carried out in the previous section (which may be compared to \cite{CadonidS2}, \cite{dS2Second}).  This will not be done here and is a possible avenue for future study.  \\ \\
Our interest here is simply to recover the expectation values of operators from the $SU(2)$ enveloping algebra by means of a dual description and we will show now that this can be done.  The $SU(2)$ coherent states are given in (\ref{SU2Coh}).  The dilaton model 
\begin{equation}
S_{SU(2)} = \int dz d\overline{z}\sqrt{g} (R - \frac{2}{j})\eta   \label{ds2Action}
\end{equation}
yields, up to a coordinate transformation, the metric (\ref{SU2Metric}) as the on-shell solution.  The dilatons that solve the on-shell field equation are the expectation values of the symmetry generators
\begin{eqnarray}
\langle z| J_z | z \rangle & = & j \frac{(z \overline z - 1)}{z \overline{z} + 1}, \nonumber \\
\langle z| \frac{1}{2}(J_{+} + J_{-}) | z \rangle & = & j\frac{z + \overline{z}}{1 + z \overline{z}}, \nonumber \\
\langle z| \frac{i}{2}(J_{+} - J_{-}) | z \rangle & = & i j\frac{\overline{z} - z}{1 + z \overline{z}}, \label{su2symbol}
\end{eqnarray}
with respect to the $SU(2)$ coherent states.  The trace of the energy momentum tensor is $g^{\alpha\beta} V_{\alpha\beta} = -R = -\frac{2}{j}$.  Again these expectation values are related to the Killing vectors (\ref{su2Killing}) by the symplectic structure (\ref{OperatorVector}).  The potential associated with the operators $u J_z + \frac{v}{2}(J_{+} + J_{-}) + 
\frac{i w}{2}(J_{+} - J_{-})$ is
\begin{equation}
M = \frac{j}{2}(u^2 + v^2 + w^2).  \label{ds2Potential}
\end{equation}
This is constant, as it should be for a generator of symmetry.  Furthermore, it is always positive, unlike for the $SU(1,1)$ duals, so that we do not get geometries that are different globally.  As before operators that share the same mass are related by a unitary transformation.  \\ \\
It is worthwhile to reflect on the difference between this case and that found in \cite{CadonidS2} where three globally different solutions are in fact found.  We suspect that this is due to considering the representation of $SU(2)$ related to the infinite dimensional $SU(1,1)$ representation by a complexification of operators.  The consequence of this is that additional factors of $i$ are introduced.  If we allow the coefficient $u, v$ and $w$ to assume complex values then the sign of $M$ can be altered.  The dilaton solutions are then rather the expectation values of non-hermitian operators.  The precise link between these two analyses warrant further investigation. \\ \\ 
For our current purposes we only wish to calculate the expectation values of $su(2)$ algebra elements and for this the use of (\ref{ds2Potential}) is sufficient.  One can proceed to find the appropriate energy momentum tensors for the enveloping algebra from this point in similar fashion to before.  The Laplacian is, as before, a representation of the Casimir and we may classify the expectation values of $su(2)$ eveloping algebra operators in terms of their eigenvalues with respect to the differential operator representation of the Casimir and $J_z$.  The procedure is analogous to that performed in section \ref{CasimirSection}.  One finds that
\begin{eqnarray}
\nabla^2 \langle z | (J_{+})^n | z \rangle & = & -\frac{R n (n+1)}{2}\langle z | (J_{+})^n | z \rangle \\
M_{(J_{+})^n} & = & -\frac{R(n+1)(n-1)}{4} (\langle z | (J_{+})^n | z \rangle)^2
\end{eqnarray} 
This allows one to calculate the expectation value of any element of the eveloping algebra of $su(2)$ in the dual dilaton gravity description.  As before this is done by including the appropriate energy momentum tensor.   

\section{The dual of the Glauber Coherent States}

The metric constructed from the Glauber coherent states (\ref{GlauberMet}) is flat.  This presents a very unique situation since the trick we employed to get rid of the constant $c$ in (\ref{EOMFinal}) is no longer applicable.  Indeed, it can be verified that the expectation values 
\begin{equation}
\langle z | c_1 a^\dag + c_2 a | z\rangle = c_1 \overline{z} + c_2 z \label{CAOp}
\end{equation}
satisfy the equations of the previous sections $\nabla_\mu \nabla_\nu \eta = 0$ with $R = 0$ and that the mass is given by the product $c_1 c_2$. However, the expectation value of the third generator of symmetry
\begin{equation}
\langle z | a^\dag a | z\rangle = z \overline{z} \label{numberEx}
\end{equation}  
rather satisfies the equation
\begin{equation}
\nabla_\mu \nabla_\nu \eta = g_{\mu\nu}
\end{equation}
so that $c = 1$ in (\ref{EOMFinal}).  The mass associated with this equation and solution is zero.  It is thus not possible to capture all the generators of symmetry into a single action and distinguish them up to the choice of mass.  For the operators (\ref{CAOp}) the action
\begin{equation}
S = \int dz d\overline{z} \sqrt{g} R \eta
\end{equation}
works but for the number operator expectation value (\ref{numberEx}) we need a different action of the form
\begin{equation}
S = \int dz d\overline{z} \sqrt{g} (R\eta + 2\eta).  
\end{equation}
Due to the fact that we cannot eliminate the constant $\lambda$ in (\ref{EOMFinal}) we cannot find an action of which the field equations are solved by the expectation values of the algebra elements simultaneously.  The expectation values of the algebra elements is thus better served by considering it as a large $j$ limit of the $su(2)$ algebra elements.  \\ \\
This indicates that for a neat and sensible dictionary a non-zero curvature of the manifold is essential.

\section{Summary}

In this chapter we investigated $2d$ gravity duals.  We managed to develop a systematic dictionary between quantum mechanical models and a JT-model of dilaton gravity.  We investigated the details of this dictionary for the Glauber coherent states, the $SU(2)$ Hamiltonians and, in particular, the $SU(1,1)$ quantum models.  We were able to show how numerous existing results in the $AdS_2/CFT_1$ literature come about in a very natural way from our construction.  We were also able to expand upon some of their results and identified several avenues of possible future investigation.  \\ \\
These results comprise the core results of this thesis.  They should not be viewed as a calculational procedure but rather a proof of concept.  It shows that quantum mechanical systems can be repackaged as gravity duals, at least in two dimensions, and that this can be done systematically.  The number of existing results from the $AdS_2/CFT_1$ literature that are reproduced in a natural way from our construction seems to indicate that the construction repackages these quantum mechanical systems in an appropriate way.  Furthermore, the systematic construction gives us direct and full access to the quantum mechanics / dilaton gravity dictionary.  This provides a significant level of clarity to the $AdS_2/CFT_1$ correspondence.  \\ \\
In the remaining chapters we will investigate the higher dimensional manifolds, the difficulties that arise when one wishes to build gravitational duals and proposed solutions to these difficulties.  We will again see existing results from the literature come about in a very natural way but our investigations will not be as well-developed as they have been in this chapter.

\chapter{A Look at Higher Dimensional Models}

\label{HighDim}

Our focus thus far has been on the two-dimensional manifolds of states.  The reason for this is clear - it is the simplest example of our construction.   Our focus has, in particular, been on time-evolved states where the time-coordinate is complexified.  Existing results of the $AdS_2/CFT_1$ correspondence were shown to follow from the construction in a natural way and we were also able to extend these results.  We further showed that more general dual gravitational models, that do not have an $AdS_2$ geometry, can be constructed in a systematic way, starting from an appropriate choice of quantum states.  \\ \\
We will now turn to higher dimensional models.  The most natural extension to higher dimensions is to include a position label for the states.  As before, in order to utilise results the geometric formulation of quantum mechanics of Ashtekar and Schilling, discussed in section \ref{GeoReform}, we will consider states labelled by complex coordinates.  This will present a difficulty since, unlike in two dimensions, the K\"ahler manifolds in higher dimensions are not maximally symmetric.  This will lead to additional terms in the equations of motion for the expectation values of operators.  \\ \\
Our study will thus almost immediately specialise to the simplest higher dimensional models, those that have full Schr\"odinger group symmetry (such as the free particle and harmonic oscillator).  Though this is a significant simplification it serves as a starting point to understanding the higher dimensional duals.  \\ \\
We will motivate, after analysis of these models, why a central extension of the Schr\"odinger algebra is a beneficial procedure.  In chapter \ref{McGChap} we will analyse the metrics that are produced from states where this central extension is considered as a dynamic quantity.  

\section{The Equations of Motion}

\label{EOMHigherSection}

We assume that the family of states are parametrised by $n$ complex parameters i.e. $|z_1, ..., z_n)$.  Note that the dimension of the state manifolds is $2n$.  We have already derived the equations of motion for the expectation value, $\Phi$ of an arbitrary operator in section \ref{EOMSection}, but restate it here for convenience
\begin{equation}
\sigma_{\gamma}^{\ \epsilon} \nabla_\alpha \nabla_\beta \nabla_\epsilon \Phi = R_{\gamma \beta \alpha}^{\ \ \ \  \delta}\sigma_{\delta}^{\ \epsilon} \nabla_{\epsilon} \Phi - \frac{1}{2}\tau_{\alpha \beta \gamma}.   \label{EOMn}
\end{equation}
In two dimensions we were able to use the simple form for the Riemann curvature tensor (\ref{2DRiem}) to derive, from (\ref{EOMn}), simple equations of motion with at most second order derivatives.  In the higher dimensional case the Riemann tensor is more complicated (see Appendix \ref{AppGeo}) which adds additional terms to the equations.  For $2n > 2$, we have that \cite{Blau}
\begin{eqnarray}
R_{\alpha \beta \gamma \delta} & = & \ \ \ W_{\alpha \beta \gamma \delta} \nonumber \\
 & & + \frac{1}{2n-2}(R_{\alpha \gamma}g_{\beta\delta} + R_{\beta\delta}g_{\alpha\gamma} - R_{\alpha \delta} g_{\beta\gamma} - R_{\beta \gamma} g_{\alpha \delta} ) \nonumber \\
& & - \frac{R}{(2n-1)(2n-2)} (g_{\alpha\gamma}g_{\beta\delta} - g_{\alpha\delta}g_{\beta\gamma}) \label{RiemannN}
\end{eqnarray}
where $W_{\alpha\beta\gamma\delta}$ is the Weyl tensor.  The Weyl tensor encodes information on the conformal properties of the metric in that the Weyl tensor is zero if and only if the metric is conformally flat \cite{Blau}. This more complicated form for the Riemann tensor makes the higher-dimensional analysis more intricate.  In this thesis we will try to simplify the equations as much as possible.  \\ \\
As a first simplification we note that the contraction of the Weyl tensor under any two indices is zero so that we first consider only the trace of the field equations (\ref{EOMn}).  We then arrive at
\begin{equation}
\sigma_{\gamma}^{\ \epsilon}g^{\alpha\beta} \nabla_\alpha \nabla_\beta \nabla_\epsilon \Phi \Rightarrow \nabla_\epsilon \nabla^2 \Phi = -\sigma_\epsilon^{\ \gamma}R_{\gamma}^{\ \delta}\sigma_{\delta}^{\ \alpha}\nabla_\alpha \Phi - R_{\epsilon}^{\ \delta}\nabla_\delta \Phi.  \label{ContEOMn}
\end{equation}
Note that if we are working with an Einstein manifold i.e. $R_{\mu\nu} = \frac{R}{2n} g_{\mu\nu}$ this simplifies to
\begin{equation}
\nabla_\epsilon \nabla^2 \Phi = -\frac{ R}{n} \nabla_\epsilon \Phi 
  \label{ScalarFieldn}
\end{equation}
and, if in addition the scalar curvature is constant, we find 
\begin{equation}
 \nabla^2 \Phi = -\frac{R}{n} \Phi .  \label{LaplaceFEn}
\end{equation}
The trace of the field equations, at least for an Einstein metric of constant scalar curvature, is of the simple form (\ref{LaplaceFEn}).  We will see in section \ref{SchrModels} that even for the free particle states the resulting metric is not Einstein.  Indeed, we will only produce metrics that are Einstein (in more than two dimensions) in chapter \ref{McGChap}.  \\ \\
If the metric is not Einstein then the trace of the field equations are more complicated.  We express the Ricci tensor as $R_{\mu\nu} = \frac{R }{2n} g_{\mu\nu} + \tilde{R}_{\mu\nu}$ where $g^{\alpha\beta} \tilde{R}_{\alpha\beta}$ = 0 and take the trace of (\ref{ContEOMn}) to find
\begin{eqnarray}
\nabla_\epsilon \nabla^2 \Phi & = & - \frac{ R }{n} \nabla_{\epsilon} \Phi  -\sigma_\epsilon^{\ \gamma}\tilde{R}_{\gamma}^{\ \delta}\sigma_{\delta}^{\ \alpha}\nabla_\alpha \Phi - \tilde{R}_{\epsilon}^{\ \delta}\nabla_\delta \Phi   \nonumber \\
& = & - \nabla_{\epsilon} \left( \frac{ R }{n} \Phi \right) + \left( \frac{\Phi}{n} \nabla_\epsilon R -\sigma_\epsilon^{\ \gamma}\tilde{R}_{\gamma}^{\ \delta}\sigma_{\delta}^{\ \alpha}\nabla_\alpha \Phi - \tilde{R}_{\epsilon}^{\ \delta}\nabla_\delta \Phi   \right). \label{RicciExpand}
\end{eqnarray}
If the second term of (\ref{RicciExpand}) is zero then we recover (\ref{LaplaceFEn}).  If not then our equations of motion will look quite different.  If, in addition we are not sourcing the expectation value of a symmetry generator then this will add an additional term in (\ref{RicciExpand}).  Note that the second term has contributions from a scalar curvature that is not constant and a Ricci tensor that is not Einstein.  The best situation we can hope for is that we can contain the deviation from (\ref{LaplaceFEn}) in a single scalar function
\begin{equation}
 \nabla^2 \Phi = -\frac{R}{n} \Phi + V(\Phi)  \label{Vmass}
\end{equation}
where $V$ gets contributions if the scalar curvature is not constant, the manifold is not Einstein and the expectation value being sourced is not a generator of symmetry.  
\\ \\
The trace of the field equations (\ref{ContEOMn}) thus only assumes the form of an eigenvalue problem for the expectation values of symmetry generators (\ref{LaplaceFEn}) if both the scalar curvature is constant and the metric is Einstein.  If we were to consider the full set of field equations (\ref{ContEOMn}) we would have to account for a non-zero Weyl tensor as well.  In this thesis we will not study the latter so that the analysis of the full set of field equations is a possible avenue of future investigation.  Note, importantly, that since we are only considering the one field equation we will need to specify additional boundary conditions in order to source the appropriate expectation value.  We will speak to this more in section \ref{GDDMFP}.  

\subsection{Massive Scalar Field}

\label{MassScSection}

Before we proceed to look at a specific higher dimensional example we will show how the field equation (\ref{LaplaceFEn}) may result from the variation of an action.  We have already introduced this action in section \ref{ScalarFieldSection} but we restate it here for higher dimensions
\begin{equation}
S = \int d^d x \sqrt{g} \left( (\nabla \Phi)^2 + m^2 \Phi^2 + 2 \tilde{V}( \Phi)  \right).  \label{MSFA}
\end{equation}
The action (\ref{MSFA}) is defined on a fixed background.  The field equation is found by varying with respect to the field $\Phi$
\begin{equation}
\frac{\delta S}{\delta \Phi} = 0 \Rightarrow \nabla^2 \Phi = m^2\Phi + \partial_\Phi \tilde{V} 
\end{equation} 
which is exactly of the form (\ref{Vmass}) if $m^2 = -\frac{2 R}{d}$ and $\partial_\Phi \tilde{V}  = V(\Phi)$.  \\ \\
Along with the field equation we will have to specify boundary conditions.  This is best illustrated by example e.g. in section \ref{GDDMFP}.  

\section{The $Schr_{d+1}$ Models }

\label{SchrModels}
  
The Schr\"odinger algebra for two dimensions (time and one spatial dimension) is given in (\ref{Schr11Alg}) and the extension to higher dimensions is straightforward (see Appendix \ref{HolsteinApp}).  Instead of a single position operator $X$ and momentum operator $P$ we have $d$ i.e. $X_1, X_2, ..., X_d$ and $P_1, P_2, ..., P_d$ related by rotation.  These then satisfy
\begin{equation}
 [X_i, P_j] = i \delta_{i j}.  \label{XPComm}
\end{equation}
For the sake of illustration we restrict ourselves to a concrete example namely the free particle states
\begin{equation}
e^{i t H} e^{i \vec{x} \cdot \vec{P} } | x=0 
) \ \rightarrow \ e^{i(t + i\beta) H} e^{i(\vec{x} + i \vec{y})\cdot \vec{P} }| x=0 ) \equiv e^{i \tau H} e^{i \vec{z} \cdot \vec{P}} | x=0 )  \label{FreePartState}
\end{equation}
though, as discussed in the context of the $2d$ metrics, any Hamiltonian from the $schr_{d+1}$ algebra, such as the harmonic oscillator, will yield states that are related by a coordinate transformation and normalisation to the states (\ref{FreePartState}).  Note the definition of the coordinates $t, \beta, x_j$ and $y_j$ in (\ref{FreePartState}).  We will work with $\left\{\beta, t, x_1, y_1, ..., x_d, y_d \right\}$ coordinates and $\left\{ \tau, \overline{\tau}, z_1, \overline{z}_1, ..., z_d, \overline{z}_d \right\}$ coordinates interchangeably.  Note that we have $n = d+1$ complex coordinates.  The overlap of these states (\ref{FreePartState}) is given by
\begin{equation}
(\tau, \vec{z} | \tau, \vec{z}) = (i\tau - i\overline{\tau})^{-\frac{d}{2}} e^{-i \frac{(\vec{z} - \vec{\overline{z}})^2}{2(\tau - \overline{\tau} )}} \label{KahlFreeOL}
\end{equation}
which is simply the complexified and unnormalised free particle propagator.  The integer $d$ refers to the number of spatial dimensions.  From the overlap (\ref{KahlFreeOL}) the metric may be calculated to be
\begin{equation}
ds^2 = \frac{d\beta + 4 \vec{y}\cdot\vec{y} }{8\beta^3}\left(d\beta^2 + dt^2 \right) - \frac{1}{\beta^2}\left( \vec{y}\cdot d\vec{y}d\beta + \vec{y}\cdot d\vec{x} dt \right) + \frac{1}{2\beta}\left( dy^2 + dx^2 \right).  \label{schrMet}
\end{equation}
The scalar curvature is constant, $R = -\frac{8(d+2)}{d}$ but the manifold is not Einstein.  Specifically there are only two non-zero entries for the Ricci tensor
\begin{equation}
R_{\tau \overline\tau} = R_{\overline\tau \tau} = -\frac{d+2}{2\beta^2}.  
\end{equation}
The equations of motion for the expectation values of symmetry generators are thus not of the form (\ref{LaplaceFEn}) so that these expectation values cannot be found as eigenfunctions of the Laplacian.    

\subsection{Examining the Laplacian}

In section \ref{CasimirSection} we were able to derive the trace of the energy momentum tensor by observing that the Laplacian is a differential operator representation of the $su(1,1)$ Casimir.  This was because the Killing vectors formed a vector field representation of the $su(1,1)$ algebra elements and all the Killing vectors commute with the Laplacian.  \\ \\
The situation for the $schr_{d+1}$ algebra is slightly different.  A dynamical symmetry (\ref{DynamicSymm}) only implies a non-zero Killing vector if it is not simply a change of phase for the family of states.  The commutator of $X_i$ and $P_i$ (\ref{XPComm}) will simply change the phase of the wave function.  While the operators $X$ and $P$ thus satisfy (\ref{XPComm}) their corresponding vector fields satisfy
\begin{equation}
[\chi_{P_i}, \chi_{X_j}] = 0 \ \ \ \forall \ i, j.  
\end{equation}
This implies that the Killing vectors no longer close on the Schr\"odinger algebra.  As an example, the Laplacian in four dimensions is given by
\begin{equation}
\nabla^2 =  8\beta^2(\partial_\beta^2 + \partial_t^2) +  16\beta y (\partial_\beta\partial_y + \partial_t \partial_x) + 2( \beta + 4y^2)(\partial_y^2 + \partial_x^2)   \label{4Lap}
\end{equation}
and commutes with all the Killing vectors, $\chi_H, \chi_D, \chi_K, \chi_{P_i}, \chi_{X_j}$ but is not the Schr\"odinger algebra Casimir (see e.g. \cite{schrCasimir}) in differential operator form.  The procedure of section \ref{CasimirSection}, where we classified the expectation values of operators of the enveloping algebra by their eigenvalues with the Casimir and scaling operator, will thus not work here.  \\ \\
One of the natural questions to ask is whether the eigenfunctions of the Laplacian can still be interpreted as the expectation values of operators from the enveloping algebra of $schr_{d+1}$.  To begin answering this question we first state the expectation values of the generators of symmetry explicitly
\begin{eqnarray}
\langle \tau, \vec{z} | H |\tau, \vec{z} \rangle & = & \frac{d}{4\beta} + \frac{\vec{y}^2}{2\beta^2}, \nonumber \\
\langle \tau, \vec{z} | D |\tau, \vec{z} \rangle & = & \frac{dt}{4\beta} + \frac{ \vec{y}\cdot(t\vec{y} -\beta \vec{x})}{2 \beta^2}, \nonumber \\
\langle \tau, \vec{z} | K |\tau, \vec{z} \rangle & = & \frac{d(t^2 + \beta^2)}{4 \beta} + \frac{(t\vec{y} -\beta \vec{x})\cdot(t\vec{y} -\beta \vec{x})}{2 \beta^2}, \nonumber \\
\langle \tau, \vec{z}| P_j |\tau, \vec{z} \rangle & = & \frac{y_j}{\beta},  \nonumber \\
\langle \tau, \vec{z} | X_j |\tau, \vec{z} \rangle & = & \frac{t y_j -\beta x_j}{\beta} .   \label{schrExp}
\end{eqnarray}
Note that the $SU(1,1)$ generators are all of the form
\begin{equation}
\langle \tau, \vec{z} | A |\tau, \vec{z} \rangle = \langle \tau, \vec{\overline{0}} | A |\tau, \vec{0} \rangle + \frac{1}{2}\sum_{i=1}^d \langle \tau, \vec{z} | B_{1,i} |\tau, \vec{z} \rangle \langle \tau, \vec{z} | B_{2,i} |\tau, \vec{z} \rangle  \label{SU11split}
\end{equation}
where $B_{1,i} = B_{2,i} = P_i$, $B_{1,i} = B_{2,i} = K_i$ and $B_{1,i} = P_i, B_{2,i} = K_i$ for $H, K$ and $D$ respectively.  On the $\vec{z} = \vec{\overline{z}} = \vec{0}$ submanifold the expectation values, metric and Ricci tensor thus look identical to the two-dimensional $su(1,1)$ case.  This may be understood intuitively since the underlying algebra, the Schr\"odinger algebra, is the semi-direct sum of $su(1,1)$ and $d$ Heisenberg algebras (related by rotation) i.e. some combination of an $AdS_2$ and $d$ flat space geometries.  The $\vec{z} = \vec{\overline{z}} = \vec{0}$ submanifold is precisely the $AdS_2$ submanifold.  \\ \\
Based on this one might venture a guess that a subset of the eigenfunctions of the Laplacian are the expectation values of operators restricted to the $AdS_2$ submanifold.  This turns out to be the case and, for the operators $L_n^q$ (defined in section \ref{CasimirSection}) we find the relation
\begin{equation}
\nabla^2 \langle \tau, \vec{0} | L_n^q |\tau, \vec{0} \rangle = -\frac{R}{d+2} n(n+1) \langle \tau, \vec{0} | L_n^q |\tau, 0 \rangle    \label{TensorEign}
\end{equation}
where $|\tau, 0\rangle = e^{i \tau H}|\vec{z} = 0)$.  In addition to this the expectation value of the position and momentum operators are eigenfunctions of the Laplacian
\begin{equation}
\nabla^2 \langle \tau, \vec{z} | P_j |\tau, \vec{z} \rangle = 0 = \nabla^2 \langle \tau, \vec{z} | X_j |\tau, \vec{z} \rangle \label{LapExpSub}
\end{equation}

\subsection{Eigenfunctions of the Laplacian}

Of course, the expressions (\ref{TensorEign}) are not expectation values with respect to the $|\tau, \vec{z}\rangle$ basis.  Using the relations (\ref{schrExp}), (\ref{SU11split}) and (\ref{TensorEign}) we may deduce that the expectation value
\begin{equation}
\langle \tau, \vec{z} | \delta A_i \delta B_i |\tau, \vec{z} \rangle = \langle \tau, \vec{z} | A_i B_i |\tau, \vec{z} \rangle -  \langle \tau, \vec{z} | A_i |\tau, \vec{z} \rangle \langle \tau, \vec{z} | B_i |\tau, \vec{z} \rangle
\end{equation}
where $A_i, B_i \in \left\{ X_i, P_i \right\}$ is an eigenfunction of the Laplacian.  This can be written in a much more suggestive form following eq. (\ref{antiComm})
\begin{equation}
\langle \tau, \vec{z} | \delta A_i \delta B_i |\tau, \vec{z} \rangle = \frac{1}{2}\langle \tau, \vec{z} | \{\delta A_i, B_i \} |\tau, \vec{z} \rangle = \frac{1}{2 }\chi^\mu_{i A_i} \partial_\mu \langle \tau, \vec{z} | B_i |\tau, \vec{z} \rangle   
\end{equation}
which suggest that the vector fields $\chi_{i O}$ may be used in some ladder operator scheme to find the eigenfunctions of the Laplacian.  Indeed, this turns out to be case. \\ \\
The eigenfunctions of the Laplacian (\ref{4Lap}), $f^q_{j, n}$ may be labelled by three indices $n \in N, q \leq |n|$ and $j \in \left\{ 0, 1, ..., d\right\}$.  The explicit meaning of these indices will become clear shortly.  The starting point of the laddering scheme is one of the $d$ expectation values of momentum.  One can then ladder up in $n$ by means of $\chi_{iH}$ immediately or first apply $\chi_{i P}$ and then ladder up in $n$ by means of $\chi_{iH}$ i.e.
\begin{eqnarray}
f^{-n}_{\ \ j, n} & = & \left( \chi^\mu_{i H} \partial_\mu \right)^{n-1} \langle \tau, \vec{z} | P_j |\tau, \vec{z} \rangle,   \nonumber \\
f^{-n}_{\ \ 0, n} & = & \left( \chi^\mu_{i H} \partial_\mu \right)^{n-1} \left( \chi^\nu_{i P_j} \partial_\nu \right) \langle \tau, \vec{z} | P_j |\tau, \vec{z} \rangle.  \label{LowestInq}
\end{eqnarray}
At this point the explicit forms of these functions may be useful for the sake of following the argument
\begin{equation}
f^{-n}_{\ \ j, n} \propto \frac{y_j}{\beta^n} \ \ \ ; \ \ \ f^{-n}_{\ \ 0, n} \propto \frac{1}{\beta^n} \ \ \ ; \ \ \ \chi^\mu_{i H} \partial_\mu = \partial_\beta \ \ \ ; \ \ \ \chi^\nu_{i P_j} \partial_\nu = \partial_{y_j}.   \label{ExplicitiVec}
\end{equation}
The expressions (\ref{ExplicitiVec}) may be used to verify the statements we make.  We prefer to write the expressions with (\ref{LowestInq}) since these expressions do not make use of a choice of coordinates.  The eigenfunctions (\ref{LowestInq}) are the lowest tiers in terms of laddering in the index $q$ since $\chi^\nu_H \partial_\nu f^{-n}_{\ \ j, n} = 0$.  We may ladder up in the index $q$ by means of $\chi_K$ i.e.
\begin{equation}
f^{q+1}_{\ \ j, n} = \chi_K^\mu \partial_\mu f^{q}_{ j, n} \label{Ladderq}
\end{equation}
so that the index $q$ is related to the eigenvalue with respect to $\chi_D$.  The index $n$ is related to the eigenvalue with respect to the Laplacian.  Explicitly, most easily verified for the functions (\ref{LowestInq}), one has
\begin{eqnarray}
\chi^\nu_D \partial_\nu f^{q}_{j, n} = (q + \frac{1}{2}) f^{q}_{j, n} \ \ \ ; \ \ \ \nabla^2 f^{q}_{j, n} = -\frac{R}{d+2} n(n-1) f^{q}_{j, n} \label{Eig1} \\
\chi^\nu_D \partial_\nu f^{q}_{0, n} = q f^{q}_{0, n} \ \ \ ; \ \ \ \nabla^2 f^{q}_{0, n} = -\frac{R}{d+2} n(n+1) f^{q}_{0, n}.   \label{Eig2}
\end{eqnarray}
The index $j$ thus divides the eigenfunctions into two classes, those with eigenvalues (\ref{Eig1}) if $j$ is non-zero and (\ref{Eig2}) if $j = 0$.  Note that the functions with non-zero $j$ are related simply by rotation.  

\subsection{Eigenfunctions Expressed as Operator Expectation Values}

The expressions (\ref{LowestInq}) and (\ref{Ladderq}) now express the eigenfunctions of the Laplacian as some combination of expectation values of operators with respect to the states $|\tau, \vec{z}\rangle$.  A few examples of these are
\begin{eqnarray}
f^0_{j, 0} & = & \langle \tau, \vec{z} | P_j |\tau, \vec{z} \rangle, \nonumber \\ 
f^{-1}_{\ \ j, 1} & = & \langle \tau, \vec{z} |H P_j |\tau, \vec{z} \rangle - \langle \tau, \vec{z} | H |\tau, \vec{z} \rangle \langle \tau, \vec{z} | P_j |\tau, \vec{z} \rangle, \nonumber \\
f^{-2}_{\ \ j, 2} & = & \langle \tau, \vec{z} |\delta H H P_j |\tau, \vec{z} \rangle - \langle \tau, \vec{z} |\delta H H |\tau, \vec{z} \rangle \langle \tau, \vec{z} | P_j |\tau, \vec{z} \rangle \nonumber \\ & & - \langle \tau, \vec{z} | H |\tau, \vec{z} \rangle \langle \tau, \vec{z} |\delta H P_j |\tau, \vec{z} \rangle.  
\end{eqnarray}
Though the eigenfunctions of the Laplacian are not interpretable directly as the expectation values of operators (instead they are combinations of products of expectation values), they may be used to calculate a specific expectation value of an operators.

\subsection{The Structure of the Laplacian in Terms of the $schr_{d+1}$ Algebra}

We suspect that these results may be better understood if the Laplacian can be recast in terms of the $schr_{d+1}$ operators.  Unfortunately, our studies thus far have not revealed how this may be done and this is an avenue of future investigations.  \\ \\
What we find intriguing is the fact that the eigenfunctions form two classes - those that are generated by a series of ladder operators on $\langle P\rangle$ and on $\langle \delta P \delta P \rangle$ (\ref{LowestInq}).  The eigenvalues for these two classes are different, but the difference between eigenvalues is the same.  \\ \\
This is reminiscent of selecting the even or odd sector in the $su(1,1)$ representation i.e. the $k=\frac{1}{4}$ or $k=\frac{3}{4}$ representation.  More explicitly, the $su(1,1)$ operators (\ref{LadderBase}) may be represented in terms of creation and annihilation operators
\begin{equation}
K_0 = \frac{1}{2}a^\dag a + \frac{1}{4} \ \ \ ; \ \ \  K_+ = \frac{1}{2}a^\dag a^\dag \ \ \ ; \ \ \ K_{-} = \frac{1}{2}a a
\end{equation}
where $[a, a^\dag] = 1$.  The coherent states for $su(1,1)$ will be of the form $e^{z K_+}|\phi_0)$ for two choices of $|\phi_0)$ namely the eigenstates $|0\rangle$, $|1\rangle$ of $a^\dag a$ i.e. $K_0 |0\rangle = \frac{1}{4}$ and $K_0 |1\rangle = \frac{3}{4}$.  The coherent states will then be a combination of the even states $|2n\rangle$ or the odd states $|2n+1\rangle$.  The operators that relate the even and odd sectors are $a$ and $a^\dag$.  \\ \\
The states we are considering (\ref{FreePartState}) are the result of the Schr\"odiger group acting on the reference state $|0\rangle$.  The Schr\"odinger group, in turn, can be viewed as the semi-direct sum of $SU(1,1)$ and $d$ Heisenberg groups (see appendix \ref{BCHApp}) related by rotation.  In a certain sense we are thus mixing the $k=\frac{1}{4}$ and $k=\frac{3}{4}$ representations.  If we only restrict ourselves to (a specific combination of) the operators with even powers of $P$ and $X$ we get the expressions (\ref{Eig2}) while odd powers yield (\ref{Eig1}).  These are the expressions we would expect if we simply calculated the eigenvalues for the $su(1,1)$ enveloping algebra elements in the two different representations, see section \ref{CasimirSection}.  \\ \\
These statements can only be made rigorous once the Laplacian is recast in terms of $schr_{d+1}$ algebra elements.  We hope to do this in future.  

\section{Notable Submanifolds}

In chapter \ref{McGChap} we will seek to put together a dual description that is simpler than that of section \ref{SchrModels}.  One may anticipate what we will do when it is observed that the Laplacian is not a differential operator representation of the Schr\"odinger algebra Casimir.  \\ \\
The reason the Laplacian is not a representation of the Schr\"odinger algebra Casimir is because the Killing vectors do not close on the $schr_{d+1}$ algebra which in turn is the consequence of the commutator between $X_j$ and $P_j$ only changing the phase of the states.  In order for it to be represented as a non-zero Killing vector we will have to ensure that it represents a non-trivial transformation of the states. \\ \\
Before we continue to do this we briefly highlight some of the interesting manifolds that result if not all of the coordinates of the states (\ref{FreePartState}) are complexified.  We will introduce only a single complex coordinate. Of course, this will spoil the relation between the Killing vectors and the expectation values of operators.  In order to develop a dictionary for these manifolds a lot more work will be required, a discussion undertaken in section \ref{SubmanDict}.  Nonetheless, these metrics are interesting because they are studied in the literature and illustrate some properties of the construction we have not yet encountered.  

\subsection{Complexified Time}

 Consider the regularisation scheme where we only complexify time i.e. $t \rightarrow t + i\beta$ in (\ref{FreePartState}).  This breaks all symmetries generated by $X_i$ and $K$ in the bulk ($\beta > 0$) and produces the metric
\begin{equation}
ds^2 = \frac{k}{2\beta^2}\left(d\beta^2 + dt^2 \right) + \frac{k}{\beta}d\vec{x}^2. \label{complexTimeMet}
\end{equation}
After the transformation $\beta =  \frac{1}{r^2}$ this becomes
\begin{equation}
ds^2 = k\left( \frac{4 dr^2}{r^2} + \frac{r^4}{2}dt^2 + r^2 d\vec{x}^2   \right)
\end{equation}
which is the metric studied in \cite{Liu} when considering Lifshitz-like fixed points.  A Lifshitz-like fixed point is a scale invariant fixed point where space and time scale as $x \rightarrow \lambda x$ and $t \rightarrow \lambda^z t$.  In \cite{Liu} the authors focus on the case $z = 2$.  This scale invariance is reflected in the metric as an isometry.  Note that the states on the boundary $\beta \rightarrow 0$ are still the states (\ref{FreePartState}).  This illustrates that the bulk for a dual theory is not unique and, in particular, depends on the regularisation scheme chosen.  

\subsection{Momentum Regularisation}

We may regularise the states in another way, by means of the absolute value of momentum.  We consider the states
\begin{equation}
|r, t, \vec{x} ) = e^{i t H}e^{i \vec{x}\cdot\vec{P}} e^{-r |\vec{P}|}|\vec{x} = \vec{0} ). \label{MomState}
\end{equation}
There are other schemes that use functions of powers of $|P_j|$ but the scheme (\ref{MomState}) is chosen so that rotational symmetry remains unbroken.  However, as with (\ref{complexTimeMet}) the symmetries generated by $X_i$ and $K$ are broken.  For $d=1$, for example, this yields the metric
\begin{equation}
ds^2 = \frac{1}{4 r^2}dr^2 -\frac{1}{2 r^3}dt dr + \frac{5}{16 r^4}dt^2  + \frac{1}{2 r^2}dx^2,  \label{MomRegMet}
\end{equation}
which is clearly a different metric to (\ref{complexTimeMet}).  Even though the symmetries we retain in the bulk are the same and the states on the boundary of the manifold of states are identical the bulk geometry is different.  This illustrates that the regularisation scheme we choose has a significant effect on the geometry.  \\ \\
Note that the interpretation of the radial coordinate is also different.  In (\ref{complexTimeMet}) it has the interpretation of an energy scale while in (\ref{MomRegMet}) it has the dimension of length.  There is thus not a general interpretation of the radial coordinate.  

\section{Discussion}

In this chapter we sought to extend the two-dimensional gravitational duals of chapter \ref{GravChap} to higher dimensions. As one may expect the equations of motion become significantly more complicated in higher dimension.  The Riemann curvature has contributions from the scalar curvature, the Ricci tensor and the Weyl tensor.  These cannot be avoided altogether since no K\"ahler manifold in four dimensions or higher is maximally symmetric.  \\ \\
In order to simplify matters we focused solely on the trace of the field equations.  The Weyl tensor is zero under any contraction of indices so that the trace of the equations of motion are not affected by the Weyl tensor.  The trace of the equations yield a Laplacian.  We took note of the fact that, if the manifold is Einstein with constant scalar curvature, the expectation values of symmetry generators are eigenfunctions of the Laplacian.  \\ \\
We then proceeded to analyse a specific example - states with Schr\"odinger symmetry which is the natural extension of the $su(1,1)$ states to higher dimensions.  The resulting manifold is, however, not Einstein and the Laplacian is not a representation of the Schr\"odinger algebra Casimir.  We managed to develop a scheme by which the expectation values of operators may be extracted from the eigenfunctions but further work is needed to understand this structure properly.  \\ \\
In the next chapter we will seek to restore the status of the Laplacian as a representation of the Casimir.  This will allow us to interpret its eigenfunctions as the expectation values of operators so that the dictionary of chapter \ref{GravChap} may be extended to higher dimensions more simply.

\chapter{Free Particle Metrics with Dynamical Mass}

\label{McGChap}

In this chapter we will proceed to centrally extend the Schr\"odinger algebra in order to construct dual gravitational theories that are as simple as possible for the Schr\"odinger Hamiltonian models.  Specifically we hope that this will restore the Laplacian to a differential operator representation of the Casimir.  The dictionary we will be able to develop should then be closely related to that of chapter \ref{GravChap}.  \\ \\
We note that metrics containing the Schr\"odinger symmetry have received interest before.  In \cite{McGreevyGravDual}, \cite{Son} the authors aim to generalize the $AdS/CFT$ correspondence to the case of non-relativistic field theories, or $AdS/NRCFT$.  As is the usual case for dual descriptions the symmetries play a crucial role.  The symmetry generators they consider, that of $NRCFT$, close on the conformal Galilei algebra \cite{McGreevyGravDual} of which the Schr\"odinger algebra of chapter \ref{HighDim} is an example.  The conformal Galilei group will thus be our starting point in this chapter.  \\ \\
This chapter can be divided into three parts.  In the first we will show how considering the mass (the central extension of the conformal Galilei group) as a dynamical variable will allow for the definition of a set of quantum states that have a simple gravitational dual.  Specifically we will find that the expectation values of symmetry generators may be sourced by a traceless energy momentum tensor, precisely the desired result.  We will conclude this part with a dictionary applicable to the complex free particle states in higher dimensions.  \\ \\
In the second we will show how metrics studied in the $AdS/NRCFT$ correspondence literature \cite{McGreevyGravDual}, \cite{Son} may be derived using our construction.  This can be done by either an appropriate choice of quantum states or viewing the metrics as defined on submanifolds of a K\"ahler manifold.  This will supplement their works by identifying the explicit quantum states that constitute the quantum theory of the duality.  These observations, along with the systematic nature of the construction, holds the promise of future generalisations to these studies.  \\ \\
These submanifolds will in general not be K\"ahler themselves and thus the duality between the expectation values of operators and Killing vectors breaks down.  In the last section we will speculate as to how a dictionary for the submanifolds may be developed.  We would like to warn the reader beforehand that the discussion contained therein will be qualitative at best.  Therein we wish only to convey some of our ideas on how one may progress beyond the work done in this thesis. 

\section{The Conformal Galilei Algebra}

We start by defining the $d+1$ dimensional conformal Galilei algebra (see Appendix \ref{HolsteinApp}) which consists of $\frac{1}{2}d(d-1)$ rotations, $d+1$ translations, $d$ Galilean boosts, dilatations and the rest mass.  These are generated by $M_{ij}$, $P_i$+$H$, $G_i$, $D$ and $N$ which satisfy the (non-zero) commutation relations
\begin{eqnarray}
\left[ M_{ij}, M_{kl} \right] &=& i( g^0_{ik} M_{j l} + g^0_{j l} M_{i k} - g^0_{i l} M_{j k} - g^0_{j k} M_{i l}   )  \nonumber \\
\left[ G_{i}, M_{k l} \right] = i(g^0_{i k}X_l - g^0_{i l}K_k) \ \ \ & ; & \ \ \ \left[ P_{i}, M_{k l} \right] = i(g^0_{i k}P_l - g^0_{i l}P_k)  \nonumber \\
\left[G_{i} , P_{j} \right] = i \delta_{i j} N \ \ \  & ; & \ \ \ \left[D, P_{i} \right] =  -\frac{i}{2} P_{i} \nonumber \\
\left[ D, G_i \right] = \frac{i}{2}(z-1)G_i \ \ \ & ; & \ \ \ \left[ D, N \right] = \frac{i}{2}(z-2)N \nonumber   \\
\left[ H, G_i \right]  =  -i P_i \ \ \ & ; & \ \ \  \left[ D, H \right] = -\frac{z}{2} i H \label{SzComm}
\end{eqnarray}
where $z$ is the so-called dynamical exponent and the (Latin) indices run from $1 \rightarrow d$.  The operator $N$ represents the central extension of the algebra and we will refer to it as the mass operator or mass for short.  This is the $d+1$-dimensional conformal Galilei algebra with dynamical exponent $z$, \cg{d+1} \cite{Duval}.  The dynamical exponent characterises the different scaling behavior of time and spatial coordinates.  The tensor $g^0_{\mu\nu}$ is the flat space metric and we immediately specialise to Euclidean signature i.e. $g^0_{\mu\nu} = \delta_{\mu\nu}$.  When $z=2$ there is also a special conformal generator $K$ which satisfies
\begin{equation}
 [K, P_i] = i G_i \ \ \ ; \ \ \ [D, K] =  i K \ \ \ ; \ \ \ [H, K] =  2 i D
 \end{equation}
 and the algebra is the $d+1$ dimensional Schr\"odinger algebra \ssc{d+1} \cite{Henkel2}.  In this chapter we will show how metrics that encode the \cg{d+1} transformations as isometries may be constructed from appropriately chosen quantum states via (\ref{Prov2}).  We will also provide a second perspective namely that they may be viewed as the metric of a submanifold of the K\"ahler manifold.   

\subsection{Representation in Terms of Quantum Mechanical Operators}

The operator $N$ is the central extension of the algebra which is a first step to restoring the status of the Laplacian as the Casimir.  The central extension by itself is not sufficient, however.  This is because the central extension still commutes with all elements of the Schr\"odinger algebra (see (\ref{SzComm})) so that the transformation it induces on the quantum states is still only a phase shift.  In order for it to induce a non-trivial transformation we will consider the mass, $N$,  as a dynamical variable and introduce a new operator $Z$ such that
\begin{equation}
[Z, N] = i.  
\end{equation}
We will refer to $Z$ as the conjugate mass.  The inclusion of $Z$ also allows us to represent the elements of algebra (\ref{SzComm}) in terms of $Z$, $N$ and position and momentum operators (we were able to do this previously for $z=2$ but not for general $z$).  \\ \\
The momentum operators are simply the $P_i$'s and the position operators can be recovered as $X_i = N^{-1}G_i$.   The operators $H$, $M_{ij}$ and $K$ can be represented in terms of momentum, $P_i$, and position, $X_i$, operators as $\frac{1}{2 N}\sum_i P_i^2$, $\frac{1}{N}\left(P_i X_j - P_j X_i\right)$ and $\frac{1}{2N}\sum_i G_i^2$ respectively.  Note that these representations are independent of the dynamical exponent and thus hold for any one of the \cg{d+1} algebras.  The dilatation operator $D$ on the other hand does depend on the dynamical exponent and may be represented as $D = -\frac{1}{4N}\sum_i \left(X_i P_i + P_i X_i \right)  + \frac{(z-2)}{4}\left( Z N + N Z \right)$.  Clearly we recover the representations in terms of position and momentum (\ref{Schr11Comm}) for the Schr\"odinger algebra when $z=2$.   \\ \\
The sets of operators $\left\{ Z, N \right\}$ and $\left\{ X_i, P_i\right\}$ are such that operators from different sets commute with each other and may thus be combined in a tensor product structure in a simple way.  This will be done shortly. 

\section{The Dynamical Mass Tensor Product States}

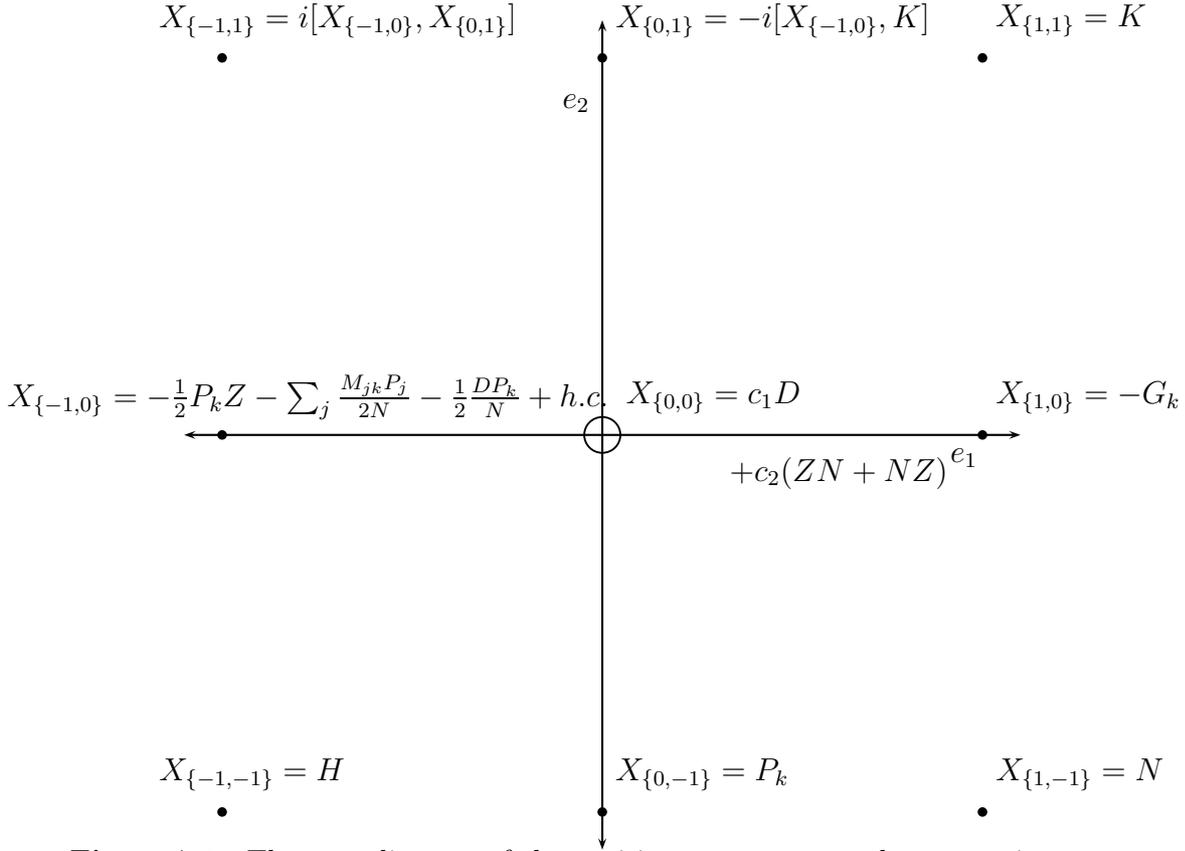
\begin{figure}[ht]
\centering
\begin{pspicture}(11,11)
%\psaxes{->}(0,0)(-1.2,-1.2)(1.2,1.2)[$x$,0][$y$, 90]	%creates axes
\psdot(10,10)	 \uput[0](10, 10.5){$X_{\left\{ 1, 1\right\} } = K$}
\psdot(10, 5) \uput[0](10, 5.5){$X_{\left\{ 1, 0\right\} } = -G_k$}
\psdot(10, 0) \uput[0](10, 0.5){$X_{\left\{ 1, -1\right\} } = N$}
\psdot(5, 10) \uput[0](5, 10.5){$X_{\left\{ 0, 1\right\} } =  -i[X_{ \left\{-1, 0 \right\} }, K ]$}
\psdot[dotstyle=o, dotsize=0.5](5, 5) \uput[0](5, 5.5){ $X_{\left\{0, 0\right\}} = c_1 D  $} \uput[0](6.5, 4.5){$+ c_2 (Z N + N Z)$} 
\psdot(5, 0) \uput[0](5, 0.5){$X_{\left\{ 0, -1\right\} } = P_k$}
\psdot(0, 10) \uput[0](-1, 10.5){$X_{\left\{ -1, 1\right\} } = i[X_{ \left\{-1, 0 \right\} }, X_{ \left\{0, 1 \right\} }]$}
\psdot(0, 5) \uput[0](-3., 5.5){$X_{\left\{ -1, 0\right\} } = -\frac{1}{2}P_k Z - \sum_j \frac{M_{j k} P_j}{2N} - \frac{1}{2} \frac{D P_k}{N} + h.c.$}
\psdot(0, 0) \uput[0](-1, 0.5){$X_{\left\{ -1, -1\right\} } = H$}
 \psline[linecolor=black]{<->}(-0.5,5)(10.5,5)\uput[0](9.4, 4.7){$e_1$}
\psline[linecolor=black]{<->}(5,-0.5)(5, 10.5)\uput[0](4.3, 9.4){$e_2$}
\end{pspicture}
\caption{The root diagram of the position-momentum and mass-conjugate mass algebra.  The rotation operators are not included but are formally part of $X_{\left\{0, 0 \right\}}$.  This constitutes a real form of the complex conformal algebra in $d+2$ dimensions.}
\label{cgRoot}
\end{figure}

We will now construct the quantum states that incorporate the dynamical mass.  The aim here is for the operator $N$ to induce a non-trivial transformation on the quantum states so that its action induces a coordinate transformation on the resulting metric. \\ \\
We note that this procedure of considering the mass as a dynamical variable corresponds to the prescription of Giulini \cite{Giulini} to introduce the Bargmann superselection rule in quantum mechanics.  Good discussions on the requirement for and the role played by the superselection rule can be found in \cite{Giulini}, \cite{Bargmann1}, \cite{Hernandez}.  Note, however, that the states we have worked with in the previous chapters are mathematically consistent and physically relevant and one may consider quantum mechanics with or without the superselection rule imposed. \\ \\
We will, in order to keep close contact with the physics, work with the representation of the operators in terms of $\left\{X_i, P_i, N, Z \right\}$.  See Appendix \ref{AppConfAlg} for the examination of the relevant algebras in terms of abstract elements.  The commutation relations, stated explicitly here for convenience, involving the conjugate mass are
\begin{eqnarray}
\left[Z, G_i \right] & = & i N^{-1} G_i \nonumber \\
\left[Z, P_i \right] & = & 0 \nonumber \\
\left[Z, H \right] & = & -i N^{-1} H \nonumber \\
\left[Z, D \right] & = & \frac{i}{2}(z-2) Z \nonumber \\
\left[Z, K \right] & = & i N^{-1} K
. \label{XPzetaComm}
\end{eqnarray}
These are useful to take note of for computational purpose but we will not be adding $Z$ directly to the algebra (\ref{SzComm}).  We will rather add the product of $Z$ with some (\ref{SzComm}) elements to the algebra.  We will show this algebra shortly.  As mentioned we will be considering the tensor product space of position-momentum and mass-conjugate mass.  The identity operator may be resolved in terms of these eigenvectors and specifically 
\begin{equation}
\hat{I} = \int d\vec{a} d n |\vec{a}, n)(\vec{a}, n|. \ \ \ \textnormal{and} \ \ \ (\vec{a}', n'| \vec{a}, n) = \delta(\vec{a} - \vec{a}') \delta(n - n')  \label{Identities}
\end{equation}    
where $\vec{a} \in \left\{ \vec{x}, \vec{p} \right\}$, $n \in \left\{m, \zeta \right\}$ and, for example, $|\vec{x}, m) \equiv e^{i \vec{x}\cdot \vec{P}}e^{i m Z}|x=0, m=0)$.  It may also be verified that
\begin{equation}
(\vec{p}, m| \vec{x}, \zeta) = \frac{1}{(2\pi)^{\frac{1}{2}(d+1)}} e^{i \vec{x}\cdot \vec{p}}e^{i m \zeta}.  
\end{equation}
Note that, like position and momentum are related by Fourier transform, mass and conjugate mass are also related by a Fourier transform.  This transformation between the mass and conjugate mass is used in \cite{Wen}, \cite{Henkel} to calculate the correlation functions for non-relativistic conformal field theories.  The non-relativistic overlaps can be found by considering the dynamical mass overlap and then simply restricting to constant mass.    \\ \\
By introducing the conjugate mass operator and building the tensor product Hilbert space we are enlarging the set of possible symmetry generators.  This is reasonably evident from the additional commutation relations we have to accommodate (\ref{XPzetaComm}).  The conjugate mass extends the \cg{d+1} algebra to one where the root diagram reads as in Fig. (\ref{cgRoot}).  The conformal Galilei algebras may be recovered as subdiagrams of this root diagram.  We will show later how the additional symmetries may be filtered out by an appropriate redefinition of the metric or equivalently, by considering an appropriate submanifold of the K\"ahler manifold.  \\ \\
%\begin{center}
%\begin{tabular}{ccc}
%$X_{\left\{ -1, 1\right\} } = -i[X_{ \left\{-1, 0 \right\} }, X_{ \left\{0, 1 \right\} }]$ & $ X_{\left\{ 0, 1\right\} } = -i[X_{ \left\{-1, 0 \right\} }, K ]$ & $X_{\left\{ 1, 1\right\} } = K$ \\ 
%  \begin{tabular}{ccc} $X_{\left\{ -1, 0\right\} } $ & = & $ N^\alpha P_kZ$ \\
%  & & $+ (\alpha-\frac{1}{2} )N^{\alpha-1} \sum_j M_{jk}P_j   + h.c$ \end{tabular} & $X_{\left\{ 0, 0\right\} } = a D + b(ZN + N Z)$ & $ X_{\left\{ 1, 0\right\} } =  N^{-\alpha} X_k $ 
% \\ $X_{\left\{ -1, -1\right\} } = H $ & $ X_{\left\{ 0, -1\right\} } = P_k $ & $X_{\left\{ 1, -1\right\} } = N^{1-2\alpha}$
%\end{tabular}
%\end{center}
The operators in the root diagram satisfy
\begin{eqnarray}
& &-i\left[ X_{\left\{ i, j\right\} }, X_{\left\{ i', j'\right\} }  \right] \nonumber \\
& = & \left\{ \begin{tabular}{cc} $ X_{\left\{ i+i', j+j'\right\} } $ & if \ \ \ $| i + i'| \leq 1 $ \ ; \ $|j+j'| \leq 1 \ ; \ i'+j' \geq i+j $ \\ $ -X_{\left\{ i+i', j+j'\right\} } $ & if \ \ \ $| i + i'| \leq 1 $ \ ; \ $|j+j'| \leq 1 \ ; \ i'+j' < i+j $ \\$0$  &  \textnormal{otherwise} \end{tabular} \right.  \label{DiagDef}
\end{eqnarray}
and is a real form of the complex conformal algebra in $d+2$ dimensions, $(conf_{d+2})_{\mathbb{C}}$.  Taking the mass as dynamical effectively adds another position label and completes conformal symmetry, from there the change from $d+1$ dimensions to $d+2$ dimensions.  See Appendix \ref{AppConfAlg} where the analysis is done explicitly. \\ \\
The explicit expressions of $X_{\left\{ -1, 1\right\} } $ and $X_{\left\{ 0, 1\right\} } $ are not only suppressed for the purpose of conciseness but also to emphasise that if $X_{\left\{ -1, 0\right\} } $ satisfies the appropriate commutation relations with the subalgebra, $X_{\left\{ i, j \leq i\right\} } $, then both $X_{\left\{ -1, 1\right\} } $ and $X_{\left\{ 0, 1\right\} } $ as defined above will satisfy their required commutation relations.  \\ \\
We again generate the appropriate quantum states, with the whole group as symmetries, by acting with a general group element on a reference state.  The states obtained by using the reference state
\begin{equation}
|\psi_0 ) \equiv f(N) |\vec{x}=\vec{0}, \zeta = 0)
\end{equation}
are an extension of the states used in (\ref{FreePartState}) and is, as such, an eigenstate of $K$, $D$, $X_i$ and $M_{ij}$ and in addition an eigenstate of $X_{\left\{ 0, 1 \right\}}$.  For the choices $f(N) = N^{\alpha}$ it is an eigenstate of $Z N + N Z$ and, finally, for $\alpha = -\frac{1}{2}$ it is also an eigenstate of $X_{\left\{-1, 0 \right\}}$ and consequently of $X_{\left\{ -1, 1 \right\}}$.  The reference state $N^{-\frac{1}{2}} |\vec{x}=\vec{0}, \zeta = 0)$ is thus the ``best" choice of reference state, see coherent states in section \ref{CohSection}.  In order to avoid difficulties with the prefactor $N^{-\frac{1}{2}}$ we will also add a factor that restricts to the positive eigenvalues of $N$.  We thus consider the states 
\begin{equation}
|t, \vec{x}, \zeta) \equiv e^{i t H} e^{-i \vec{x}\cdot\vec{P}} e^{i \zeta N} N^{-\frac{1}{2}} \theta(N) |\vec{x} = \vec{0}, \zeta = 0 ).  \label{zetaStates}
\end{equation}
These possess full conformal symmetry, as generated by the elements of Fig. (\ref{cgRoot}), which contains the \cg{d+1} symmetry as a subset.  See Appendix \ref{AppConfAlg} for an explicit representation of the operator action on these states in terms of differential operators in the coordinates $\left\{t, \vec{x}, \zeta \right\}$.  \\ \\
The overlap of the states (\ref{zetaStates}) may be calculated by inserting the momentum-mass identity states (\ref{Identities}).  Two things should be noted.  First the propagator is only well-defined for $Im(t)>0$ so that we need to add an infinitesimal complex number to time $t \rightarrow t+i\epsilon$ and take the $\epsilon \rightarrow 0$ limit afterwards.  Also, even though we are restricted to the positive eigenvalues of $N$ i.e. $m > 0$ there is a singularity at $m$ = 0.  The appropriate way to deal with the mass-integration is to integrate from a small, non-zero mass $m_0$ and take the limit where $m_0 \rightarrow 0$ afterwards.  We thus find
\begin{eqnarray}
& & (t', \vec{x}', \zeta'| t, \vec{x}, \zeta )  \nonumber \\
&=& \lim_{m_0 \to 0} \lim_{\epsilon \rightarrow 0}  \int^\infty_{m_0} dm  \int \frac{d^d p}{(2\pi)^{d+1}} \ e^{i \frac{\tilde{p}^2}{2 m} (t-t' + 2 i \epsilon)} \ e^{i (\vec{x}-\vec{x}')\cdot \vec{p}} \ e^{i(\zeta - \zeta') m} \ m^{-1} \nonumber \\
& = &  \lim_{m_0 \to 0} \lim_{\epsilon \rightarrow 0} \ \frac{1}{2} \pi^{-\frac{d}{2} - 1} \left[ (\vec{x}-\vec{x}')^2 - 2 (\zeta - \zeta')(t-t' + 2 i \epsilon) \right]^{\frac{d}{2}} \times \nonumber \\  & &\Gamma\left[\frac{d}{2} \ , \ m_0\left(  -(\zeta - \zeta') + \frac{(\vec{x}-\vec{x}')^2}{2(t-t' + 2 i \epsilon) }\right) \right] \nonumber \\
& = & \frac{1}{2} \pi^{-\frac{d}{2} - 1} \  \Gamma\left( \frac{d}{2} \right) \  ( (\vec{x} - \vec{x}')^2 - 2(t-t')(\zeta-\zeta'))^{-\frac{d}{2}}.  \label{ConfOLap}
\end{eqnarray}
where $\Gamma[a, b]$ is the incomplete gamma function with power $a$ and lower bound $b$ \cite{Abram}.  The limit process from the second last line to the last line needs to be handled with a little care.  Note that the last step no longer possesses the property that $(0, \vec{x}', \zeta'| 0, \vec{x}, \zeta ) \propto \delta(\vec{x} - \vec{x}') $.  This is due to the $m_0 \rightarrow 0$ limit.  Since we are integrating over all positive values of the mass and the time-scale is connected to the inverse mass this limit thus renders the $t=t'$ limit inaccessible.  For our construction and the discussion of symmetries the final expression for the overlap (\ref{ConfOLap}) will be used throughout.  \\ \\
The form of the overlap is reminiscent of the 2-point functions for conformal fields \cite{PolyakovCorr}, \cite{Freedman} and is, in fact, related to the 2-point functions by a complex coordinate transformations.  Indeed, as pointed out, this is due to the fact that the symmetry generators of (\ref{ConfOLap}) and the symmetry generators of the 2-point functions both constitute real forms of the complex algebra $(conf_{d+2})_{\mathbb{C}}$.  \\ \\ 
As expected, the overlaps (\ref{ConfOLap}) diverge when we put $\vec{x}' = \vec{x}$, $t' = t$, $\zeta' = \zeta$.  The states (\ref{zetaStates}) are thus non-normalisable and need to be regularised.  

\section{The Gravitational Dual of the Dynamical Mass Free Particle}

\label{GDDMFP}

We are now in a position to investigate possible dual gravitational descriptions of the Schr\"odinger algebra Hamiltonians with dynamical mass.  We will here work with the free particle but, as before, another choice of Hamiltonian (from the conformal Galilei algebra) will only entail a coordinate transformation. We can recover regularised states from (\ref{zetaStates}) by complexifying all the coordinates
\begin{equation}
t \ \rightarrow \ \tau \equiv t + i \beta  \ \ \ ; \ \ \ \vec{x} \ \rightarrow \vec{z} \equiv \vec{x} + i\vec{y} \ \ \ ; \ \ \ \zeta \ \rightarrow \ \theta \equiv \zeta + i\alpha.   
\end{equation}
We may derive the K\"ahler potential by simply complexifying the coordinates in the overlap (\ref{ConfOLap}) and taking its logarithm.  The inclusion of complex time, $\beta > 0$, in fact makes the integrals in the derivation of (\ref{ConfOLap}) more well-behaved and the calculation carries through without alteration.  As an example we give here the metric for $d=1$ 
\begin{eqnarray}
ds^2 & = & \frac{1}{2(y^2 - 2\alpha\beta)^2}\left( \alpha^2(dt^2 + d\beta^2) + \frac{y^2 + 2\alpha\beta}{2}(dx^2 + dy^2) + \beta^2(d\zeta^2 + d\alpha^2) \right. \nonumber \\
& & \ \ \ \ \ \ \ \ \  - y\alpha(d\beta dy + dyd\beta + dt dx + dx dt ) -y\beta (dyd\alpha + d\alpha dy + dxd\zeta + d\zeta)  \nonumber \\
 & & \ \ \ \ \ \ \ \ \  \left. + \frac{y^2}{2}(d\beta d\alpha + d\alpha d\beta + dt d\zeta + d\zeta dt)  \right) \label{DynMassMet}
\end{eqnarray}
The higher dimensional manifolds look very similar with the $y$'s and $x$'s replaced by the appropriate vectors (and dot products added).  It may be verified that these manifolds (for any $d$) are Einstein and the scalar curvature is constant $R = -\frac{72}{d}$.  \\ \\
This implies that the expectation values of symmetry generators satisfy (\ref{LaplaceFEn}) i.e.
\begin{equation}
\nabla^2 \Phi = -\frac{R}{d+2} \Phi.  \label{SFE}
\end{equation} 
The Killing vectors, which are the vector fields related to these expectation value (\ref{OperatorVector}) close on the algebra of Fig. (\ref{cgRoot}).  The Laplacian commutes will all the Killing vectors and is a second order differential operator.   It must thus be a differential operator representation of the Casimir.  \\ \\
We are only considering the trace of the field equations in (\ref{SFE}).  Though the metric (\ref{DynMassMet}) is Einstein with constant scalar curvature it is not conformally flat and thus the Weyl tensor is still non-zero.  As explained we may consider this one field equation (\ref{SFE}) if we add more boundary conditions.  We will do this explicitly for this example.  \\ \\
The expectation values of the symmetry generators are not the only solutions of (\ref{SFE}), however.  This is a consequence of the fact that we are only considering the trace of the field equations.  We have to specify additional boundary conditions in order for the field equation solution to be that of the desired expectation value.  \\ \\
The symmetry generator expectation values may be distinguished from the other solutions by verifying that their vector fields indeed solve the Killing equation.  This is our first boundary condition.  The symmetry generators may then be distinguished from each other (up to unitary transformation) by their eigenvalue w.r.t. the vector field of $D$ and $Z N + N Z$.  The eigenvalues classify the symmetry generators such as in the diagram (\ref{cgRoot}).  This constitutes the second boundary condition.  The choice of coordinates and identification of $D$ and $Z N + N Z$ constitute the final boundary conditions. \\ \\
For this very fortuitous example of the free particle we may thus identify the expectation values of the symmetry generators by these additional boundary conditions (the Killing vector requirement and the eigenvalues with respect to the Cartan subalgebra).  From these we may apply the procedure of section \ref{SymmCalc} to calculate the expectation value of an arbitrary string of symmetry generators.  \\ \\
The equation (\ref{SFE}) can be obtained as the field equation of a massive scalar field defined on the background (\ref{DynMassMet}).  The discussion is contained in section \ref{MassScSection}. The dictionary thus reads as
 \begin{center}
  \begin{tabular}{ | c | c | }
    \hline
    $Schr_{d+1}$ Hamiltonian  & Massive scalar field  \\ 
		with dynamical mass & on the background (\ref{DynMassMet}) \\ \hline \hline
		Quantum state & Point on the manifold of states \\ \hline
		Symmetries of states & Isometries of the metric \\ \hline
    Hamiltonian & Choice of coordinates  \\ \hline
    Representation label, $k$ & Scalar curvature $R = -\frac{4}{k}$ \\  \hline
		Operator expectation value & Field equation solution \\ \hline
		Operator $\in$ $(conf_{d+2})_{\mathbb{C}}$  & Vacuum solution vector \\ 
		 & fields satisfy Killing's equation \\ \hline
		Eigenvalues w.r.t. & Eigenvalues w.r.t.  \\ 
		Cartan subalgebra & vector fields $\chi_D$ and $\chi_{Z N + N Z}$ \\ \hline
		Specific unitary operator from solution set & Boundary condition on the dilaton \\ \hline
  \end{tabular} 
\end{center}

\subsection{The Argument for Considering Submanifolds}

Though the emergence of the massive scalar field (\ref{MassScSection}) as the dual to conformally symmetric states is reminiscent of calculations performed in $AdS/CFT$ (see e.g. \cite{Skenderis}) it is important to emphasise that the geometry (\ref{DynMassMet}) is not an $AdS$ geometry. Specifically, the geometry is not conformally flat line $AdS$.  The dictionary we have developed here is thus rather different from the conventional. \\ \\
What distinguishes the construction we have made here further from the more conventional $AdS/CFT$ is the fact that we have added an extra dimension for every real coordinate of the quantum states (\ref{zetaStates}).  In conventional $AdS/CFT$ only a single dimension is added namely the $AdS$ radial coordinate.  Even in works where non-relativistic field theories are analysed \cite{McGreevyGravDual}, \cite{Son}, only two additional coordinates are added.  The number of extra dimensions we have to add, in the context of existing literature, is certainly an oddity.  Note that this was not the case for the two-dimensional manifolds studied in chapter \ref{GravChap} (where we only needed to add a single complex coordinate).  \\ \\
One may suspect that, if we regularise the states differently (where we do not require so many complex coordinates to be added) then we might produce the geometries that are relevant in the literature.  Certainly, this will spoil the crucial link between the expectation values of operators and vector fields (\ref{OperatorVector}) which has formed the foundation of our dictionary.  This is definitely a big price to pay if our only gain is to reproduce metrics that are relevant in the literature.  \\ \\
We believe that a dictionary could possibly be developed between these states, no longer parametrised by complex coordinates, and models of gravity.  Indeed, it is possible that the dictionary for the complex coordinates states, that we have started to develop in this thesis, could be used in order to do this.  This is the topic of section \ref{SubmanDict}.  \\ \\
Before that we will show that many of the metrics that are studied in gauge/gravity dualities in the literature may be recovered from our construction.  These metrics will be closely related to the K\"ahler manifolds we have studied up to this point in the sense that they may be identified as submanifolds.  

\section{Restricting the Analysis to a Submanifold of the K\"ahler Manifold}

Throughout our investigations we have considered states parametrised by complex coordinates.  Even if our starting family of states was not, like the time-evolved states, we regularised these states in such a way that they were.  There were two major motivations for doing this.  Firstly this allowed us to utilise the many features of the geometric reformulation of quantum mechanics of \cite{Ashtekar} which formed the foundation of the dictionaries presented.  For the cases we considered this also ensured that the states were properly regularised. \\ \\
Secondly, possibly even more importantly, the regularisation scheme ensured that the symmetries of the quantum states were all encoded as bulk isometries.  The central role of symmetries in the dualities is thus respected.  If we want a metric to reflect a certain set of symmetries we start with a family of states with the desired symmetry and construct the metric from it.  Fixing a coordinate i.e. considering a submanifold typically excludes symmetries.  Thus, though we were certainly at rights to consider submanifolds previously, the symmetries that remained were not of interest to us.  \\ \\
This is not the case for the dynamical mass manifold (\ref{DynMassMet}) - it contains maximally symmetric submanifolds.  The metric (\ref{DynMassMet}) has full $(conf_{d+2})_{\mathbb{C}}$ symmetry and the submanifold $x_d = 0$, $y_1 = 0$, $...$, $y_{d-1} = 0$, $\beta = 0$, $\alpha = 0$  i.e. one complexified position and the rest real, has $(conf_{d+1})_{\mathbb{C}}$ symmetry.  Since the submanifold is $d+2$-dimensional it is maximally symmetric. \\ \\
 In the sections that follow we will show some of the interesting submanifolds of (\ref{DynMassMet}) that are studied in the literature.  In deriving them we will present two perspectives.  First we will show how the metrics of the submanifolds may be constructed explicitly from quantum states.  This will identify the dual quantum states (and thus specifically the boundary states) of these existing works explicitly.  Second we will derive these metrics by restricting the K\"ahler manifold to a submanifold.  We believe that this second perspective is key to furthering the higher dimensional dictionaries in a systematic way, our topic of discussion in section \ref{SubmanDict}.  

\subsection{The Maximally Symmetric Submanifold}

To emphasise its importance and its role as the radial coordinate on the submanifolds we now relabel $y_d \rightarrow r$.  We first consider the maximally symmetric submanifold.  The submanifold metric may be constructed from the states
\begin{equation}
|r, t, \vec{x}, \zeta) \equiv e^{-r P_{d}} e^{i t H} e^{i \vec{x}\cdot\vec{P} } e^{i \zeta N} N^{-\frac{1}{2}}\theta(N)|\vec{x} = \vec{0}, \zeta = 0) \label{RegOverlap}
\end{equation}
where $r \neq 0$ and the components of $\vec{P}$ run from $1$ to $d-1$.  These states only have one additional coordinate (compared to (\ref{zetaStates})) for regularisation and possess the full set of symmetries generated by the operators in the root diagram, Fig. (\ref{cgRoot}) dimension $d+1$ i.e. $(conf_{d+1})_{\mathbb{C}}$.  The overlap of these states can be calculated, in identical fashion to (\ref{ConfOLap}), as
\begin{equation}
(r', t', \vec{x}', \zeta'|r, t, \vec{x}, \zeta) = \frac{1}{2}\pi^{-\frac{d}{2} - 1} \  \Gamma\left( \frac{d}{2}\right) \  (-(r+r')^2 + (\vec{x} - \vec{x}')^2 - 2(t-t')(\zeta-\zeta'))^{-\frac{d}{2}}.  \label{MaxSymOver}
\end{equation}
Using the overlap (\ref{MaxSymOver}) we find the metric
  \begin{equation}
 ds^2 = \frac{d}{4 r^2}( dr^2 - d\vec{x}^2 + dt d\zeta  ).   \label{MaxSymmSub}
 \end{equation}
As mentioned, this $d+2$-dimensional metric possess $\frac{1}{2}(d+2)(d+3)$ isometries so that it is maximally symmetric.  This, together with its negative scalar curvature, implies that this manifold is $AdS_{d+2}$. \\ \\
The calculation from (\ref{RegOverlap}) is a direct calculation of the metric (\ref{MaxSymmSub}) from the family of quantum states.  It is useful to think from another perspective namely that the metric (\ref{MaxSymmSub}) is the metric of a submanifold of the K\"ahler manifold (\ref{DynMassMet}).  It is the $x_d = 0$, $y_1 = 0$, $...$, $y_{d-1} = 0$, $\beta = 0$, $\alpha = 0$ submanifold of (\ref{DynMassMet}).  
 
 \subsection{The $z=2$-Symmetric Metrics}
 
 The symmetry of (\ref{MaxSymmSub}) is conformal.  If we are interested in metrics that only possess a smaller set of symmetry, e.g. only the Schr\"odinger symmetry, we have to restrict the dynamical symmetry in some way.  We define
 \begin{equation}
|r, t, \vec{x}, \zeta)_\alpha =  e^{-r P_{d}} e^{i t H} e^{i \vec{x}\cdot\vec{P} } e^{i \zeta N} e^{-\alpha N} N^{-\frac{1}{2}}\theta(N)|\vec{x} = \vec{0}, \zeta = 0)  \label{alphaStates2}
 \end{equation}
where we do not consider $\alpha$ as a coordinate.  It is thus only a parameter and fulfils the role of a symmetry filter, similar to the role played by the density matrix in section \ref{DMatSymmFilter}.  The rationale for doing this is that only the \Sc{d} operators of Fig (\ref{cgRoot}) commute with $N$, see Fig. (\ref{cgRoot}).  The factor of $e^{-\alpha N}$ thus breaks all the undesired conformal symmetries.  The overlap can be calculated using the steps in (\ref{ConfOLap}) and is given by
 \begin{eqnarray}
 & &\left._{\alpha} ( r', t', \vec{x}', \zeta'| r, t, \vec{x}, \zeta )_\alpha \right. \nonumber \\
& = &\frac{\pi^{-\frac{d}{2} - 1}}{2} \  \Gamma\left( \frac{d}{2} \right) \left(-(r+r')^2 + (\vec{x} - \vec{x}')^2 - 2(t - t')(\zeta - \zeta' + 2 i \alpha) \right)^{-\frac{d}{2}} \label{RegOverlap2}
 \end{eqnarray}
 which produces the metric
  \begin{equation}
 ds^2 = \frac{d}{4 r^2}(dr^2 - d\vec{x}^2 + dt d\zeta  ) + \frac{d\alpha^2}{2 r^4}dt^2.   \label{McGreevyMet}
 \end{equation}
 This is, up to rescaling of the coordinates and a change of signature, precisely the metric studied by \cite{McGreevyGravDual} and \cite{Son} for $z=2$. The procedure we have employed here is pretty much identical to that of \cite{Son} namely to start from something with full conformal symmetry and break it in an appropriate way to only retain the Schr\"odinger symmetry.  In their study the symmetries are broken on the level of the metric while here we provide the analogous procedure on the level of the quantum states.  \\ \\
  The resulting metric (\ref{McGreevyMet}) is, of course, already known so that it does not present a new result itself.  What we have managed to do with the construction is to provide an explicit and systematic procedure to construct these metrics from the quantum states of the free particle.  This strengthens the dictionary developed by these works and allows us to test aspects of this dictionary very directly.  As an example of this we provide our own perspective on the apparent two additional dimensions which is a novelty of this $AdS/NRCFT$ correspondence. \\ \\
	As is stated in \cite{McGreevyGravDual} the additional dimension $\zeta$ is the conjugate mass.  Unlike the conventional wisdom of $AdS/CFT$ the regularisation parameter $r$ is not associated with an energy scale but rather with a length scale.  The reason for the two additional dimensions (when comparing the $d+2$-dimensional manifold to the \ssc{d} algebra) is also very clear - the Schr\"odinger algebra in $d$ dimensions needs to be viewed, in this construction, as a subalgebra of the $d+1$ dimensional conformal algebra after introducing the conjugate mass.  To the conformal algebra one may add, in line with the standard wisdom, one additional dimension to regularise the metric in the bulk.  These two steps then comprise the apparent two dimensions one needs to add to the non-relativistic theory. \\ \\
	We may once more view the metric (\ref{McGreevyMet}) as being constructed directly from the quantum states (\ref{alphaStates2}) or simply as the metric on a submanifold of the K\"ahler manifold (\ref{DynMassMet}).  Here it is the $\left\{ x_d = 0, y_1 = 0, ..., y_{d-1} = 0, \beta = 0, \alpha = \textnormal{fixed but non-zero} \right\}$ slice of the K\"ahler manifold (\ref{DynMassMet}).  
 
 \subsection{The $z\neq2$ Symmetric Metrics}
 
 The $z \neq 2$ conformal Galilei algebra (\ref{SzComm}) is another subalgebra of the complex conformal algebra.  An important difference to the $z=2$ algebra is that the special conformal generator, $K$, is not included.  We may again alter the states appropriately in order to filter out only the conformal Galilei symmetry as dynamical symmetry.  From the states (\ref{alphaStates2}) we have to filter out the special conformal symmetry generated by $K$. \\ \\
In order to break the special conformal symmetry of (\ref{alphaStates2}) it is useful to understand exactly why the states still possess special conformal symmetry.  The factor $e^{-r P_{d}}$ in (\ref{alphaStates2}) commutes will all the position, momentum and rotation operators and with $H$.  The operators $D$ and $K$ are thus the only Schr\"odinger algebra generators that induce non-trivial transformations on $r$.  If we thus change the factor $e^{\alpha}N$ to $e^{\alpha f(r) N}$ we could thus break the $D$ (scaling) and $K$ (special conformal) symmetry.  We must pick $f(r)$ in such a way that it breaks the special conformal symmetry but retains the scaling symmetry.   We thus make the ansatz for the states 
\begin{equation}
|r, t, \vec{x}, \zeta)_{\alpha_r} = e^{-r P_{d}} e^{i t H} e^{i \vec{x}\cdot\vec{P} } e^{i \zeta N} e^{-\alpha r^{2-z} N} N^{-\frac{1}{2}}\theta(N)|\vec{x} = \vec{0}, \zeta = 0).
\end{equation}
Following the step of (\ref{ConfOLap}) this leads to the overlap
\begin{eqnarray}
& & \left._{\alpha_r}( r', t', \vec{x}', \zeta' | r, t, \vec{x}, \zeta)_{\alpha_r} \right. \nonumber \\
& = & \left(-(r+r')^2 + (\vec{x} - \vec{x}')^2 - 2(t - t')(\zeta - \zeta' + i \alpha(r^{2-n} + r'^{2-n} ) \right)^{-\frac{d}{2}}
 \end{eqnarray}
 and then the metric
 \begin{equation}
  ds^2 = \frac{d}{4 r^2}(dr^2 - d\vec{x}^2 + dt d\zeta  ) + \frac{d\alpha^2}{2 r^{2z}}dt^2.  \label{zn2Met}
 \end{equation}
 This is the metric studied by \cite{McGreevyGravDual} for arbitrary $z$ and is symmetric under the \cg{d} transformations.  \\ \\
Once more the metric (\ref{zn2Met}) may be viewed as the metric of a submanifold of the K\"ahler manifold (\ref{DynMassMet}), but one where the slice is now dependent on the $AdS$ radial coordinate.  Specifically we have to consider the $\left\{ x_d = 0, \chi_1 = 0, ..., \chi_{d-1} = 0, \beta = 0, \right.$ $\left. \alpha = \alpha' r^{2-z} \right\}$ submanifold.  

\subsection{Discussion of the Submanifold Metrics}

In the works \cite{McGreevyGravDual}, \cite{Son}, \cite{Wen} the metrics are taken as the starting point to their analysis which involves the calculation of the correlation functions.  Their calculations invariably involve considering the massive scalar field on the background of interest and applying the conventional $AdS/CFT$ dictionary i.e. that the functional derivative with respect to the boundary value of the fields yields the correlation functions.  \\ \\
In the dictionaries we have developed for the families of states with complex coordinates this is, of course, not the interpretation we have attached to the fields.  In our dictionaries the fields are interpretable directly as the expectation values of operators.  It thus seems that in order to recast the works \cite{McGreevyGravDual}, \cite{Son}, \cite{Wen} in the mould of a systematic set of tools we will have to undertake the study of a whole new dictionary.  \\ \\
Nonetheless, the fact that we could derive the metrics (\ref{MaxSymmSub}), (\ref{McGreevyMet}) and (\ref{zn2Met}) systematically, starting from a family of quantum states achieves two things.  Firstly, it supplements these works by identifying the quantum states on the boundary explicitly and makes clear the interpretations of the coordinates and the scalar curvature.  \\ \\
Secondly, it hints that a systematic construction of these dual systems may, in fact be possible.  This will certainly require a lot of additional work, but we believe that this can be done.  In the final section of this chapter we will discuss a possible strategy for doing this that utilises the existing dictionaries for the complex coordinate states.  We warn the reader that this is simply speculation and must not be viewed as a rigorous discussion.   
 
\section{Dictionary on the Submanifold?}

\label{SubmanDict}

The two perspectives on the metrics we have constructed in this chapter is to either construct them explicitly from the quantum states or consider the metric as the metric of a submanifold of a K\"ahler metric.  It can be summarised in diagrammatic form as follows \\
\begin{center}
\begin{tikzpicture}
  \matrix (m) [matrix of math nodes,row sep=3em,column sep=3em,minimum width=3em]
  {
     \left( \begin{array}{c} \textnormal{Family of States}, \\ \textnormal{Complex coordinates} \\  |z_1, ..., z_n) \end{array} \right) & \left(\begin{array}{c} \textnormal{K\"ahler Geometry} \\ \textnormal{Existing Dictionary}  \end{array} \right) \\
       \left( \begin{array}{c} \textnormal{Family of States} \\  |s_1, ..., s_m) \end{array} \right) &   \left( \begin{array}{c} \textnormal{Metric of Lower Dimension} \\  \textnormal{No Current Dictionary} \end{array} \right) \\};
  \path[-stealth]
    (m-1-1) edge node [left] {fixed} (m-2-1)
						edge node [right] {coordinates} (m-2-1)
            edge node [below] {} (m-1-2)
    (m-2-1.east|-m-2-2) edge node [below] {}
            node [above] {} (m-2-2)
    (m-1-2) edge node [right] {submanifold} (m-2-2);
\end{tikzpicture}
\end{center}
The important observation is that we have an existing dictionary for the complex coordinate states and the K\"ahler geometry.  As discussed, the expectation values of operators are intricately linked to vector fields on the manifold in this case.  This formed the basis of the dictionaries we developed.  \\ \\
The submanifold is not necessarily K\"ahler so that this very useful relation no longer holds.  However, it is useful to recall what quantities we are interested in.  Our goal is to calculate the expectation value of an arbitrary string of operators.   Suppose then we are given a family of states $|\vec{s})$ that can be obtained from a family of states of complex coordinates $|\vec{z})$ by fixing some its coordinates.  Suppose further that we wish to calculate the expectation values of some string of operators $\Phi = \langle \vec{s} | A_1, ..., A_j|\vec{s}\rangle$.  \\ \\
We cannot at this point say anything regarding the equations of motion for $\Phi$.  What we have derived are the equations of motion for $\Phi' = \langle\vec{z} | A_1, ..., A_j|\vec{z}\rangle$ in section \ref{EOMHigherSection}.  Though there is certainly some work still needed to deal with the full set of equations of motion (which entails understanding the role played by the Weyl tensor) this should be possible for these complex coordinates.  \\ \\
What one will have to do, in order to write down the appropriate equations of motion for $\Phi$, is to take the equations of motion for $\Phi'$ and restrict these to the submanifold.  One should then be able to recover the appropriate equations of motion for $\Phi$ from these and then try to match them up to an appropriate theory of gravity.  This is certainly an avenue of future research that begs pursuing after which, hopefully, one may start to develop systematic dictionaries for the dual of quantum systems and theories of gravity where the metric is given by (\ref{McGreevyMet}), for instance.   

\section{Discussion}

Our investigations in this chapter has met with partial success.  By considering the central extension of the Schr\"odinger algebra (the mass) as a dynamical variable we were able to sidestep some of the difficulties we encountered in chapter \ref{HighDim} for the higher dimensional duals.  The field equations were simple enough that we could package them in a simple dictionary involving the massive scalar field.\\ \\
One stumbling block that remains is that the manifolds are not conformally flat which adds terms to the equations of motion.  This can be remedied by only considering one field equation - the trace of the field equations - which has no contribution from the Weyl tensor.  The discarded field equations have to be substituted with boundary conditions.  \\ \\
The dictionary departs from the conventional gauge/gravity duality in that the bulk dimensions are numerous and not just one or two.  We identified that submanifolds of these K\"ahler manifolds are precisely the manifolds investigated in the literature \cite{McGreevyGravDual}, \cite{Son} in the context of the $AdS/NRCFT$ and we also identified an $AdS$ submanifold.  Our construction thus recovers the metrics and supplements these existing works by identifying the dual quantum mechanical states explicitly.  However, because the properties we discussed in section \ref{GeoReform} are no longer applicable on these submanifolds, we could not, as yet, develop a systematic, working dictionary.  \\ \\
We concluded the chapter with a speculative discussion of how one may proceed to develop such a dictionary.  Two ingredients for this development will prove essential - firstly understanding how the equations of motion may be restricted to a submanifold and secondly, understanding not just the trace of the equations of motion but the full set.  The reason for this is simple - the Weyl tensor contributions on the submanifold may be significantly less complex.  The full set of field equations may thus be tractable on the submanifold, though they are not on the K\"ahler manifold.  We hope to address these questions in future.

\chapter{Conclusion and Outlook}

In this thesis we managed to develop a systematic procedure to repackage a given quantum mechanical model as a semi-classical theory of gravity.  The most important part of this construction was the identification of a metric - a way to build a geometry from a given family of quantum states.  The construction we chose was relatively easy to work with but also ensured that the dynamical symmetries are encoded as isometries of the metric and anti-symmetric two-form.  \\ \\
Many aspects of this construction, before theories of gravity even enter the discussion, are intriguing for the purposes of the $AdS/CFT$ correspondence.  For non-normalisable reference states it is necessary to regularise the quantum states by some means.  This gives rise, naturally, to the idea of a bulk, where the states are normalisable, and a boundary where the original non-normalisable states are defined.  In the case of time-evolved states (with complexified time), an asymptotically anti-de Sitter geometry results very generally.  In addition, the $AdS_2$ radial coordinate has the interpretation of an energy scale.  \\ \\
We showed that in the case of normalisable reference states the resulting metric can be much more general.  In particular the $SU(2)$ coherent states and Glauber coherent states resulted in de Sitter and flat space metrics respectively.  Though these geometries were not studied in the same detail as the $AdS$ examples they are, especially in two dimensions, a topic of future research that may yield interesting results in the context of, for instance, the $dS_2/CFT_1$ correspondence.  \\ \\
The procedure from the geometry to the dual gravitational description is not as clear cut precisely because there are several ways in which to do this.  One essentially has to make the first few entries in the quantum mechanics / theory of gravity dictionary - a choice that determines the dictionary one will develop subsequently.  The investigation of choices other than the one we made can well lead to other interesting dictionaries.  For the sake of simplicity we proposed that metric and expectation values of operators should be associated with the metric and fields that solve the field equations of some model of gravity.  After using some results from section \ref{GeoReform} we managed to write down a set of equations of motion for the expectation values of operators (with respect to complex coordinate states).  The gravitational action should thus be chosen so that its field equations are the same as these equations of motion.  \\ \\
Though this was a simple choice, it met with success for especially the two-dimensional manifolds.  When we focused on the $SU(1,1)$ Hamiltonian models, which all produce an $AdS_2$ geometry, we were able to provide clarity to and extend the $AdS_2/CFT_1$ correspondence proposed in \cite{Jackiw}.  We could identify the quantum states of the $AdS_2$ dual explicitly, provide the mapping between quantum state and geometry (and vice versa) and clarify some of the puzzles they discuss in their work.  Specifically, we could explain why the appropriate form of the $2$- and $3$-point correlation functions are produced despite the quantum states being non-normalisable and despite the absence of a conformally invariant state in the Hilbert space.  We showed that this is simply a consequence of the dynamical symmetries.  \\ \\
We went beyond this correspondence of \cite{Jackiw}, which is roughly a dual between the geometry of $AdS_2$ and $CFT_1$ and thus devoid of gravitational content, and identified the model of dilaton gravity as our appropriate gravitational dual.  As it turns out, this model features prominently in the context of the $AdS_2/CFT_1$ correspondence \cite{Cadoni1}-\cite{CadoniCQM}.  Since our construction is systematic and explicit we have direct access to the quantum mechanics / dilaton gravity dictionary.  This enabled us to fill in some details of these existing works pertaining to the interpretation of the dilaton, the interpretation of the dilaton black hole mass, the explicit quantum mechanical model on the boundary and the scalar curvature.  Also, we were able to extend to the dual description of operators that are not generators of symmetry.  Of significance is how natural the analysis follows from our construction.  This holds the promise of generalisations to models that do not possess so many symmetries.  \\ \\
We concluded with an explicit dictionary between conformal quantum mechanics and $2d$ JT-model dilaton gravity.  We showed that the correct expression for the entropy of the dilaton gravity black hole can be recovered by simply considering the conformal transformations of the model.  Our analysis here is still qualitative in that we need address issues pertaining to the appropriate normalisation of the dilaton.  \\ \\
We also briefly looked at the $SU(2)$ Hamiltonians which are dual to theories on de Sitter space.  We showed here that the values of operators can be recovered in the dual gravitational description in an almost identical way to the $SU(1,1)$ calculation.  The Glauber coherent states, which produce a flat space geometry, were also considered.  We commented that the Glauber coherent states are thus best served as a large $j$ limit of the $SU(2)$ coherent states.  \\ \\
Our attention moved, in the last two chapters, to the higher dimensional duals.  Our results here are less developed than the two-dimensional examples we explored previously.  The procedure and considerations for constructing a dual in higher dimensions are identical to the two-dimensional case.  A calculational difficulty is that the Riemann curvature tensor (which determines the equations of motion) for K\"ahler manifolds always has a non-zero Weyl tensor.  This adds terms into the equations of motion that we do not yet know how to handle generally.  The treatment of these terms is, of course, an important avenue of future study.  We thus made a sensible simplification as a first step - we consider only the trace of the equations of motion and consign the information of the other equations of motion to additional boundary conditions.  We identified the massive scalar field as a model that produces the appropriate field equation.\\ \\
The natural generalisation of the $su(1,1)$ coherent states to the higher-dimensional case is to add position.  This extension to the higher-dimensional case caused a problem.  The Killing vectors do not close on the Schr\"odinger algebra because one of the operators of the algebra only generates phase shifts of the quantum states.  Its associated Killing vector is thus zero.  Despite this we managed to develop a scheme for calculating the expectation values of operators.  Understanding the algebraic content of the Laplace operator will give great insight into this matter.  This we postpone to future study.  \\ \\
The situation is a lot more favourable if one considers the central extension of the Schr\"odinger algebra (the mass) as dynamical.  This allows us to enlarge the Schr\"odinger algebra to a real form of the complex conformal algebra.  The Killing vectors are now a differential operator representation of the algebra and the analysis can be done more simply.  \\ \\
Unlike its two-dimensional counterpart these higher dimensional duals do not resemble many similar works in the literature.  The primary reasons for this are that an $AdS$ geometry cannot be achieved and that there are more than one additional dimensions.  The lack of comparable examples in the literature is precisely why this case is not investigated to the same level of detail as its two-dimensional counterpart.  For this case we did provide a simple dictionary. \\ \\
At first glance it may appear as if this construction for the higher dimensions, though possessing a sensible and systematic dictionary, departs from the traditional $AdS/CFT$ approach too much to provide insight therein.  However, we showed explicitly that several important metrics, often the starting point of analysis e.g. \cite{McGreevyGravDual}, \cite{Son}, \cite{Wen}, are easily identifiable submanifolds of the K\"ahler manifold.  We thus supplemented their study by identifying the quantum states on the boundary explicitly, providing the explicit origins of the extra dimensions and an interpretation for the scalar curvature.  In terms of the question of establishing a gravitional dual (as opposed to just calculating the metrics) we currently have no answer.  Since we lose the K\"ahler structure on the submanifold we also lose the dictionary we have developed throughout the rest of the thesis.  We proposed that it is possible that one may utilise the K\"ahler dictionary to learn from the dictionary on the submanifold.  Developing this submanifold dictionary is, in the opinion of the writer, the most pressing question that emerges from the investigation in this thesis.  \\ \\
As was mentioned in the introductory section of this thesis, the power of the construction does not lie in its ability to mimic all the aspects of the gauge/gravity duality but rather in its systematic nature.  It is the hope of this author that this work and its future generalisations can create a systematic framework in which some examples of the gauge/gravity duality may be investigated directly.

\begin{appendices}

\chapter{The Geometric Quantities of Relevance} \label{AppGeo}

In this thesis we will construct a metric and anti-symmetric two-form from a family of quantum states and match these to a semi-classical theory of gravity in some way.  The curvature properties of these metrics, as one would expect, play a key role and we define all the quantities used in the thesis in this appendix.  A good reference for the formulas provided here is \cite{Blau}. \\ \\
Whenever tensor indices are used a chosen geometry is implicitly defined.  The metric tensor $g_{\mu\nu}$ is used to raise and lower indices of tensors e.g.
\begin{equation}
y_{\mu} = g_{\mu\alpha} y^{\alpha} \label{RaiseL}
\end{equation}
where, also throughout the thesis, the Einstein summation convention is used i.e. repeated indices are summed over.  We have illustrated it in (\ref{RaiseL}) for a vector but it is true of any tensor index. The covariant derivative of a vector field $x^\mu$ is defined as
\begin{equation}
\nabla_\nu x^\mu \equiv \partial_\nu x^\mu + x^\alpha \Gamma^\mu_{ \ \alpha\nu} \ \ \ ; \ \ \ \nabla_\nu x_\mu \equiv \partial_\nu x_\mu - x_\alpha \Gamma^\alpha_{ \ \mu\nu} \label{covDer}
\end{equation}
where $\Gamma^{\alpha}_{\ \mu \nu}$ are the Christoffel symbols of the second type defined as
\begin{equation}
\Gamma^{\alpha}_{\ \mu \nu} \equiv \frac{1}{2}g^{\alpha \beta}\left( \partial_\mu g_{\beta \nu} + \partial_\nu g_{\beta \mu} - \partial_\beta g_{\mu\nu}   \right).
\end{equation}
The covariant derivative has the interpretation of how a given vector field changes as its transported around the manifold.  The Christoffel symbol part (\ref{covDer}) indicates that it may also change orientation if the geometry is curved.  Note that, by definition, we have that
\begin{equation}
\nabla_\mu g_{\alpha \beta} = \nabla_{\mu} g^{\alpha \beta} = \nabla_{\mu} g^{\alpha}_{\  \beta} = 0  \label{CompatMet}
\end{equation}
i.e. the metric tensor is compatible with the covariant derivative.  Indeed, technically, metric compatibility or incompatibility defines the covariant derivative and not the other way around.  Throughout the thesis we use the compatible definition of the covariant derivative.  \\ \\ 
The Riemann curvature tensor, $R_{\mu\nu\alpha }^{\ \ \ \ \beta}$, is defined as
\begin{equation}
(\nabla_\mu \nabla_\nu - \nabla_\nu \nabla_\mu) x_\alpha \equiv R_{\mu\nu\alpha }^{\ \ \ \ \beta} x_\beta \label{RiemannTensor}
\end{equation}
and has the interpretation of by how much the orientation of a vector is going to differ if one moves it from point A to point B, infinitesimally close to one another, along two different paths.  The curvature tensor (\ref{RiemannTensor}) contains all curvature information of the manifold.  It can be shown that the curvature tensor possesses the following symmetries
\begin{equation}
R_{\mu\nu\alpha\beta} = -R_{\nu\mu\alpha\beta} = -R_{\mu\nu \beta \alpha} = R_{\alpha\beta \mu\nu}.  
\end{equation}
Consequently there is only one unique, non-trivial contraction of the Riemann tensor
\begin{equation}
R_{\mu\nu} = R^{\alpha}_{ \ \mu \alpha \nu }
\end{equation}
which is the Ricci tensor.  It certainly contains less curvature information than the Riemann tensor but contains sufficient information for some physical applications such as in the Einstein field equations.  The scalar curvature is defined as the trace of the Ricci tensor 
\begin{equation}
R = R^{\alpha}_{\ \alpha}
\end{equation}
while an Einstein manifold has the property that the Ricci tensor is proportional to the metric i.e.
\begin{equation}
R_{\mu\nu} = \frac{R}{d} g_{\mu\nu} \label{EinsteinMetric}
\end{equation}
where $d$ is the dimension of the manifold.  Note that this does not imply that the scalar curvature is necessarily constant.  \\ \\
It can be shown that all two-dimensional ($d=2$) metrics are Einstein (\ref{EinsteinMetric}) while all three-dimensional metrics satisfy
\begin{equation}
R_{\alpha \beta \gamma \delta} =  (R_{\alpha \gamma}g_{\beta\delta} + R_{\beta\delta}g_{\alpha\gamma} - R_{\alpha \delta} g_{\beta\gamma} - R_{\beta \gamma} g_{\alpha \delta} ) - \frac{R}{2} (g_{\alpha\gamma}g_{\beta\delta} - g_{\alpha\delta}g_{\beta\gamma}).  
\end{equation}
These identities make the $2$- and $3$-dimensional case quite novel.  For higher dimensions one has that 
\begin{eqnarray}
R_{\alpha \beta \gamma \delta} & = & \ \ \  W_{\alpha \beta \gamma \delta} \nonumber \\
& & + \frac{1}{n-2}(R_{\alpha \gamma}g_{\beta\delta} + R_{\beta\delta}g_{\alpha\gamma} - R_{\alpha \delta} g_{\beta\gamma} - R_{\beta \gamma} g_{\alpha \delta} ) \nonumber \\
& & -\frac{R}{(n-1)(n-2)} (g_{\alpha\gamma}g_{\beta\delta} - g_{\alpha\delta}g_{\beta\gamma})
\end{eqnarray}
where $W_{\alpha \beta \gamma \delta}$ is the Weyl tensor, which contains information pertaining the conformal properties of the metric.  If the metric is conformally flat i.e. $g_{\mu\nu} = f(x_1, x_2, ..., x_d) \delta_{\mu\nu}$ then the Weyl tensor is zero.  Two- and three-dimensional metrics are thus always conformally flat while the two-dimensional metrics are in addition Einstein metrics.  These simplifications will prove useful in the analysis in the thesis.

\chapter{Field Equations for Einstein- and Dilaton Gravity} \label{FieldEqApp}

\section{Einstein Gravity}

See the works \cite{Wald}, \cite{Hartle} and \cite{Martin} for good discussions on Einstein gravity i.e. general relativity.  In this thesis we will be primarilly interest in the action derivation of the field equations.  We start with the Einstein-Hilbert action (with cosmological constant included)
\begin{equation}
S_{EH} = \int{ d^d x \sqrt{-g}\left( R - 2\Lambda + L_M \right) }.   \label{EHAction}
\end{equation}
Here $g$ is the determinant of the metric, $R$ is the scalar curvature, $\Lambda$ the cosmological constant and $L_M$ the matter content.  The field equations can be derived from (\ref{EHAction}) by varying the action with respect to the inverse metric i.e.
\begin{equation}
\frac{1}{\sqrt{-g}}\frac{\delta S_{EH}}{\delta g^{\mu\nu}} = 0 = \frac{1}{\sqrt{-g}} \int{ d^dx \left( \frac{\delta (\sqrt{-g} R)}{\delta g^{\mu\nu}} -  2\Lambda \frac{\delta \sqrt{-g} }{\delta g^{\mu\nu}} + \frac{\delta (\sqrt{-g} L_M)}{\delta g^{\mu\nu}} \right)    }  \label{FieldDer}
\end{equation}
where the factor of $\frac{1}{\sqrt{-g}}$ is for convenience.  Note that we have are taking a functional derivative.  In order to avoid confusion we make the following remark.  The functional derivative act on functionals such as the action.  Whenever we write 
\begin{equation}
\frac{\delta}{\delta f(x)} \int dx' F(f(x')) = \int dx' \frac{\delta F(f(x'))}{\delta f(x)}
\end{equation}
we mean 
\begin{equation}
\int dx' \frac{\delta F(f(x'))}{\delta f(x)} = \int dx' \partial_f F(f(x')) \delta(x-x') = \partial_f F(f(x)).  
\end{equation}
In other words, the functional derivative on a function introduces a delta function which is integrated out.  Typically when a term is begin integrated over we use the notation for functional derivatives but if not we use the notation of partial derivative.  \\ \\
Returning to the field equations of (\ref{FieldDer}) we first require
\begin{eqnarray}
\frac{\partial \sqrt{-g} }{\partial g^{\mu\nu} } & = & \frac{1}{2\sqrt{-g}} \frac{\partial (-g) }{\partial g^{\mu\nu}} \nonumber \\
& = & -\frac{1}{2\sqrt{-g}} \frac{\partial \det{(g_{\mu\nu})}}{\partial g^{\mu\nu}} \nonumber \\
& = & -\frac{1}{2\sqrt{-g}} \frac{\partial \det^{-1}{(g^{\mu\nu})}}{\partial g^{\mu\nu}} = \frac{1}{2\sqrt{-g}\det^2{(g^{\mu\nu})}} \frac{\partial \det{(g^{\mu\nu})}}{\partial g^{\mu\nu}}.   \label{dsqrtg}
\end{eqnarray}
In order to present the derivation as cleanly as possible we temporarily substitute the variation of the metric as $\partial g^{\mu\nu} \equiv a^{\mu\nu}$.  
By now using properties of the determinant we have that
\begin{eqnarray}
\det{(g^{\mu\nu} + a^{\mu\nu})} & = & \det{(g^{\mu\alpha})}\det{(\delta^{\nu}_\alpha + g_{\alpha\beta}a^{\beta\nu})} \nonumber \\
&=& \det{(g^{\mu\alpha})} \exp{\left(tr(\ln(\delta^{\nu}_\alpha + g_{\alpha\beta}a^{\beta\nu}))\right)} \nonumber \\
& = & \det{(g^{\mu\alpha})}\exp{\left(tr(g_{\alpha\beta}a^{\beta\nu}) + O(a^2)\right)} \nonumber \\
& = & \det{(g^{\mu\alpha})}\left(1 + \det{(g^{\mu\nu})}tr(g_{\alpha\beta}a^{\beta\nu}) + O(a^2) \right)\nonumber \\
& = & \det{(g^{\mu\alpha})} + \det{(g^{\alpha\beta})} g_{\mu\nu}a^{\mu\nu} \nonumber \\
\Rightarrow \partial \det(g^{\mu\nu}) & = & \det{(g^{\alpha\beta})} g_{\mu\nu}a^{\mu\nu}  = \det{(g^{\alpha\beta})} g_{\mu\nu}\partial g^{\mu\nu} \label{dgpa}
\end{eqnarray}
where we have used the fact that $\det(g^{\mu\nu}) = \det(g^{\alpha\beta})$ since it is merely a relabeling. By now combining (\ref{dsqrtg}) and (\ref{dgpa}) we find that
\begin{equation}
\frac{1}{\sqrt{-g}}\frac{\partial \sqrt{-g} }{\partial g^{\mu\nu} }  = \frac{g_{\mu\nu}}{2(-g)\det(g^{\mu\nu})} = \frac{g g_{\mu\nu}}{2(-g)} = -\frac{1}{2} g_{\mu\nu}. \label{Lambdaterm}
\end{equation}
This result is the same for any signature of the metric.  We have derived it here for $|g| = -g$ but an almost identical derivation can be done for $|g| = g$.  Using this we now define the energy momentum tensor $T_{\mu\nu}$ as
\begin{equation}
-T_{\mu\nu} \equiv \frac{1}{\sqrt{-g}}\frac{\partial(\sqrt{-g} L_M)}{\partial g^{\mu\nu}} = \frac{1}{\sqrt{-g}}\frac{\delta \sqrt{-g}}{\partial g^{\mu\nu}}L_M + \frac{\partial L_M}{\partial g^{\mu\nu}} = -\frac{1}{2}g_{\mu\nu} L_M + \frac{\partial L_M}{\partial g^{\mu\nu}}.  \label{TVar}
\end{equation}
The last bit we need to complete the field equations (from (\ref{FieldDer})) is the partial derivative of the scalar curvature
\begin{equation}
\frac{\partial (\sqrt{-g} R)}{\partial g^{\mu\nu}} = \frac{\partial (g^{\alpha\beta}R_{\alpha\beta})}{\partial g^{\mu\nu}}  = R_{\mu\nu} + g^{\alpha\beta}\frac{\partial R_{\alpha\beta}}{\partial g^{\mu\nu}}  \label{RVar}
\end{equation}  
which we will be handling slightly differently.  The Ricci scalar is given by tracing over indices of the Riemann tensor which in turn is given in terms of the Christoffel symbols
\begin{equation}
R_{\mu\nu} = R^\alpha_{\ \mu\alpha\nu} \ \ \ ; \ \ \ R^{\alpha}_{\ \mu\beta\nu} = \partial_{\beta}\Gamma^{\alpha}_{\mu\nu} - \partial_{\nu}\Gamma^{\alpha}_{\mu\beta} + \Gamma^{\alpha}_{\beta\lambda}\Gamma^{\lambda}_{\mu\nu} - \Gamma^{\alpha}_{\nu\lambda}\Gamma^{\lambda}_{\mu\beta}.
\end{equation}
By applying the chain rule we calculate $\frac{\partial R_{\mu\nu}}{\partial g^{\mu\nu}} = \frac{\partial R_{\mu\nu}}{\partial \Gamma^\rho_{\mu\nu}} \frac{\partial \Gamma^\rho_{\mu\nu}}{\partial g^{\mu\nu}}$.  We will show that the contribution of this term is merely a surface term so we will not calculate this explicitly.  Instead it will suffice to calculate $\partial R_{\mu\nu} = R_{\mu\nu}(\Gamma^{\rho}_{\mu\nu} + \gamma^{\rho}_{\mu\nu}) - R_{\mu\nu}(\Gamma^{\rho}_{\mu\nu})$ where $\gamma^{\rho}_{\mu\nu} \equiv \delta \Gamma^{\rho}_{\mu\nu}$ is defined as the variation of the Christoffel symbol (and is assumed to be small).  We begin with
\begin{eqnarray}
\partial R^{\alpha}_{\ \mu\beta\nu} & = & \partial_{\beta}\gamma^{\alpha}_{\mu\nu} - \partial_{\nu}\gamma^{\alpha}_{\mu\beta} +  \gamma^{\alpha}_{\beta\lambda}\Gamma^{\lambda}_{\mu\nu} + \Gamma^{\alpha}_{\beta\lambda}\gamma^{\lambda}_{\mu\nu} - \gamma^{\alpha}_{\nu\lambda}\Gamma^{\lambda}_{\mu\beta} - \Gamma^{\alpha}_{\nu\lambda}\gamma^{\lambda}_{\mu\beta} \nonumber \\
& = & \left( \partial_{\beta}\gamma^{\alpha}_{\mu\nu} + \Gamma^{\alpha}_{\beta\lambda}\gamma^{\lambda}_{\mu\nu} - \gamma^{\alpha}_{\nu\lambda}\Gamma^{\lambda}_{\mu\beta}    \right) - \left( \partial_{\nu}\gamma^{\alpha}_{\mu\beta} + \Gamma^{\alpha}_{\nu\lambda}\gamma^{\lambda}_{\mu\beta} - \gamma^{\alpha}_{\beta\lambda}\Gamma^{\lambda}_{\mu\nu}   \right) \nonumber \\
& = &\left( \partial_{\beta}\gamma^{\alpha}_{\mu\nu} + \gamma^{\lambda}_{\mu\nu}\Gamma^{\alpha}_{\beta\lambda} - \gamma^{\alpha}_{\nu\lambda}\Gamma^{\lambda}_{\mu\beta} - \gamma^{\alpha}_{\mu\lambda}\Gamma^{\lambda}_{\nu\beta}   \right) \nonumber \\
& & - \left( \partial_{\nu}\gamma^{\alpha}_{\mu\beta} + \gamma^{\lambda}_{\mu\beta}\Gamma^{\alpha}_{\nu\lambda} - \gamma^{\alpha}_{\beta\lambda}\Gamma^{\lambda}_{\mu\nu} - \gamma^{\alpha}_{\mu\lambda}\Gamma^{\lambda}_{\nu\beta} \right) \nonumber \\
& = & \nabla_\beta\gamma^{\alpha}_{\mu\nu} - \nabla_{\nu}\gamma^{\alpha}_{\mu\beta}
\end{eqnarray}
from which it follows that
\begin{equation}
g^{\mu\nu}\partial R_{\mu\nu} = g^{\mu\nu}(\partial R^{\alpha}_{\mu\alpha\nu}) = g^{\mu\nu} \nabla_\alpha\gamma^{\alpha}_{\mu\nu} - \nabla_{\nu}\gamma^{\alpha}_{\mu\alpha}  = \nabla_{\alpha}( g^{\mu\nu} \gamma^{\alpha}_{\mu\nu} - g^{\mu\alpha}\gamma^{\beta}_{\mu\beta} ). \label{Riccidgamma}
\end{equation}
The metric is compatible with the covariant derivative (\ref{CompatMet}).  This should hold before as well as after the variation i.e.
\begin{equation}
\nabla_\sigma g^{\mu\nu} = 0 = \nabla'_\sigma (g^{\mu\nu} + a^{\mu\nu})
\end{equation}
where $a^{\mu\nu}$ is assumed to be small and the covariant derivatives are varied as $\nabla_\sigma \rightarrow \nabla'_\sigma$.  The variation is assumed small so that we can derive, up to first order
\begin{equation}
\nabla_\sigma a^{\mu\nu} = -g^{\mu\lambda} \gamma^{\nu}_{\sigma\lambda} - g^{\lambda\nu}\gamma^{\mu}_{\sigma\lambda}
\end{equation}
from which we can derive two identities
\begin{eqnarray}
\nabla_\sigma g_{\mu\nu} a^{\mu\nu} & = & -2 \gamma^{\alpha}_{\sigma\alpha} \nonumber \\
\nabla_\sigma a^{\mu\sigma} & = & -g^{\mu\lambda} \gamma^{\sigma}_{\sigma\lambda} - g^{\lambda\sigma}\gamma^{\mu}_{\sigma\lambda }.  
\end{eqnarray}
By combining these 
\begin{eqnarray}
 \nabla_\alpha  g^{\alpha\lambda} \nabla_\lambda g_{\mu\nu} a^{\mu\nu} - \nabla_\alpha\nabla_\sigma a^{\alpha\sigma} &  = & \nabla_\alpha( g^{\mu\nu}\gamma^{\alpha}_{\mu\nu} - g^{\alpha\lambda}\gamma^{\sigma}_{\lambda\sigma}  ) \nonumber \\
 \Rightarrow g^{\mu\nu} \delta R_{\mu\nu} & = & \nabla_\alpha \left( g_{\mu\nu} \nabla^\alpha a^{\mu\nu} - \nabla_\sigma a^{\alpha\sigma}  \right) \equiv \nabla_\alpha v^\alpha.  \label{RicVar}
\end{eqnarray}
By inserting the explicit expressions (\ref{Lambdaterm}), (\ref{TVar}), (\ref{RVar}), (\ref{RicVar}) into (\ref{FieldDer}) we find that
\begin{equation}
0 = \frac{1}{\sqrt{-g}}\frac{\delta S_{EH}}{\delta g^{\mu\nu}} = R_{\mu\nu} - \frac{1}{2}R g_{\mu\nu} + \Lambda g_{\mu\nu} - T_{\mu\nu} +  \int d^d x \sqrt{-g} \frac{\delta}{\delta g^{\mu\nu}} \nabla_a  v^a(\delta g^{\mu\nu}).  \label{EHFieldE1}
\end{equation}
The integral over the divergence of a vector, by Stokes' theorem, can yield at most a boundary term so that the integral does not contribute to the field equations.  The remainder of (\ref{EHFieldE1}) is the Einstein field equations with cosmological constant.

\section{Dilaton Gravity Field Equations}

Another model of gravity that will be of interest is dilaton gravity \cite{Gegenberg3}.  In these models we have included additional fields into the action which can change the field equations significantly.  We will be interested in actions of the following form
\begin{equation}
S = \int d^d x\sqrt{-g}\left( H(\eta) g^{\mu\nu}\nabla_\mu \eta \nabla_\nu \eta +  R D(\eta) + V(\eta, L_M)    \right).  
\end{equation}  
For the field equations resulting from varying with respect to the metric we can apply the results above almost imediately
\begin{eqnarray}
0 & = & \frac{1}{\sqrt{-g}}\frac{\delta S}{\delta g^{\mu\nu}} \nonumber \\
  & = & D(\eta)\left(R_{\mu\nu} - \frac{1}{2} R g_{\mu\nu}\right) + H(\eta)\left(\nabla_\mu \eta \nabla_\nu \eta - \frac{1}{2}g_{\mu\nu}g^{\alpha\beta}\nabla_\alpha \eta \nabla_\beta \eta    \right) \nonumber \\
  &   & - \frac{1}{2} g_{\mu\nu} V(\eta, L_M) + \frac{\partial V(\eta, L_M)}{\partial g^{\mu\nu}} \nonumber \\
	& & + \frac{1}{\sqrt{-g}} \int d^dx \sqrt{-g} \frac{\delta }{\delta g^{\mu\nu}} \left( D(\eta) \nabla_\alpha v^\alpha \right).  \label{MannField}
\end{eqnarray}
The integral term is now no longer a simple divergence of a vector precisely because of the presence of the dilaton.  This means we cannot apply Stokes' theorem and it does not simply contribute a boundary term.  Examining the integral more carefully yields
\begin{eqnarray}
& & \frac{1}{\sqrt{-g}} \int d^dx \sqrt{-g} \frac{\delta }{\delta g^{\mu\nu}} \left( D(\eta) \nabla_\alpha v^\alpha \right) \nonumber \\
& = &  \frac{1}{\sqrt{-g}} \int d^dx \sqrt{-g}  \frac{\delta }{\delta g^{\mu\nu}} \left( g_{\mu\nu}D(\eta) \nabla^2 a^{\mu\nu} - D(\eta) \nabla_\alpha \nabla_\sigma a^{\alpha\sigma}  \right) \nonumber
\end{eqnarray}
which will produce a number of delta functions in the integral.  The metric variations $a^{\mu\nu}$ are contravariant and the gamma functions are thus added.  Integrating by parts will keep the double derivatives unchanged but will switch the sign of the single derivatives.  This will correspond to a covariant derivative.  This then implies that
\begin{equation}
\frac{1}{\sqrt{-g}} \int d^dx \sqrt{-g} \frac{\delta }{\delta g^{\mu\nu}} \left( D(\eta) \nabla_\alpha v^\alpha \right)  = g_{\mu\nu} \nabla^2 D(\eta) - \nabla_\mu \nabla_\nu D(\eta) \label{BoundTerm}
\end{equation}
so that we find the field equations also found in \cite{Mann} after combining (\ref{MannField}) and (\ref{BoundTerm}).  \\ \\
Of particular interest for the $2d$ models is the Jackiw-Teitelboim model which has $H[\eta] = 0$ and $D[\eta] = \eta$ and, due to two dimensions, $R_{\mu\nu} = \frac{R}{2}g_{\mu\nu}$ which yields the field equations
\begin{equation}
- \frac{1}{2} g_{\mu\nu} V(\eta, L_M) + \frac{\partial V(\eta, L_M)}{\partial g^{\mu\nu}} + g_{\mu\nu} \nabla^2 \eta - \nabla_\mu \nabla_\nu \eta = 0.  
\end{equation}

\chapter{Algebras That Feature in This Thesis}  \label{HolsteinApp}

In this appendix we summarise the various algebras that feature in the thesis.  We start with the Heisenberg algebra spanned by the operators $P, X$ and $I$
\begin{equation}
[X, P] = i I \ \ \ ; \ \ \ [X, I] = [P, I] = 0.  \label{HeisenbergAlg}
\end{equation}
The algebra can also be represented in terms of creation and annihilation operators $a^\dag = \frac{1}{\sqrt{2}}(X + i P)$ and $a = \frac{1}{\sqrt{2}}(X - i P)$ which satisfies
\begin{equation}
[a, a^\dag] = I \ \ \ ; \ \ \ [a, I] = [a^\dag, I] = 0.   \label{HeisenbergAlg2} 
\end{equation}
Two of the special unitary groups will also be of importance.  The $su(2)$ algebra reads
\begin{equation}
[J_{z}, J_\pm] = \pm J_{\pm} \ \ \ ; \ \ \ [J_{+}, J_{-}] = 2 J_{z}  \label{su2Algebra}
\end{equation}  
and may be expressed in terms of creation and annihilation operators as \cite{Holstein1} 
\begin{equation}
J_z = a^\dag a - \frac{N}{2} \ \ \ ; \ \ \ J_{+} = \sqrt{N} a^\dag \sqrt{1 - \frac{a^\dag a}{N}} \ \ \ ; \ \ \ J_{-} = \sqrt{N} \sqrt{1 - \frac{a^\dag a}{N}} a   \label{HP1}
\end{equation}
where $N$ is the related to the representation label, the lowest eigenvalue of $J_z$.  In the $N\rightarrow \infty$ limit it is clear that $\frac{J_{+}}{\sqrt{N}} \rightarrow a^\dag$ and $\frac{J_{-}}{\sqrt{N}} \rightarrow a$.  \\ \\
The $su(1,1)$ algebra reads
\begin{equation}
\left[H, D \right] = i H \ \ \ ; \ \ \  \left[K, D \right] = -i K \ \ \ ; \ \ \ \left[H, K \right] = 2 i D.  \label{su11Alg}
\end{equation}
The Cartan-Weyl basis for the $su(1,1)$ algebra is given by
\begin{eqnarray}
& & K_0 = \frac{1}{2}(K + H) \ \ \ ; \ \ \ K_\pm = \frac{1}{2}(K - H) \pm i D \nonumber \\
& & [K_{0}, K_{\pm}] = \pm K_{\pm} \ \ \ ; \ \ \ [K_{-}, K_{+}] = 2 K_{0}. \label{CWBasis}
\end{eqnarray}
The $su(1,1)$ and Heisenberg algebras may be combined to form the Schr\"odinger algebra.  This algebra derives its name from the generators of dynamical symmetry for the free Schr\"odinger equation \cite{NiedererFree} and is the semi-direct sum of the $su(1,1)$ algebra (\ref{su11Alg}) and $d$ copies of the Heisenberg algebra (\ref{HeisenbergAlg}) (related by rotation operators $M_{jk}$).  The commutation relations are
\begin{eqnarray}
 \left[X_j, P_k \right] = -i \delta_{j,k} & ; & \ \ \ \left[K, H\right] = -2 i D   \nonumber \\
\left[ P_j, D \right] = \frac{i}{2} P_j \ \ \  & ; & \ \ \ \left[ X_j, D \right] = -\frac{i}{2} X_j \nonumber \\
\left[ P_j, K  \right] = iX_j  \
 \ \ & ; & \ \ \ \left[ X_j, H \right] = -iP_j \nonumber \\
 \left[K, D \right] = -iK \ \ & ; & \ \ \ \left[ H, D\right] = i H \nonumber \\
\left[ X_j, M_{k l} \right] = i(g^0_{j k} X_l - g^0_{j l} X_k) & ; & \ \ \ \left[ P_j, M_{k l} \right] = i(g^0_{j k} P_l - g^0_{j l} P_k) \nonumber \\
0 & & \textnormal{otherwise}
\end{eqnarray}
where $g^0_{kl}$ is the flat space metric in arbitrary signature.  In this thesis Euclidean signature is used i.e. $g^0_{kl} = \delta_{k l}$.  \\ \\
The conformal Galilei algebra, \cg{d+1}, is a generalisation of the Schr\"odinger algebra where one has included a dynamical exponent $z$. The algebra is also centrally extended and reads
\begin{eqnarray}
\left[ M_{ij}, M_{kl} \right] &=& i( g^0_{ik} M_{j l} + g^0_{j l} M_{i k} - g^0_{i l} M_{j k} - g^0_{j k} M_{i l}   )  \nonumber \\
\left[ G_{i}, M_{k l} \right] = i(g^0_{i k}X_l - g^0_{i l}K_k) \ \ \ & ; & \ \ \ \left[ P_{i}, M_{k l} \right] = i(g^0_{i k}P_l - g^0_{i l}P_k)  \nonumber \\
\left[G_{i} , P_{j} \right] = i \delta_{i j} N \ \ \  & ; & \ \ \ \left[D, P_{i} \right] =  -\frac{i}{2} P_{i} \nonumber \\
\left[ D, G_i \right] = \frac{i}{2}(z-1)G_i \ \ \ & ; & \ \ \ \left[ D, N \right] = \frac{i}{2}(z-2)N \nonumber   \\
\left[ H, G_i \right]  =  -i P_i \ \ \ & ; & \ \ \  \left[ D, H \right] = -\frac{z}{2} i H
\end{eqnarray}
where $N$ is the central extension.  The dynamical exponent $z$ characterises the different scaling behavior of time and position captured by the different scaling for $H$ and $G_i$. \\ \\
 The final algebra that is of importance is the conformal algebra given by
\begin{eqnarray}
\left[\widetilde{D}, \widetilde{K}_\mu \right] & = & i \widetilde{K}_\mu \nonumber \\ 
\left[\widetilde{D}, \widetilde{P}_\mu \right] & = & -i \widetilde{P}_\mu \nonumber \\ 
\left[ \widetilde{P}_\mu, \widetilde{K}_\nu \right] & = & 2 i \widetilde{M}_{\mu\nu} - 2 i g^0_{\mu\nu} \widetilde{D} \nonumber \\
\left[\widetilde{K}_{\alpha}, \widetilde{M}_{\mu\nu} \right] & = & i(g^0_{\alpha\mu}\widetilde{K}_\nu - g^0_{\alpha\nu}\widetilde{K}_\mu) \nonumber \\
\left[\widetilde{P}_{\alpha}, \widetilde{M}_{\mu\nu} \right] & = & i(g^0_{\alpha\mu}\widetilde{P}_\nu - g^0_{\alpha\nu}\widetilde{P}_\mu) \nonumber \\ 
\left[ \widetilde{M}_{\alpha\beta}, \widetilde{M}_{\mu\nu} \right] & = & i( g^0_{\alpha\mu} \widetilde{M}_{\beta\nu} + g^0_{\beta\nu} \widetilde{M}_{\alpha\mu} - g^0_{\alpha\nu} \widetilde{M}_{\beta\mu} - g^0_{\beta\mu}\widetilde{M}_{\alpha\nu}   ). 
\end{eqnarray}
In Appendix \ref{AppConfAlg} it is shown that the conformal Gaililei algebra is a subalgebra of the conformal algebra of one dimension higher.  

\chapter{Baker-Campbell-Hausdorff Formula}

\label{BCHApp}

In this thesis the calculation of quantum state overlaps and transformation induced by a unitary transformation is critical.  For this purpose the Baker-Campbell-Hausdorff formula \cite{Lie1}, \cite{Lie2} is used extensively.  In a nutshell the formula is used to split an arbitrary group element as the product of a desired set of ``basis" group elements.  In this appendix we show two examples of this formula that features in this thesis. 
 
\section{BCH formula for the Heisenberg Group}

The most well-known example of the BCH formula is its application to the Heisenberg group.  The relevant algebra is given in (\ref{HeisenbergAlg}) and (\ref{HeisenbergAlg2}).  Consider the following Heisenberg algebra group element
\begin{equation}
U = e^{z a^\dag - \overline{z} a}.  \label{HeissGroup}
\end{equation}
This group element may be split as the product of the exponent of a creation operator and the exponent of an annihilation operator as follows
\begin{equation}
e^{z a^\dag + \overline{z} a} = e^{z a^\dag} e^{-\overline{z} a} e^{-\frac{1}{2}[z a^\dag, \overline{z} a]} = e^{z a^\dag} e^{\overline{z} a} e^{\frac{I}{2}z \overline{z}}  \label{GlauberExp}
\end{equation} 
following (\ref{HeisenbergAlg2}).  This expansion is possible precisely because $[a^\dag, a] = constant$.  The expansion (\ref{GlauberExp}) is particularly useful when the the group element (\ref{HeissGroup}) acts on the state $|0\rangle$ annihilated by $a$.  In this case
\begin{equation}
e^{z a^\dag - \overline{z} a}|0\rangle = e^{z a^\dag} e^{\overline{z} a} e^{\frac{I}{2}z \overline{z}} |0\rangle = e^{\frac{I}{2}z \overline{z}}e^{z a^\dag}|0\rangle
\end{equation}
so that the $e^{\overline{z} a}$ part of the expansion induces no transformation of $|0\rangle$.  Another useful example is if a Heisenberg group element acts on the state $|x=0)$ which is such that $X|x=0) = 0$.  We then have that
\begin{equation}
e^{i a P + i b X}|x = 0) = e^{i a P} e^{i b X} e^{-\frac{1}{2}a b [P, X]}|x = 0) = e^{i \frac{a b}{2} I} e^{i a P}|x = 0). \label{xpSplit}
\end{equation}

\section{BCH formula for $SU(1,1)$}

As a second example we consider the $BCH$ formulae for the $SU(1,1)$ group.  A simple way to derive the formula is to consider the $2\times 2$ matrix representation of $SU(1,1)$ given by
\begin{equation}
K_{+} = \left(\begin{array}{cc} 0 & 1 \\ 0 & 0 \end{array} \right) \ \ \ ; \ \ \ K_{-} = \left(\begin{array}{cc} 0 & 0 \\ -1 & 0 \end{array} \right) \ \ \ ; \ \ \ K_0 = \left(\begin{array}{cc} \frac{1}{2} & 0 \\ 0 & -\frac{1}{2} \end{array} \right).  \label{2b2Rep}
\end{equation}
It can be verified that the matrices (\ref{2b2Rep}) satisfies the correct commutation relations (\ref{CWBasis}).  The BCH formula is representation independent so that, even though we are considering a simple $2\times2$ matrix representation, the formula is valid for any representation.  Given a general matrix element of $SL(2)$ in the $2\times2$ matrix representation
\begin{equation}
\left( \begin{array}{cc} \alpha & \beta \\  \gamma & \delta \end{array}  \right) \ \ \ ; \ \ \ \alpha\delta - \beta\gamma = 1   \label{sl2Gen}
\end{equation}
it may be broken up as
\begin{eqnarray}
\left( \begin{array}{cc} \alpha & \beta \\  \gamma & \delta \end{array}  \right) & = & \exp{\left(\frac{\beta}{\delta} K_{+}\right)} \exp{\left(-2\log{(\delta)}K_{0}\right)}\exp{\left(-\frac{\gamma}{\delta}K_{-}\right)} \ ; \ \ \ \delta \neq 0 \label{KpKmSplit} \\
 & = & \exp{\left(-\frac{\gamma}{\alpha}K_{-}\right)} \exp{\left( 2\log{(\alpha)} K_0 \right)} \exp{\left(\frac{\beta}{\alpha}K_{+}\right)} \ \ \ ; \ \ \ \alpha \neq 0.  \label{KmKpSplit}
\end{eqnarray}
$SU(1,1)$ is a subgroup of $SL(2)$ so that the above formulas holds for $SU(1,1)$ elements also.  As an application of the above we calculate the overlap of the harmonic oscillator states
\begin{equation}
|t) \equiv e^{i t (H + \omega^2 K)}|x = 0) = e^{i t (H + \omega^2 K)}e^{-K_{+}}|0\rangle   \label{BCHHO}
\end{equation}
where $|0\rangle$ is the state such that 
\begin{equation}
K_{-}|0\rangle = 0 \ \ \ ; \ \ \ K_0 |0\rangle = k   \label{ZerProps}
\end{equation} 
where $k$ is the representation label.  The overlap of the states (\ref{BCHHO}) is given by
\begin{equation}
(t'| t) = \langle 0 | e^{-K_{-}}e^{i (t-t') (H + \omega^2 K)}e^{-K_{+}} |0\rangle.  \label{SU11BCHOp}
\end{equation}
The $SU(1,1)$ element wedged between the states $\langle 0|$ and $|0\rangle$ may be split up using (\ref{KpKmSplit}).  The $2\times 2$ matrix representation of (\ref{SU11BCHOp}) is given by
\begin{eqnarray}
& & e^{-K_{-}}e^{i (t-t') (H + \omega^2 K)}e^{-K_{+}} \nonumber \\
& = & \left(\begin{array}{cc} \cos(\omega(t - t')) + \frac{i(1+\omega^2)\sin(\omega(t - t'))}{2\omega} & -\cos(\omega(t - t')) - \frac{i \sin(\omega(t- t'))}{\omega} \\ +\cos(\omega(t - t')) + \frac{i \sin(\omega(t- t'))}{\omega} & -\frac{2 i \sin(\omega(t-t'))}{\omega}  \end{array}\right).   \nonumber
\end{eqnarray}
Now, after applying (\ref{KpKmSplit}) we will have that $\langle 0 |\exp{\left(\frac{\beta}{\delta} K_{+}\right)} = 1$ and $\exp{\left(-\frac{\gamma}{\delta}K_{-}\right)}|0\rangle = 1$.   Only the $\exp{\left(-2\log{(\delta)}K_{0}\right)}|0\rangle = \exp{\left(-2 k \log{(\delta)}\right)}|0\rangle$ factor contributes to the overlap.  We thus find that
\begin{equation}
(t'| t) = \left(-i\frac{2 \sin(\omega(t-t'))}{\omega} \right)^{-2k}  \label{HOLapCalc}
\end{equation}
We may also break the element (\ref{sl2Gen}) as
\begin{eqnarray}
& &\left( \begin{array}{cc} \alpha & \beta \\  \gamma & \delta \end{array}  \right) \nonumber \\
& = & \exp{\left(\frac{\alpha - \delta - \beta + \gamma}{\alpha + \delta - \beta - \gamma} H\right)} \exp{\left(-2 i \log{\left(\frac{1}{2}(\alpha + \delta - \beta - \gamma)\right)} D\right)} \times \nonumber \\
& &\exp{\left(\frac{\alpha - \delta + \beta - \gamma}{\alpha + \delta - \beta - \gamma}K\right)} \label{KTDsplit} \\
& = & \exp{\left(\frac{\alpha - \delta + \beta - \gamma}{\alpha + \beta + \gamma + \delta}K\right)}\exp{\left(2i\log{\left(\frac{1}{2}(\alpha + \beta + \gamma + \delta) \right)}D\right)} \times \nonumber \\
& & \exp{\left(\frac{\alpha - \delta - \beta + \gamma}{\alpha + \beta + \gamma + \delta}H \right)} \label{TKDsplit}
\end{eqnarray}
with the restrictions $\alpha + \delta - \beta - \gamma \neq 0$ and $\alpha + \delta + \beta + \gamma \neq 0 $ respectively. \\ \\ 
The formula (\ref{KTDsplit}) is particularly useful when the $SU(1,1)$ element acts on the state
\begin{equation}
|x) \equiv e^{i x P}|x=0) = e^{i x P}e^{-K_{+}}|0\rangle.  
\end{equation} 
Two properties of the state $|x=0)$ are useful
\begin{equation}
K|x=0) = 0 \ \ \ ; \ \ \ e^{i D} |x=0) = e^{k}|x=0)
\end{equation}
where $k$ is the representation label.  These may be derived using the BCH formulas and the properties (\ref{ZerProps}).  One then has that
\begin{equation}
e^{i a D} |x) = e^{i a D} e^{i x P} e^{-i a D} e^{i a D}|x = 0) = e^{ a k} e^{i e^{\frac{a}{2}} x P} |x=0) = e^{a k} |e^{\frac{a}{2} }x )  \label{kTrans}
\end{equation}
and 
\begin{equation}
e^{i a K} |x) = e^{i x (P + a X)} e^{i a K}|x = 0) = e^{i \frac{1}{2}x^2 a} e^{i x P}|x) = e^{i \frac{1}{2}x^2 a} |x)  \label{dTrans}
\end{equation}
after using (\ref{xpSplit}).  \\ \\
As an application of the above we now present a procedure for calculating the transformation induced by an arbitrary $SU(1,1)$ group element on the free particle state
\begin{equation}
|t, x) \equiv e^{i t H} e^{i x P}|x = 0).   \label{txStatesDef}
\end{equation}
Note that, in the terminology of the $su(1,1)$ discussion, $k=\frac{1}{4}$.  Consider a general $SU(1,1)$ group element $U$ acting on the state (\ref{txStatesDef}).  Split the term $U e^{i t H}$ according to (\ref{KTDsplit}).  This yields a term of the form $e^{i t' H} e^{i a_1 D} e^{i a_2 K}$.  The action of $e^{i a_1 D} e^{i a_2 K} $ on $e^{i x P} |x=0)$ can then be calculated using (\ref{kTrans}) and (\ref{dTrans}).  This will always yield a term of the form
\begin{equation}
|t', x') = (f(t', x'))^{-1} e^{i t' H} e^{i x' P} |x = 0)
\end{equation}
from which the induced transformation can be read off. As an explicit example consider the special conformal transformation generated by $K$ on the free particle state.  We start by factorising 
\begin{equation}
e^{i \alpha K} e^{i t H} = \exp\left(i\frac{t}{1 - \alpha t} H\right) \ \exp\left(-2i\log(1 - a t)D\right) \ \exp\left( i \frac{\alpha}{1 - \alpha t}K \right)  
\end{equation} 
according to (\ref{KTDsplit}).  By acting with the last two operators on the state $|x)$ we have from (\ref{kTrans}) and (\ref{dTrans}) that
\begin{eqnarray}
& &\exp\left(-2i\log(1 - a t)D\right) \ \exp\left( i \frac{\alpha}{1 - \alpha t}K \right) |x) \nonumber \\
& = & e^{i \frac{\alpha x^2 }{2(1 - \alpha t)}} e^{-2k\log(1 - a t)} \left| \frac{x}{1 - a t} \right)  \label{SCTrans}
\end{eqnarray}
which is the expression that appears in (\ref{Schr11Coord}) when we set $k=\frac{1}{4}$.  

\chapter{The Subalgebras of the Complex Conformal Algebra}  \label{AppConfAlg}
 
 In this appendix we will show in particular how the \cg{d+1} and \Sc{d+1} algebras may be viewed as real forms of the complex conformal group in $d+2$ dimensions $(conf_{d+2})_{\mathbb{C}}$.  The discussion herein borrows greatly from \cite{Henkel}.

\section{Relating the Wave Equations}

This connection between the complex conformal group and conformal Galilei group can already be seen on the level of the relevant wave equations (for $d>1$), the free Schr\"odinger equation and the free Klein-Gordon equation.  We start with the free particle Schr\"odinger equation in $(d+1)$ dimensions
\begin{equation}
-2m i \partial_t \phi + \partial^2_{\vec{x}} \phi = 0 \label{SchrW}
\end{equation}
and apply the prescription by Giulini \cite{Giulini} which is to treat the mass as a dynamical variable.  This can be achieved on the level of the wavefunction by performing a Fourier transform.  We define the new function $\psi(\zeta, t, \vec{r})$ by
\begin{equation}
\psi(\zeta, t, \vec{r}) \equiv \int_0^\infty dm e^{i m \zeta}\phi_m( t, \vec{r})  \label{GiuliniState}
\end{equation}
where $\phi_m( t, \vec{r}) $ is the function in (\ref{SchrW}) but now has the subscript $m$ to remind us of its explicit dependence on mass.  The free Schr\"odinger equation now becomes
\begin{eqnarray}
0 & = & -2 \int^\infty_0 dm e^{i m \zeta} (i m) \partial_t\phi_m +  \int^\infty_0 dm e^{i m \zeta} \partial^2_{\vec{x}} \phi_m \nonumber \\
  & = & -2 \partial_t \partial_\zeta \psi + \partial^2_{\vec{x}}\psi. \label{SchrConf}
\end{eqnarray}
The full set of dynamical symmetry generators of equation (\ref{SchrConf}) is a real subalgebra (which contains the \Sc{d+1} algebra) of the complex conformal algebra in $d+2$ dimensions, $(conf_{d+2})_\mathbb{C}$  \cite{Henkel}.  This can already by anticipated by performing the complex coordinate transformation $y_0 = i(t + \zeta)$, $y_1 = t - \zeta$, $y_{\mu} = \sqrt{2} x_\mu$ on (\ref{SchrConf}).  This implies that $\partial_t = i\partial_{y_0} + \partial_{y_1}$, $\partial_\zeta = i\partial_{y_0} - \partial_{y_1}$ and $\partial_{x_\mu} = \sqrt{2}\partial_{y_\mu}$.  The equation (\ref{SchrConf}) now becomes
\begin{equation}
2 (\partial_{y_0}^2 + \partial_{y_1}^2 + \partial_{y_\mu}^2 ) \psi(\vec{y}) = 0 \label{FreeKG}
\end{equation} 
which can be identified as the free Klein-Gordon equation in Euclidean flat space, the symmetry generators of which form the conformal algebra.  Though this matching of the wavefunction is instructive, we will start our analysis with the $(conf_{d+2})_\mathbb{C}$ algebra in its abstract form and show the subalgebra structure explicitly. 

\section{The Conformal Algebra}

The $(d+2)$-dimensional conformal algebra consists of $d+2$ translations, $d+2$ Lorentz transformations, $\frac{1}{2}(d+1)(d+2)$ rotations and one dilitation/scaling.  These are generated by $\widetilde{P}_\mu$, $\widetilde{K}_\mu$, $\widetilde{M}_{\mu\nu}$ and $\widetilde{D}$ respectively and satisfy the commutation relations (\ref{ConformalComm}). 

\section{Classification in Terms of Scaling Properties}

\begin{figure}[ht]
\centering
\begin{pspicture}(11,10)	
%\psaxes{->}(0,0)(-1.2,-1.2)(1.2,1.2)[$x$,0][$y$, 90]	%creates axes
\psdot(9,9)	 \uput[0](9, 9.5){$X_{\left\{ 1, 1\right\} } \propto i \widetilde{K}_0 - \widetilde{K}_1$}
\psdot(9, 5) \uput[0](9, 5.5){$ X_{\left\{ 1, 0\right\} } \propto  i \widetilde{M}_{0 \mu} - \widetilde{M}_{1\mu} $}
\psdot(9, 1) \uput[0](9, 1.5){$X_{\left\{ 1, -1\right\} } \propto i\widetilde{P}_0 - \widetilde{P}_1$}
\psdot(5, 9) \uput[0](5, 9.5){$X_{\left\{ 0, 1\right\} } \propto \widetilde{K}_\mu$}
\psdot[dotstyle=o, dotsize=0.5](5, 5) 
\psdot(5, 5) \uput[0](5, 5.5){$X_{\left\{ 0, 0\right\} } = z_1 \widetilde{D} + z_2 \widetilde{M}$} 
\psdot(5, 1) \uput[0](5, 1.5){$ X_{\left\{ 0, -1\right\} } \propto \widetilde{P}_\mu $}
\psdot(1, 9) \uput[0](-1, 9.5){$X_{\left\{ -1, 1\right\} } \propto i \widetilde{K}_0 + \widetilde{K}_1$}
\psdot(1, 5) \uput[0](-1, 5.5){$X_{\left\{ -1, 0\right\} } \propto i \widetilde{M}_{0\mu} + \widetilde{M}_{1\mu}$}
\psdot(1, 1) \uput[0](-1, 1.5){$X_{\left\{ -1, -1\right\} } \propto i \widetilde{P}_0 + \widetilde{P}_1 $}
 \psline[linecolor=black]{<->}(0,5)(10,5)\uput[0](9.4, 4.7){$e_1$}
\psline[linecolor=black]{<->}(5,0)(5, 10)\uput[0](4.3, 9.4){$e_2$}
\end{pspicture}
\label{confRoot}
\caption{The complex conformal algebra elements classified in terms of their scaling behaviour with respects to $\widetilde{D}$ ($e_1$) and $\widetilde{M}$ ($e_2$).  The rotation operators are suppressed but formally form part of $X_{0, 0}$}
\end{figure}
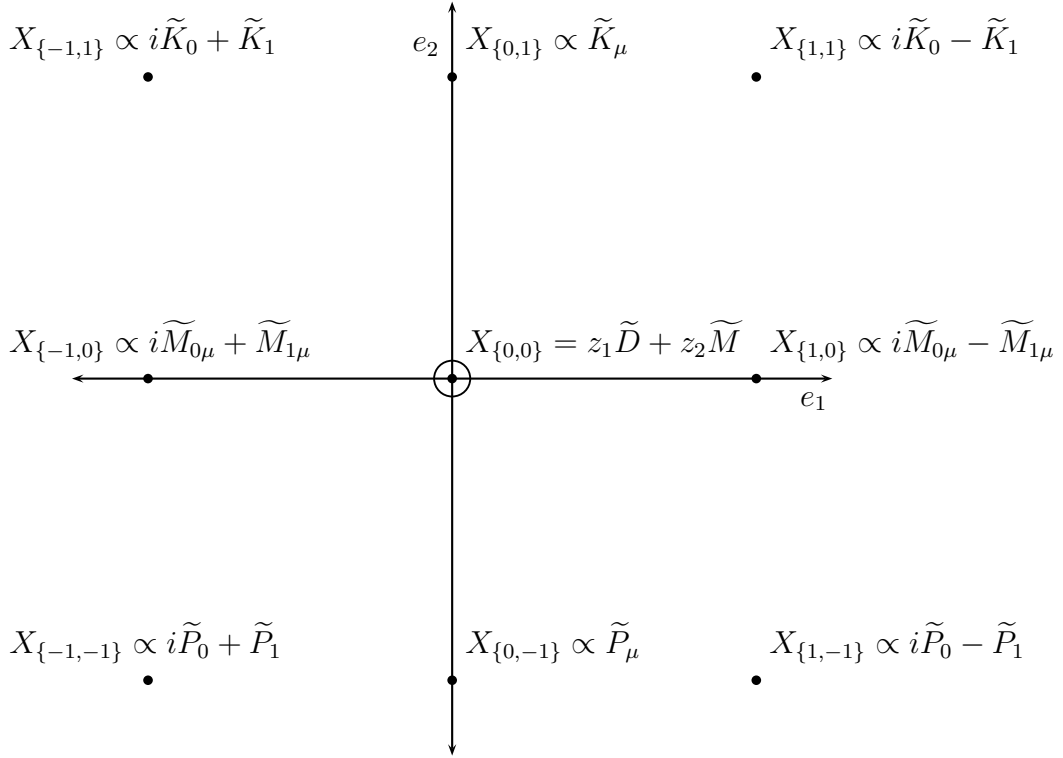

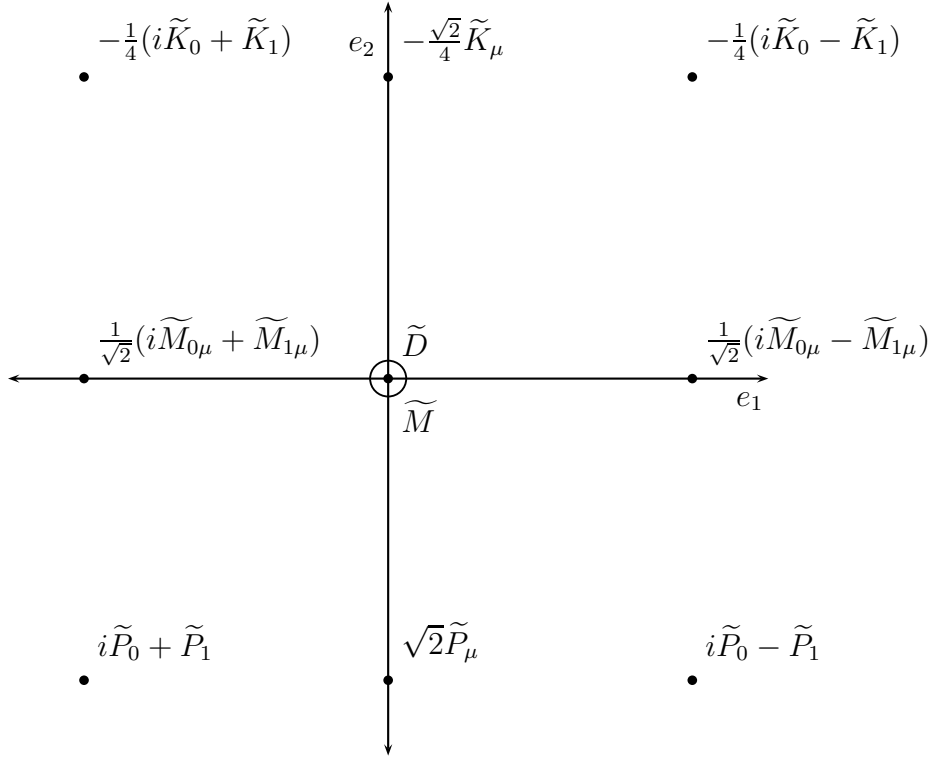
\begin{figure}[ht]
\centering
\begin{pspicture}(11,10)	
%\psaxes{->}(0,0)(-1.2,-1.2)(1.2,1.2)[$x$,0][$y$, 90]	%creates axes
\psdot(9,9)	 \uput[0](9, 9.5){$-\frac{1}{4}(i \widetilde{K}_0 - \widetilde{K}_1)$}
\psdot(9, 5) \uput[0](9, 5.5){$ \frac{1}{\sqrt{2}}(i \widetilde{M}_{0 \mu} - \widetilde{M}_{1\mu} ) $}
\psdot(9, 1) \uput[0](9, 1.5){$ i\widetilde{P}_0 - \widetilde{P}_1$}
\psdot(5, 9) \uput[0](5, 9.5){$-\frac{\sqrt{2}}{4} \widetilde{K}_\mu$}
\psdot[dotstyle=o, dotsize=0.5](5, 5) \uput[0](5, 4.5){$\widetilde{M}$}
\psdot(5, 5) \uput[0](5, 5.5){$\widetilde{D}$} 
\psdot(5, 1) \uput[0](5, 1.5){$ \sqrt{2}\widetilde{P}_\mu $}
\psdot(1, 9) \uput[0](1, 9.5){$-\frac{1}{4}(i \widetilde{K}_0 + \widetilde{K}_1)$}
\psdot(1, 5) \uput[0](1, 5.5){$\frac{1}{\sqrt{2}}( i \widetilde{M}_{0\mu} + \widetilde{M}_{1\mu})$}
\psdot(1, 1) \uput[0](1, 1.5){$ i \widetilde{P}_0 + \widetilde{P}_1 $}
 \psline[linecolor=black]{<->}(0,5)(10,5)\uput[0](9.4, 4.7){$e_1$}
\psline[linecolor=black]{<->}(5,0)(5, 10)\uput[0](4.3, 9.4){$e_2$}
\end{pspicture}
\label{FinalRoot}
\caption{A real form of the complex conformal algebra that contains the conformal Galilei algebra.  The rotation operators are suppressed but formally form part of $X_{0, 0}$}
\end{figure}

We will now proceed to classify the elements of the complex conformal algebra in terms of scaling behaviour and show that the \cg{d+1} algebra can be identified as a subset of a real form of this complex algebra.  We select and complexify one of the rotation operators which, without loss of generality, we choose as $\widetilde{M} \equiv -i \widetilde{M}_{01}$.  
\\ \\
We may now classify the elements of $(conf_{d+2})_\mathbb{C}$ in terms of their scaling with the operators $\widetilde{D}$ and $\widetilde{M}$.  We define $X_{\left\{e_1, e_2 \right\}}$ via $[\widetilde{M}, X_{\left\{e_1, e_2 \right\}}] = i e_1 X_{\left\{e_1, e_2 \right\}}$ and $[\widetilde{D}, X_{\left\{e_1, e_2 \right\}}] = i e_2 X_{\left\{e_1, e_2 \right\}}$.  Clearly any combination of momenta will have $e_2 = -1$, rotations $e_2 = 0$ and boosts $e_2 = 1$ (see (\ref{ConformalComm})).  One can furthermore verify that any operator combination $iA_0 + A_1$ will have $e_1 = -1$, while any combination of $iA_0 - A_1$ will have $e_1 = 1$.  We thus propose the root diagram of Fig. (\ref{confRoot}).
Two things are important to realise when interpreting the figure.  First, the coefficients in front of the operators $ X_{\left\{e_1, e_2 \right\}}$ may still be complex.  We will shortly be focussing on a specific real form of the complex algebra, though one is different than the real form (\ref{ConformalComm}).  Secondly, the operators satisfy the properties  
\begin{equation}
\left[ X_{\left\{ m, n\right\} }, X_{\left\{ m', n'\right\} }  \right] = \left\{ \begin{tabular}{cc} $X_{\left\{ m+m', n+n'\right\} } $ & if \ \ \ $| m + m'| \leq 1 $ \ ; \ $|n+n'| \leq 1 $ \\ $0$  &  \textnormal{otherwise} \end{tabular} \right.
\end{equation}
The first can be checked by calculating the scaling behaviour with $\widetilde{D}$ amd $\widetilde{M}$.  We may thus, after selecting a real form of the complex algebra, build an algebra out of any subset of the diagram that is closed under the horizontal, vertical and diagonal shifts that are included.  Many of these are discussed in \cite{Henkel}.  The ones that we will point out explicitly are \Sc{d+1}, \cg{d+1} and the extended Schr\"odinger algebra $\widetilde{Schr}(d+1)$.
 \\ \\
 Following \cite{Henkel}, we specialise to the real form of $(conf_{d+2})_\mathbb{C}$ packaged in Fig (\ref{FinalRoot})
and $M_{\mu\nu} = \widetilde{M}_{\mu\nu}$.  The coefficients are chosen so that they precisely satisfy the property $[X_{\left\{e_1, e_2 \right\}},  X_{\left\{e_1', e_2' \right\}}] = X_{\left\{e_1+e_1', e_2 + e_2' \right\}}$.  \\ \\
We may identify the \cg{d+1} algebra studied by \cite{McGreevyGravDual} as a sub-diagram with the necessary properties.  We find
\begin{equation}
H = X'_{\left\{ -1, -1\right\} } \ \ \ ; \ \ \ D = \frac{1}{2}\tilde{D} + \frac{1}{2}(z-1)\tilde{M}  \ \ \ ; \ \ \ N = X'_{\left\{ 1, -1\right\} } \ \ \ ; \ \ \ P_i = X'_{\left\{ 0, -1\right\} } \ \ \ ; \ \ \ X_i = X'_{\left\{ 1, 0\right\} }.
\end{equation}
Furthermore the operator $X'_{\left\{ 1, -1\right\} } = K$ when we are considering the algebra in the $z=2 $ case.  This can be checked explicitly by using the commutation relations (\ref{ConformalComm}).  From the diagram it can also be seen clearly why, in the case $z\neq 2$, there isn't a special conformal generator. The operator $X'_{\left\{ 1, -1\right\} }$ cannot be included in that case since, always, $[H, K] = [X'_{\left\{ -1, 1\right\} }, X'_{\left\{ 1, -1\right\} }] = 2i\left(\frac{1}{2}(\widetilde{D} + \widetilde{M})\right)$.  This is only the appropriate commutation relation, $[H, K] = 2 i D $ if  $z=2$.   We can still close the elements under commutation if we allow any linear combination of $\widetilde{D}$ and $\widetilde{N}$.  However, this then refers to a different algebra, named $\widetilde{Schr}_{d+1}$ by \cite{Henkel}. 

\section{Representation as Differential Operators}

To aid both the discussion and the calculation of commutation relationships in the thesis we now state the coordinate realisations (for both the wave equation (\ref{SchrConf}) and the states (\ref{zetaStates}))  of the operators in the above table explicitly.  The $(d+2)$ coordinates of the conformal algebra are indicated by $y_\mu$ while the coordinates of the Schr\"odinger group with dynamical mass are $t, \zeta$ and $d$ coordinates $x_\mu$.  First, the conformal algebra in position representation, in Euclidean flat space, acting on a state with scaling dimension $k$, can be represented as
\begin{eqnarray}
\widetilde{P}_\mu & = & -i \partial_{y_\mu} \nonumber \\
\widetilde{D}_\mu & = & i\vec{y}\cdot\vec{\partial_y} + i k \nonumber \\
\widetilde{M}_{\mu\nu} & = & i(y_\mu \partial_{y_\nu} - y_\nu \partial_{y_\mu}) \nonumber \\
\widetilde{K}_\mu & = & i(\vec{y}\cdot\vec{y} \partial_{y_\mu} - 2 y_\mu \vec{y}\cdot\vec{\partial_y} ) - 2 i k y_\mu. \label{ConfDiff} 
\end{eqnarray}
It can be verified that these differential operators satisfy the commutation relationships (\ref{ConformalComm}).  As already indicated when comparing the free Schr\"odinger equation with dynamical mass (\ref{SchrConf}) and the free Klein-Gordon equation (\ref{FreeKG}) we will be performing the following coordinate transformation
\begin{equation}
y_0 = i(t + \zeta) \ \ \ ; \ \ \ y_1 = t - \zeta \ \ \ ; \ \ \ y_\mu = \sqrt{2} x_\mu
\end{equation}
which then also implies
\begin{equation}
\partial_{y_0} = -\frac{i}{2}(\partial_t + \partial_\zeta) \ \ \ ; \ \ \ \partial_{y_1} = \frac{1}{2}(\partial_t - \partial_\zeta) \ \ \ ; \ \ \ \partial_{y_\mu} = \frac{1}{\sqrt{2}} \partial_{x_\mu}.
\end{equation}
Also note that $\vec{y}\cdot\vec{y} = -4 t \zeta + 2 \vec{x}\cdot \vec{x}$.  Expressed in terms of these coordinates one may thus find for the conformal generators (\ref{ConfDiff})
\begin{eqnarray}
i\widetilde{P}_0 & = & -\frac{i}{2}(\partial_t + \partial_\zeta) \nonumber \\
\widetilde{P}_1 & = & -\frac{i}{2}(\partial_t - \partial_\zeta) \nonumber \\
\widetilde{P}_\mu & = & -\frac{i}{\sqrt{2}}\partial_{x_\mu} \nonumber \\
\widetilde{D} & = & i(t \partial_t + \zeta \partial_\zeta + \vec{x}\cdot \vec{\partial_x}) + i k \nonumber \\
\widetilde{M} \equiv -i\widetilde{M}_{01} & = & i t\partial_t - i\zeta \partial_\zeta \nonumber \\
i\widetilde{M}_{0\mu} & = & -\frac{i}{\sqrt{2}}( t \partial_{x_\mu} + \zeta \partial_{x_\mu} + x_\mu \partial_t + x_\mu \partial_\zeta  ) \nonumber \\ 
\widetilde{M}_{1\mu} & = & \frac{i}{\sqrt{2}}( t \partial_{x_\mu} - \zeta \partial_{x_\mu} - x_\mu \partial_t + x_\mu \partial_\zeta  ) \nonumber \\
\widetilde{M}_{\mu\nu} & = & i(x_\mu \partial_{x_\nu} - x_\nu \partial_{x_\mu}) \nonumber \\
i\widetilde{K}_0 & = & i(\vec{x}\cdot\vec{x}\partial_t + \vec{x}\cdot\vec{x}\partial_\zeta + 2 t^2\partial_t + \zeta^2\partial_\zeta + 2t\vec{x}\cdot\vec{\partial}_x + 2\zeta \vec{x}\cdot\vec{\partial}_x + 2kt + 2k\zeta  ) \nonumber \\
\widetilde{K}_1 & = & i(\vec{x}\cdot\vec{x}\partial_t - \vec{x}\cdot\vec{x}\partial_\zeta - 2 t^2\partial_t + \zeta^2\partial_\zeta - 2t\vec{x}\cdot\vec{\partial}_x + 2\zeta \vec{x}\cdot\vec{\partial}_x - 2kt + 2k\zeta  )\nonumber \\
\widetilde{K}_\mu & = & i( (-2\sqrt{2} t\zeta + \sqrt{2}\vec{x}\cdot\vec{x})\partial_{x_\mu} - 2\sqrt{2} x_\mu(t\partial_t + \zeta \partial_\zeta + \vec{x}\cdot\vec{\partial}_x + k ) ).  
\end{eqnarray}
Linear combinations of these operators fill the table we have put together above and all thus form symmetries of the state $\psi(\zeta, t, \vec{x}) $ (\ref{GiuliniState}).  Explicitly, the ones that may be put together to form the \cg{d+1} group are
\begin{eqnarray}
H = X_{\left\{-1, -1 \right\} } = i \widetilde{P}_0 + \widetilde{P}_1 & = & -i\partial_t \nonumber \\ 
P_i = X_{\left\{0, -1 \right\} } = \sqrt{2} \widetilde{P}_\mu & = & -i \partial_{x_\mu} \nonumber \\
N = X_{\left\{1, -1 \right\} } = i \widetilde{P}_0 - \widetilde{P}_1 & = & -i\partial_\zeta \nonumber \\
D = X_{\left\{0, 0 \right\} } = a \widetilde{D} + b\widetilde{M} & = & i( \frac{z}{2} t \partial_t + \frac{1}{2}(2-z)\zeta\partial_\zeta + \frac{1}{2}\vec{x}\cdot \vec{\partial}_x  )\nonumber \\
X_i = X_{\left\{1, 0 \right\} } = \frac{1}{\sqrt{2}}( i \widetilde{M}_{0\mu} - \widetilde{M}_{1\mu}   ) & = & -i t \partial_{x_\mu} - i x_\mu \partial_\zeta \nonumber \\
K = X_{\left\{1, 1 \right\} } = -\frac{1}{4}(i K_0 - K_1) & = & -i t^2 \partial_t - i t \vec{x}\cdot \vec{\partial}_x - \frac{i}{2}\vec{x}\cdot\vec{x} \partial_\zeta - i k t
\end{eqnarray}
where $K$ is only included if $z=2$.

\end{appendices}

\end{document}